\documentclass[aps,nofootinbib,superscriptaddress,twocolumn,prd,longbibliography]{revtex4-2}
\usepackage{textcomp}

\usepackage{amsmath}
\usepackage{amssymb}
\usepackage{graphicx}
\usepackage{color}
\usepackage{gensymb}
\usepackage{float}
\usepackage{ulem}
\usepackage{empheq} 
\usepackage{multirow,bigdelim}
\usepackage{makecell}
\usepackage[]{xcolor}
\usepackage{hyperref}

\usepackage{threeparttable}

\newcommand{\be}{\begin{equation}}
\newcommand{\ee}{\end{equation}}
\newcommand{\bl}{\begin{align}}
\newcommand{\el}{\end{align}}
\newcommand{\bseq}{\begin{subequations}}
\newcommand{\eseq}{\end{subequations}}

\newcommand{\ve}{\varepsilon}

\newcommand{\g}{\gamma}
\newcommand{\p}{\partial}

\renewcommand{\l}{\lambda}
\newcommand{\di}{\mathrm d}
\def\d{\partial}
\newcommand{\e}{\mathrm e}

\makeatletter
\def\l@subsection#1#2{}
\def\l@subsubsection#1#2{}
\makeatother

\makeatletter

\def\@sect@ltx#1#2#3#4#5#6[#7]#8{%
    \@ifnum{#2>\c@secnumdepth}{%
        \def\H@svsec{\phantomsection}%
        \let\@svsec\@empty
    }{%
        \H@refstepcounter{#1}%
        \def\H@svsec{%
            \phantomsection
        }%
        \protected@edef\@svsec{{#1}}%
        \@ifundefined{@#1cntformat}{%
            \prepdef\@svsec\@seccntformat
        }{%
            \expandafter\prepdef
            \expandafter\@svsec
            \csname @#1cntformat\endcsname
        }%
    }%
    \@tempskipa #5\relax
    \@ifdim{\@tempskipa>\z@}{%
        \begingroup
        \interlinepenalty \@M
        #6{%
            \@ifundefined{@hangfrom@#1}{\@hang@from}{\csname @hangfrom@#1\endcsname}%
            {\hskip#3\relax\H@svsec}{\@svsec}{#8}%
        }%
        \@@par
        \endgroup
        \@ifundefined{#1mark}{\@gobble}{\csname #1mark\endcsname}{#7}%
        \addcontentsline{toc}{#1}{%
            \@ifnum{#2>\c@secnumdepth}{%
                \protect\numberline{}%
            }{%
                \protect\numberline{\csname the#1\endcsname}%
            }%
            #7}
    }{%
        \def\@svsechd{%
            #6{%
                \@ifundefined{@runin@to@#1}{\@runin@to}{\csname @runin@to@#1\endcsname}%
                {\hskip#3\relax\H@svsec}{\@svsec}{#8}%
            }%
            \@ifundefined{#1mark}{\@gobble}{\csname #1mark\endcsname}{#7}%
            \addcontentsline{toc}{#1}{%
                \@ifnum{#2>\c@secnumdepth}{%
                    \protect\numberline{}%
                }{%
                    \protect\numberline{\csname the#1\endcsname}%
                }%
                #8}%
        }%
    }%
    \@xsect{#5}}%

\makeatother

\begin{document}
\title{Renormalization group flow of projectable Ho\v rava gravity in
  (3+1) dimensions}
\author{Andrei O. Barvinsky}
\affiliation{Theory Department, Lebedev Physics Institute, Leninsky Prospect 53, Moscow 117924, Russia}
\affiliation{Institute for Theoretical and Mathematical Physics, Moscow State University, Leninskie Gory, GSP-1, Moscow, 119991, Russia}
\author{Alexander V. Kurov}
\affiliation{Theory Department, Lebedev Physics Institute, Leninsky Prospect 53, Moscow 117924, Russia}
\author{Sergey M. Sibiryakov}
\affiliation{Department of Physics \& Astronomy, McMaster University, Hamilton, Ontario, L8S 4M1, Canada}
\affiliation{Perimeter Institute for Theoretical Physics, Waterloo, Ontario, N2L 2Y5, Canada}

\begin{abstract}
We report a comprehensive numerical 
study of the renormalization group flow of marginal
couplings in $(3+1)$-dimensional 
projectable Ho\v rava gravity. First, we classify all
fixed points of the flow and analyze their stability matrices. We find
that some of the stability matrices possess complex eigenvalues and
discuss why this does not contradict unitarity. Next, we scan over the
renormalization group trajectories emanating from all asymptotically
free fixed points. We identify a unique fixed point giving rise to a
set of trajectories spanning the whole range of the kinetic coupling $\lambda$
compatible with unitarity. This includes the
region $0<\lambda-1\ll 1$ assumed in previous
phenomenological applications. The respective trajectories closely
follow a single universal trajectory, differing only by the 
running of the gravitational coupling. The latter
exhibits non-monotonic behavior along the flow, vanishing both in the
ultraviolet and the infrared limits.  
The requirement that the theory remains weakly coupled along the
renormalization group trajectory implies a natural hierarchy between
the scale of Lorentz invariance violation and a much larger value of
the Planck mass inferred from low-energy interactions.
\end{abstract}

\maketitle

\tableofcontents

\section{Introduction}
Ho\v rava gravity (HG)~\cite{Horava:2009uw} (see
\cite{Mukohyama:2010xz,Sotiriou:2010wn,Barvinsky:2023mrv,Herrero-Valea:2023zex}
for reviews), is an approach to building unitary, local and
renormalizable theory of quantum gravity in four spacetime dimensions.
It considers metric theories with an action 
classically invariant at high energies under anisotropic (Lifshitz)
scaling of time and space, 
\be
\label{scaling}
t\mapsto b^{-d} t\;, \qquad x^i\mapsto b^{-1} x^i\;, \quad i=1,\dots,d \:,
\ee
where $d$ is the number of spatial dimensions and $b$ is an arbitrary
scaling parameter. This allows one to include in the action terms with
higher spatial derivatives which lead to a faster decay of propagators
at high momenta and thereby improve convergence of the loop
integrals. On the other hand, the theory remains quadratic in time
derivatives and avoids  
Ostrogradsky instabilities arising in covariant 
higher-derivative gravity \cite{Stelle:1976gc,Stelle:1977ry}.

For $d>1$ the scaling \eqref{scaling} is incompatible with the full 
diffeomorphism invariances which needs to be restricted to
foliation preserving transformations (FDiff):
\be
\label{FDiff}
t\mapsto t'(t)\;, \qquad x^i\mapsto x'^i(t,{\bf x})\;,
\ee
where $t'(t)$ is a monotonic function.
A consequence of this property is explicit breaking of the Lorentz
invariance which can emerge only as an approximate symmetry at low
energies. While satisfying the stringent experimental constraints on
deviations from Lorentz invariance 
\cite{Kostelecky:2008ts,Liberati:2013xla} is challenging, 
there are several mechanisms that could achieve this goal
\cite{GrootNibbelink:2004za,Pospelov:2010mp,
Pujolas:2011sk,Anber:2011xf,Bednik:2013nxa,Kharuk:2015wga,Baggioli:2024vza}.   

HG appears in two main versions differring by their field content. In
the {\it non-projectable} version \cite{Horava:2009uw,Blas:2009qj} the
lapse function (the time-time component of the metric) is taken to be
fully dynamical, with both time and space dependence. Whereas in the
{\it projectable} HG the lapse is restricted to depend only on
time. Low-energy phenomenology of the non-projectable HG in a certain
region of parameter space can be sufficiently close to that of general
relativity (GR) to pass the observational tests
\cite{Blas:2014aca,EmirGumrukcuoglu:2017cfa}.  
However, its complicated structure in the ultraviolet (UV) has so far
precluded a rigorous proof of perturbative renormalizability, even
though a remarkable cancellation of non-local divergences found in 
Ref.~\cite{Bellorin:2022np} suggests that such proof is plausible.

In this paper we focus on the simpler projectable model. This has been
proven
to be perturbatively renormalizable in any number of spacetime
dimensions~\cite{Barvinsky:2015kil,Barvinsky:2017zlx} and the one-loop
beta-functions for its full set of marginal operators with respect to
the scaling (\ref{scaling}) were computed in $d=2$
\cite{Barvinsky:2017kob} and $d=3$ \cite{towards,3+1}. In both cases
the theory possesses asymptotically free UV fixed points.

The global structure of the renormalization group (RG) flow for
projectable HG in $d=2$ is rather simple and was completely characterized
in Ref.~\cite{Barvinsky:2017kob}. The situation is significantly more
involved in the case $d=3$. Here the flow occurs in a multi-dimensional
parameter space and even finding all its fixed points presents a
non-trivial task. A first step towards elucidating the global
properties of the RG flow of the three-dimensional projectable HG was
made in Ref.~\cite{Barvinsky:2023uir}. It found a family of RG
trajectories flowing from one of the asymptotically free fixed points
in the UV towards an infrared (IR) region in the parameter space where
the kinetic term in the Lagrangian of the theory approaches the same
form as in GR. This IR limit was assumed in most phenomenological
applications
\cite{Mukohyama:2009mz,Izumi:2009ry,Izumi:2011eh,Gumrukcuoglu:2011ef,
Bassani:2024lan}
and corresponds to the regime where projectable 
HG behaves (at least at the classical level) similarly to GR
supplemented with a dust-like matter.
 
In the present work we complete the study of the RG flow in several
ways. 
We focus on the marginal operators with respect to the scaling
(\ref{scaling}) since they form a closed set under renormalization and
dominate the flow in an infinite span of energies above the Lorentz
violating scale. 
First, we numerically find all fixed points of the flow and
classify their properties. Second, we characterize the local properies
of the flow around the fixed points by computing the repsective
stability matrices. Third, we study the global structure of the RG
trajectories emanating from asymptoically free fixed points. We find
that, apart from the family found in \cite{Barvinsky:2023uir}, all
trajectories quickly run into strong coupling, without reaching any
GR-like regime. In other words, the family of
Ref.~\cite{Barvinsky:2023uir} exhausts all RG trajectories connecting
consistent UV asymptotics with phenomenologically interesting IR
behavior. 

It is worth noting that the RG flow reaching a proper IR domain is
necessary but not sufficient condition for phenomenological
viability. Indeed, the full IR limit of projectable HG including the
relevant operators with respect to the scaling (\ref{scaling})
features an instability which can be suppressed only at the expense of
introducing strong coupling \cite{Koyama:2009hc,Blas:2010hb}. In this
paper we do not attempt 
to address this issue and consider the projectable HG as a toy model
for possible RG structure of more complicated versions of the theory
with stable IR dynamics, such as e.g. the non-projectable model.  

The paper is organized as follows. In Sec.~\ref{sec:projectableHG} we
review the action of the projectable HG and its one-loop
beta-functions. In Sec.~\ref{sec:FPs} we describe our algorithm for
the numerical search of fixed points and give their complete
list. Section~\ref{SectStabMat} analyzes the stability matrices of the
fixed points. In Sec.~\ref{SectRGflows} we turn to the global
properties of the RG flow and study the trajectories emanating from
asymptotically free fixed points. Scanning over their possible initial
conditions, we identify the unique family of trajectories which
exhibits interesting IR behavior. Section~\ref{sec:strength} further
explores the RG running of the gravitational coupling along the family
and points out a natural hierarchy arising between the Planck mass and
the scale of Lorentz violation, the former being much larger than the
latter. Section~\ref{sec:conclusions} is devoted to conclusions. Some
technical details are relegated to Appendices.

\section{Projectable Ho\v rava gravity} 
\label{sec:projectableHG}

HG is defined using the
Arnowitt--Deser--Misner (ADM) decomposition of the metric
\be
\di s^2 = N^2 \di t^2 - \gamma_{ij} (\di x^i + N^i \di t) (\di x^j +
N^j \di t)\,,
\ee
where $N$ is the lapse function, $N^i$ is the shift vector and
$\g_{ij}$ is the spatial metric. These fields are assigned the
following scaling dimensions with respect to
\eqref{scaling}:\footnote{One says that a field $\Phi$ has scaling
  dimension $[\Phi]$ if under (\ref{scaling}) it transforms as
$\Phi\mapsto b^{[\Phi]}\Phi$.} 
\be
\label{dims}
[N]=[\gamma_{ij}]=0~,~~~~~[N^i]=d-1\;,
\ee
and transformation under FDiff \eqref{FDiff} as,
\bseq
\label{FDiff1}
\begin{gather}
N\mapsto N\frac{\di t}{\di t'}~,~~~~~
N^i\mapsto \Big(N^j\frac{\d x'^i}{\d x^j}-\frac{\d x'^i}{\d
  t}\Big)\frac{\di t}{\di t'}~,\\
\gamma_{ij}\mapsto \gamma_{kl}\frac{\d x^k}{\d x'^i}\frac{\d x^l}{\d
  x'^j}\;. 
\end{gather}
\eseq
The action invariant under FDiff and  containing only relevant and
marginal operators 
with respect to the anisotropic scaling reads,
\be
\label{Sgen}
S=\frac{1}{2G}\int \di t\,\di^d x\,N \sqrt{\g}\big(K_{ij}K^{ij}-\l
K^2-{\cal V}\big)\;,
\ee
where $G$ and $\l$ are marginal coupling constants, 
\be
K_{ij}=\tfrac{1}{2N}\left(\p_t\gamma_{ij}-\nabla_i N_j -\nabla_j N_i\right)
\ee
is the extrinsic curvature of the foliation, $K\equiv K_{ij}\gamma^{ij}$
and $\nabla_i$ is the covariant derivative with respect to the spatial
metric $\g_{ij}$. 
The potential part ${\cal V}$ does not contain any
time derivatives. It depends on the $d$-dimensional metric $\g_{ij}$
and its spatial derivatives. In what follows we specify to $d=3$. 
Note that in the case of GR written in the ADM form the four-dimensional diffeomorphism invariance forces the parameter $\l$ in the kinetic term for the metric
to be equal to one, $\l=1$.

In the case of non-projectable model where the lapse is a full-fledged
function of time and space, ${\cal V}$ also depends on the  
acceleration vector $a_i = \p_i N/N$. The low-energy limit of the
theory corresponds to a scalar-tensor gravity, with the scalar
describing a preferred foliation of spacetime by space-like surfaces
\cite{Germani:2009yt,Blas:2009yd,Kimpton:2010xi,Blas:2010hb}. 
The dynamics of the scalar is stable and weakly coupled, its
interaction with visible matter and gravity can be suppressed by an
appropriate choice of parameters. This includes, in particular, a constraint $0<\l-1\ll1$. 
Thus, the model can reproduce the
phenomenology of GR at the scales where the latter has been tested
\cite{Blas:2014aca}. The strongest constraints on the model come from
tests of the Lorentz invariance in the matter sector
\cite{Kostelecky:2008ts,Liberati:2013xla} and gravity \cite{Monitor:2017mdv},
but still leave a viable
parameter space \cite{EmirGumrukcuoglu:2017cfa}. 

In the non-projectable model the total number of terms in the
potential is very large (order $O(100)$)
\cite{Kimpton:2013zb}.\footnote{Most of these terms are irrelevant at
  low energies and get important only in the UV.} 
This complicates the analysis of the UV properties of the
theory. Another complication comes from the fact that the propagator
of the lapse contains certain irregular terms that can potentially
give rise to divergences with nonlocal space dependence
and thus spoil renormalizability \cite{Barvinsky:2015kil}. An
important step towards overcoming the latter problem was made in
\cite{Bellorin:2022np} which showed that the dangerous divergences
cancel if one defines the path integral of the theory using a proper
integration 
measure. However, the results of \cite{Bellorin:2022np} are obtained
within the canonical approach which is inconvenient for carrying out
renormalization. Their reformulation in the Lagrangian language is
pending.  

Projectable HG is a simpler verson of the theory, in which the lapse
is assumed to be a function of time only. Using the time
reparameteraization symmetry, its value can be set to an arbitrary
constant, thereby eliminating it from the action. Upon setting $N=1$,
using    
the Bianchi identities, integration by parts and Ricci decomposition
the most general expression for the potential term in Eq.~(\ref{Sgen})
reads \cite{Sotiriou:2009gy}, 
\begin{align}\label{pot31}
&{\cal V}=2\Lambda-\eta R+\mu_1 R^2+\mu_2 R_{ij}R^{ij}+\nu_1 R^3+\nu_2 RR_{ij}R^{ij}\nonumber\\
&+\nu_3R^i_jR^j_kR^k_i +\nu_4 \nabla_i R\nabla^i R+\nu_5 \nabla_iR_{jk}\nabla^i R^{jk}\;,
\end{align}
where $R_{ij}$, $R$ are the Ricci tensor and scalar of the metric
$\gamma_{ij}$, and $\Lambda$, $\eta$, $\mu_a$, $\nu_b$ are the
couplings. 

It is straightforward to see that the
operators multiplying the couplings 
\be
\label{margin}
G\,,~~~\l\,,~~~\nu_a\,,~~{a=1,\ldots,5}
\ee
have scaling dimension $6$. This matches (with the opposite sign) the
scaling dimension of the integration measure in the action
(\ref{Sgen}), so these operators are marginal. They control the UV
properties of the theory. For unitarity and perturbative 
stability in UV $G$ and
$\nu_5$ must be positive, whereas $\l$ and $\nu_4$ must satisfy
\cite{Blas:2010hb,3+1}:
\bseq
\begin{gather}
\label{lambdaunitary}
\lambda<1/3~~\text{or}~~\lambda>1\;,\\
\nu_4>-3\nu_5/8\;.
\end{gather}
\eseq
On the other hand, the operators multiplying $\Lambda$, $\eta$ and
$\mu_a$, $a=1,2$ have lower scaling dimensions and represent relevant
(in the RG sense) deformations of the UV action.

If $\Lambda=0$, the theory possesses a flat background solution. Its
spectrum of perturbations contains a transverse-traceless (tt)
graviton and a scalar mode with the dispersion relations,
\bseq
\label{disp}
\begin{align}
\label{disp1}
&\omega_{tt}^2=\eta k^2+\mu_2 k^4+\nu_5k^6\;,\\
\label{disp2}
&\omega_s^2=\frac{1-\l}{1-3\l} \big(-\eta k^2+(8\mu_1+3\mu_2)k^4\big)+u_s^2\nu_5k^6\;,
\end{align}
\eseq
where 
\be\label{us}
u_s=\sqrt{\frac{1-\l}{1-3\l}\left(\frac{8\nu_4}{\nu_5}+3\right)}\;.
\ee
Note that the conditions (\ref{lambdaunitary}) ensure that $u_s^2$ is
positive. We see that the dispersion relation for the tt-graviton
(\ref{disp1}) exhibits a transition between the Lifshitz behavior
$\omega_{tt}\propto k^3$ at large $k$ and the relativistic scaling
$\omega_{tt}\propto k$ at small momenta. If we adjust the space and time
units to set $\eta=1$, the transition between these regimes will happen
at the momentum\footnote{For simplicity, we assume $\mu_2^2\sim \nu_5$
on dimensional grounds.}
\be
\label{MLV}
k=M_{LV}\sim \nu_5^{-1/4}\;.
\ee  
From the perspective of the low-energy theory, this momentum
corresponds to the scale where violation of the Lorentz invariance
becomes manifest. We will refer to it as the Lorentz violation scale. 

The negative sign in front of the $\eta k^2$ term in (\ref{disp2})
signals an instability of the flat background with respect to the long
scalar modes. In principle, the instability could be suppressed by
either sending $\l\to 1^+$ or ${\eta\to 0}$. The first case is assumed in the works 
\cite{Mukohyama:2009mz,Izumi:2009ry,Izumi:2011eh,Gumrukcuoglu:2011ef,Bassani:2024lan} and corresponds to the limit in which projectable HG classically behaves 
as GR supplemented with a dust-like matter. 
Quantum mechanically, however, a too close approach of $\l$ to unity is problematic 
since it
corresponds to strong coupling \cite{Koyama:2009hc,Blas:2010hb}. 
On the other hand, for $\eta\to 0$ the frequency of the tt-graviton scales
quadratically with momentum, $\omega_{tt}\propto k^2$, even at low
energies, so the relativistic dispersion relation is never
recovered. Addressing this low-energy instability is beyond the scope
of the present work. Instead, we focus on the high-energy regime of the
model and keep only the last five marginal terms in the potential
(\ref{pot31}).     

Projectable HG is renormalizable
\cite{Barvinsky:2015kil,Barvinsky:2017zlx}, with the marginal
couplings (\ref{margin}) forming a closed set under
renormalization. Their RG flow equations were derived in
\cite{towards,3+1}. They are formulated in terms of the one-loop beta-functions
for the {\it essential} couplings whose running is independent of the
gauge choice. Indeed, it is known that a change of gauge can add to
the one-loop effective action a term vanishing on the tree-level
equations of motion \cite{PhysRev.162.1195, KALLOSH1974293}. In the
model at hand this induces the following shift of the renormalized
couplings \cite{3+1},  
\be
\label{gaugeshift}
G \mapsto G - 2G^2 \epsilon, \quad \l \mapsto \l, \quad \nu_a \mapsto  \nu_a -
4G\nu_a\epsilon , 
\ee
where $\epsilon$  is an infinitesimal parameter.
Since $G$ and $\nu_a$ are not invariant under this shift, their
individual beta-functions are gauge-dependent. However, it is easy to
construct six combinations of couplings left invariant by
(\ref{gaugeshift}). We choose them to be
\be 
\label{new_couplings}
{\cal G} \equiv G/\sqrt{\nu_5},\quad \lambda,
\quad u_s, \quad v_a \equiv \nu_a/\nu_5,~~  a=1,2,3.
\ee
The coupling ${\cal G}$ plays a special role since it 
controls the overall strength of
the gravitational interaction in the model. We will refer
to it simply as `gravitational coupling'.
The respective beta-functions read \cite{towards,3+1}, 
\begin{widetext}
\begin{subequations}\label{betafun}
 \begin{align}
    \beta_\lambda  &=  {\cal G}\frac{27(1\!-\!\l)^2+3u_s(11\!-\!3\l)(1\!-\!\l)-2u_s^2(1\!-\!3\l)^2}{120\pi^2 u_s (1+u_s)(1\!-\!\l)}\;, 
    \label{beta_lam} \\
    \beta_{\cal G}  &= {\cal G}^2\frac{ 1}{26880\pi^2(1-\lambda)^2(1-3\lambda)^2
    (1+u_s)^3 u_s^3} \sum_{n=0}^7 u_s^n\, {\cal P}^{\cal G}_n[\l,v_1,v_2,v_3] \;,
    \label{betaG} \\
    \beta_\chi    &={\cal G} \frac{ A_\chi  }{26880\pi^2(1-\lambda)^3(1-3\lambda)^3
    (1+u_s)^3 u_s^5} \sum_{n=0}^9 u_s^n\, {\cal
    P}^{\chi}_n[\l,v_1,v_2,v_3] \;,
    \label{beta_chi}
\end{align}
\end{subequations}
\end{widetext}
where we have collectively denoted $\chi= \{u_s,v_1,v_2,v_3\}$ and the
coefficients $A_\chi$ are equal to 
$A_{u_s} = u_s (1-\lambda)$, $A_{v_1} = 1$, $A_{v_2} =A_{v_3} = 2$.
Note that the coupling ${\cal G}$ factorizes.  The functions ${\cal P}^{\cal
  G}_n$, ${\cal P}^{\chi}_n$ multiplying various powers of $u_s$ in the sums are themselves polynomials in $\l$ and $v_a$ with
integer coefficients. ${\cal P}^{\cal G}_n$, ${\cal P}^{u_s}_n$ and
${\cal  P}^{v_a}_n$ are up to
the fourth, fifth and sixth order in $\lambda$, respectively. Their overall
order in the couplings $v_a$ is up to two for ${\cal P}^{\cal G}_n$, ${\cal
  P}^{u_s}_n$ and up to three for  ${\cal  P}^{v_a}_n$. Explicit expressions
for these polynomials are lengthy and can be found in \cite{3+1}.

\section{Fixed points of the RG flow} 
\label{sec:FPs}

We start the investigation of the RG flow by searching for its fixed points. Of particular interest are the asymptotically free fixed points corresponding to vanishing gravitational coupling ${\cal G}$. The coupling $\l$ at a fixed point can be either finite or infinite. The latter case $\l=\infty$ was shown to represent a regular weakly coupled limit of the theory \cite{Gumrukcuoglu:2011xg,Radkovski:2023cew}. The only restriction we impose on the values of $\l$ is that they lie inside the unitary domain (\ref{lambdaunitary}). The other couplings $u_s$, $v_a$ are required to take finite values at the fixed point.

Beta-functions \eqref{betafun} are defined as derivatives of the couplings with respect to  $\log k_\star$, 
the logarithm of the sliding momentum scale.\footnote{Note that the logarithm of momentum is different from the logarithm of energy by a factor $1/3$ due to the Lifshitz scaling, $d\log k_\star=\tfrac{1}{3}d\log E_\star$. } 
It is convenient to change the parameterization of the RG trajectories by introducing new independent variable 
$\tau$ through
\be\label{tau}
 d\tau = {\cal G}\, d\log k_\star\;,
\ee  
and redefine the beta-functions,
\be
\label{newbetas}
\frac{dg_i}{d\tau}=\tilde \beta_{g_i},\qquad
g_i=\{\l,u_s,v_1,v_2,v_3\}.
\ee 
Then the RG flow 
in the subspace of the couplings $g_i$ separates from the running of the gravitational coupling ${\cal G}$.
Thus, the search for fixed points reduce to finding the roots of the system of equations 
\be
\label{fullsystem}
\tilde\beta_{g_i}  = 0\;
\ee
for the five variables $g_i$. 

The functions $\tilde\beta_{g_i}$ are, up to some non-vanishing factors, polynomial in the couplings $g_i$. Thus, in principle, all roots of the system (\ref{fullsystem}) can be found with the Buchberger algorithm \cite{Buchberger:2010}. In practice, however, this is unfeasible. Indeed, according to the 
B\'ezout's theorem \cite{Shafarevich} the number of (complex) roots of a system of polynomial equations is in general equal to the product of the degrees of the polynomials. The degree of the polynomial in $\tilde\beta_\l$ is $4$, whereas an inspection of the expressions for ${\cal P}_n^\chi$ given in \cite{3+1} yields the degrees of $\tilde\beta_{u_s}$ and $\tilde\beta_{v_a}$, $a=1,2,3$, to be $15$ and $17$, respectively. 
This gives $4 \cdot 15 \cdot 17^3= 294\,780$ as the possible number of roots. While the actual number of roots can be lower due to the special structure of the polynomials in $\tilde\beta_{g_i}$, it is still likely to be forbiddingly large.

Fortunately, we need only {\it real} roots of the system (\ref{fullsystem}). These can be found by means of a scanning algorithm which we presently describe. We separately discuss the cases of finite and infinite $\l$.

\subsection{Fixed points at finite $\boldsymbol{\l}$}
\label{sec:FPfinl}

First, we search for the roots of the system \eqref{fullsystem} lying at finite values of all couplings.
We note that $\tilde\beta_\lambda$ depends only on $u_s$ and $\l$. The solution of the equation $\tilde\beta_\lambda = 0$ relates these 
variables\footnote{We choose the positive value of the square root in (\ref{lamus}). The opposite choice would result in
 $\l$ lying in the non-unitary region $[1/3, 1]$. 
 Similarly, the positive square root in (\ref{uslam}) ensures that $u_s$ is positive.}
\be\label{lamus} 
\l (u_s) = \frac{9+7u_s-2u_s^2 + 2u_s\sqrt{10(1+u_s)}}{3(3+u_s - 2u_s^2)},
\ee
or
\be\label{uslam} 
\begin{split}
u_s(\l)= 3\bigg[&\frac{11 - 14\l +3\l^2  }{4(1-3\l)^2}\\
&+\frac{ \sqrt{5(1-\l)^2(29 - 42\l + 45\l^2)} }{4(1-3\l)^2}\bigg].
\end{split}
\ee
In what follows we use \eqref{lamus} and \eqref{uslam} to eliminate either $\l$ or $u_s$ in the system \eqref{fullsystem}.

\begin{table*}
\centering
\begin{tabular}{@{}|c| c | c | c | c | c | c |@{}}
 \hline
\makecell{Fixed point\\label}&$\lambda$  &$u_s$ & $v_1$ & $v_2$ & $v_3$ &  $\beta_{\cal G}/{\cal G}^2$ \\[0.5ex] 
 \hline\hline
F1&0.1787&  60.57 &-928.4 & -6.206 & -1.711 &  -0.1416  \\ [0.5ex]
  \hline
 F2&0.2773 & 390.6  &-19.88& -12.45 & 2.341 &  -0.2180    \\ [0.5ex]
 \hline
F3&0.3288 & 5.453$\times 10^4$ & 3.798$\times 10^8$ & -48.66 & 4.736 &  -0.8484  \\ [0.5ex]
  \hline
 F4&0.3289 & 5.732$\times 10^4$&-4.125$\times 10^8$ & -49.17 & 4.734 & -0.8784\\ [0.5ex]
 \hline   
 F5&0.333332 & 3.528$\times 10^{11}$ & -6.595$\times 10^{23}$ &-1.950$\times 10^8$ & 4.667  & -3.989$\times 10^6$ \\ [0.5ex]
 \hline   
  \end{tabular}
  \caption{ \label{tabFP1} Solutions of the system \eqref{fullsystem} corresponding to the fixed points of the RG flow at finite values of $\l$. The last column shows the coefficient in front of ${\cal G}^2$ in the beta-function of the gravitational coupling ${\cal G}$, evaluated at the fixed point. Its negative values imply that the fixed points are asymptotically free.}
\end{table*}

We start by excluding the variable $\l$ from the system \eqref{fullsystem} with the help of relation \eqref{lamus}. To avoid appearance of the square roots 
and maintain the polynomial structure of the equations, we introduce a new variable
\be\label{ut}
u_t \equiv \sqrt{10(1+u_s)}\;, \quad u_s = \frac1{10} u_t^2 -1\;, \quad u_t \in (\sqrt{10},\infty)\;.
\ee
The resulting system of four equations is still 
too complicated for standard root finder algorithms. 
To overcome this obstacle, we fix the value $u_t=u_t^*$ and remove one of the equations, say $\tilde\beta_{v_1}$. This leaves us with a system of three equations for three variables $v_a$: 
\be
\left\{
\begin{aligned}\label{reduced}
\tilde{\beta}_{u_s}(u_t^*,v_a) &=0\;,  \\
\tilde{\beta}_{v_2}(u_t^*,v_a) &=0\;, \\
\tilde{\beta}_{v_3}(u_t^*,v_a) &=0\;.
\end{aligned}\right.
\ee
The order of this system is greatly reduced --- recall that the polynomials in $\tilde\beta_{u_s}$ and $\tilde\beta_{v_a}$ are only quadratic or cubic in $v_a$. In more detail, the product of degrees of the polynomials in the system 
(\ref{reduced}) is $2\cdot3\cdot 3=18$, which, by the B\'ezout's theorem, gives the maximal possible number of its complex roots. All these roots are easily found by the 
NSolve command on Wolfram Mathematica \cite{Mathematica}.
We call them `partial roots' of the system (\ref{fullsystem}).

We pick up purely real partial roots --- let us denote them by $v_a^{(n)}$ --- and evaluate $\tilde\beta_{v_1}\big(u_t^*,v_a^{(n)}\big)$. The difference of this quantity from zero characterizes the failure of the partial root to be the solution of the full system (\ref{fullsystem}). We identify the partial root $v_a^*$ with the lowest absolute value of this residual, 
\be
\mid\tilde{\beta}_{v_1} (u_t^*,v_a^*) \mid \,= \min_n \mid \tilde{\beta}_{v_1} \big(u_t^*,v_a^{(n)}\big) \mid\;,
\ee
and define
\be
\label{minbeta}
\delta\tilde\beta(u_t^*)\equiv\tilde\beta_{v_1}(u_t^*,v_a^*)\;.
\ee
Then we scan over values of $u_t^*$ and find where $\delta\tilde\beta(u_t^*)$ crosses zero. This point corresponds to a root of the full system (\ref{fullsystem}).

We have scanned the interval $u_t^* \in (\sqrt{10},10^{8}]$ (see Appendix \ref{AppA} for details) and found five fixed points. They are listed in 
Table~\ref{tabFP1} in terms of the original variables $\{\l,u_s,v_a\}$. The fixed points F1--F4 were already reported in \cite{3+1}. Here we add the fixed point F5.
To make sure that no other fixed points exist,
we repeated the above procedure with different choices of the removed equation. We removed, in turn, $\tilde\beta_{v_2}$, $\tilde\beta_{v_3}$ and $\tilde\beta_{u_s}$ and in all cases obtained the identical set of fixed points. We also verified absence of fixed points at higher values of $u_s$ by eliminating it in favor of $\l$ using Eq.~(\ref{uslam}). Then the limit $u_s\to\infty$ corresponds to $\l\to 1/3^-$. A scan over $\l$ in the vicinity of $\l=1/3$ does not yield any new fixed points compared to those listed in Table~\ref{tabFP1} (see Appendix~\ref{AppA}).

In the last column of Table~\ref{tabFP1} we show the value of the beta-function for the gravitational coupling ${\cal G}$, divided by ${\cal G}^2$ and evaluated at each fixed point. We see that all these values are negative implying that the gravitational coupling vanishes when the RG flow approaches the fixed point in the UV. In other words, all found fixed points are asymptotically free. 

We also observe that all fixed points lie in the interval 
$\lambda\in(0,1/3)$. The latter property is in contrast to
 the projectable HG in $(2+1)$ dimensions, which possesses a fixed point with finite $\l>1$ \cite{Barvinsky:2017kob}.\footnote{More precisely, $(2+1)$-dimensional 
 projectable HG has fixed point with $\l=\frac{15}{14}$.}
Since the RG flow 
cannot cross non-unitary region, none of the trajectories starting from any of the fixed points at finite $\l$ in UV can reach the IR domain $\l\to1^+$ assumed in the phenomenological studies. 

The last three fixed points in Table \ref{tabFP1} are located at very large values of couplings $u_s$ and $v_1$. For the point F5 the value of $v_2$ is also huge. On the other hand, the value of $v_3$ is always order-one. We point out that this vast hierarchy of couplings appears in a theory without any large input parameters and is a consequence of the complicated non-linear structure of the beta-functions. Presence of very large couplings puts in question the validity of the one-loop approximation for the analysis of the RG trajectories emanating from the points F3--F5. Thus, we are not going to consider such trajectories when studying the global properties of the RG flow.
It should be noted, however, that the existence and  location of all fixed points is robust because all of them 
are asymptotically free. Hence the higher-loop contributions suppressed by ${\cal G}$ will not affect the values of couplings in the Table~\ref{tabFP1}.

\subsection{Fixed points at $\boldsymbol{\lambda = \infty}$}
\label{sec:FPinfl}

\begin{table*}
\begin{center}
\begin{tabular}{@{}|c| c | c | c | c | c | c |c |@{}}
 \hline
\makecell{Fixed point \\label}&$u_s$  & $v_1$ & $v_2$ & $v_3$ &   $\beta_{\cal G}/{\cal G}^2$&\makecell{Asymptotically\\ free?}
 &\makecell{Sign of\\ $\d\tilde\beta_\varrho/\d\varrho$}\\ [0.5ex]
\hline\hline
1&0.0195& 0.4994 & -2.498 & 2.999 &  -0.2004 &yes&$+$\\ [0.5ex]
\hline
 2&0.0418 & -0.01237 & -0.4204 & 1.321 &  -1.144&yes&$+$ \\ [0.5ex]
\hline
3&0.0553 & -0.2266 & 0.4136 & 0.7177 &  -1.079 &yes&$+$\\ [0.5ex]
 \hline
4&12.28 & -215.1 & -6.007 & -2.210 &   -0.1267 &yes&$-$\\ [0.5ex]
\hline
5 (A)&21.60 & -17.22 & -11.43 & 1.855 &   -0.1936 &yes&$-$\\ [0.5ex]
 \hline
6 (B)&440.4 & -13566 & -2.467 & 2.967 &   0.05822 &no&$-$\\ [0.5ex]
 \hline
7&571.9 & -9.401 & 13.50 & -18.25 &  -0.0745 &yes&$-$\\ [0.5ex]
 \hline
8&950.6 & -61.35 & 11.86 & 3.064 &   0.4237 &no&$-$\\ [0.5ex]
 \hline
  \end{tabular}
    \caption{ \label{tabFP2} Solutions of the system (\ref{lmsystem}) corresponding to fixed points of the RG flow at $\l=\infty$ ($\varrho=1$). The sixth column shows the value of the beta-function for the gravitational coupling ${\cal G}$ at the fixed point, divided by ${\cal G}^2$. The seventh column classifies whether the fixed point is asymptotically free or not. The last column gives the sign of the derivative of the beta-function for the coupling $\varrho$ at the fixed point.
    The points 5 and 6 are of particular importance and are also labeled as A and B.}
\end{center}
\end{table*}

In \cite{Gumrukcuoglu:2011xg} it was argued that a natural candidate for the UV point of the renormalization group flow of projectable Ho\v rava gravity is the limit $\l\to\infty$.
This limit was proved to be regular  and independent on the direction  $\l\to\pm \infty$, at least in perturbation theory  \cite{Radkovski:2023cew}.
To analyze this limit, it is convenient to introduce a new variable
\be\label{rho}
\varrho \equiv \frac{3(1-\l)}{1-3\l}~~~~\Leftrightarrow~~~~ \l = \frac{3-\varrho}{3(1-\varrho)}\;.
\ee
The coupling $\varrho$ is positive in the unitary domain \eqref{lambdaunitary}. The limit $\l=\infty$ corresponds to finite $\varrho=1$. The interval $\l\in (1,+\infty)$ maps to $\varrho\in(0,1)$, whereas $\l\in(-\infty,1/3)$ maps to $\varrho\in(1,+\infty)$.
Beta function of the new variable $\varrho$, divided by ${\cal G}$, reads
\be\label{betarho}
\tilde{\beta}_\varrho =
3(1-\varrho)\frac{2u_s^2+u_s\varrho(4-5\varrho)-3\varrho^2}{40\pi^2u_s(1+u_s)\varrho}\;.
\ee
It vanishes for $\varrho=1$. All other beta-functions \eqref{betaG}, \eqref{beta_chi} are regular in the neighborhood of the hyperplane $\varrho=1$, implying regularity of the RG flow in the parameterization $(\varrho, u_s, v_a)$. 

The hyperplane $\varrho=1$ is an invariant manifold of the RG flow and we look for fixed points inside it. These correspond to solutions of the system of four equations in four variables,
\be\label{lmsystem}
\tilde{\beta}_\chi (\chi) \Big|_{\varrho=1} = 0\;, \quad \chi = \{u_s,v_a\}\;.
\ee
We find all solutions of this system by a similar scanning procedure, as in the case of finite $\l$. The scan is performed over the variable $u_s$ in the interval $0<u_s<10^{15}$, the details are given in Appendix~\ref{AppA}. The search yields eight solutions listed 
in Table \ref{tabFP2}. These solutions were found in \cite{3+1,Barvinsky:2023uir}; here we have shown that there are no more solutions beyond this list.

We note that the hierarchy between the values of different couplings for the fixed points in Table~\ref{tabFP2} is milder than for the fixed points at finite $\l$ (cf. Table~\ref{tabFP1}). The strongest hierarchy occurs at the fixed point $6$, which has the value of $v_1$ a few thousands times larger than the values of $v_2$ and $v_3$.  

In the sixth column of Table~\ref{tabFP2} we list the values of the beta-function of the gravitational coupling, divided by ${\cal G}^2$. These values are negative for the points 1 to 5 and 7, implying that these points are asymptotically free. The point 5 will be the most important one for our later analysis and we give it a special notation as `point A'. The two remaining points 6 and 8 are not asymptotically free and thus cannot serve as consistent UV limits of the RG flow. Nevertheless, we are going to see that the point 6 plays a pivotal role in the structure of the RG flow. For his reason, we give it a special name `point B'.

The last column of Table~\ref{tabFP2} shows the sign of the derivative $\d\tilde\beta_\varrho/\d\varrho$. As will be discussed in the next section, this sign determines whether RG trajectories can ($-$) or cannot ($+$) flow out of the hyperplane $\varrho=1$. We see that the latter case is realized for the first three fixed points, so no RG trajectories starting from them can reach the IR domain $\varrho\to0^+$ ($\l\to1^+$). On the other hand, this possibility remains open for the fixed points 4,5 and 7. Below we will see that this indeed happens for trajectories starting at the point 5.  
Moreover, the RG trajectories emanating from this point
cover the whole unitary domain \eqref{lambdaunitary} (corresponding to $\varrho\in(0,\infty)$).

\section{Local analysis. Stability matrix }
\label{SectStabMat}

In the vicinity of a fixed point, the linearized RG flow is controlled by the stability matrix $B_i^{\:\:j}$,
\be
\label{stabilityB}
\tilde{\beta}_{g_i} \cong \sum_j B_i^{\:\:j} ( g_j- g_j^\star), \quad B_i^{\:\:j} \equiv   \left(\frac{\partial \tilde{\beta}_{g_i}}{\partial g_j}\right)\Bigg|_{g_i=g_i^\star},
\ee
where $g_i^\star$ are fixed point values of the coupling constants. The eigenvalues $\theta^J$ of the stability matrix determine whether the RG flow is repelled from (${\rm Re}\, \theta^J <0$) or attracted to (${\rm Re}\, \theta^J >0$) the fixed point along the corresponding eigendirection when the energy scale is lowered from UV to IR.\footnote{Note that the situation is opposite when the energy increases: the flow is attracted to the fixed point if ${\rm Re}\, \theta^J <0$ and repelled from it if ${\rm Re}\, \theta^J >0$. When discussing repulsion or attraction we stick to the Wilsonian perspective of the RG flow starting in UV and running towards IR.} 
We now discuss the eigenvalues of the stability matrices for the fixed points found in the previous section.

\subsection{Fixed points at finite $\boldsymbol{\l}$}


Eigenvalues of the stability matrices for the fixed points F1--F5 are provided in Table~\ref{EVfull}.
\begin{table*} 
\centering
\begin{tabular}{@{}|c| c | c | c | c | c |c|@{}}
 \hline
\makecell{Fixed point\\label}& $\theta^{1}$ &$\theta^{2}$ &$\theta^{3}$ &$\theta^{4}$&$\theta^{5}$ \\[0.5ex] 
 \hline\hline
F1  & -0.3416 &-0.06495&0.002639 &\multicolumn{2} {c|}{0.1902 $\pm$ 0.1760 $i$} \\ [0.5ex]
  \hline
 F2 &-0.06504 &0.001944& 0.02859& 0.2647& 0.2751 \\ [0.5ex]
 \hline
F3&-1.970&  -0.1448& -1.142 & -0.1142& 0.1289   \\ [0.5ex]
  \hline
 F4 & -1.976& -1.202& -0.2046&\multicolumn{2} {c|}{0.01148 $\pm$ 0.1006 $i$} \\ [0.5ex]
 \hline   
 F5  & -7.979$\times10^6$& -7.979$\times10^6$& -3.040$\times10^6$& -0.3161&0.3159\\ [0.5ex]
 \hline   
  \end{tabular}
  \caption{ Eigenvalues $\theta^J$ of the stability matrix for the fixed points with finite $\l$, ordered by the values of their real parts. Eigenvalues $\theta^{1}$, $\theta^{2}$ 
  for the point F5 coincide up to four significant digits but we checked that they differ at higher orders.}
  \label{EVfull}
\end{table*}
Note that some fixed points have complex conjugate eigenvalues. 
This is surprising since the eigenvalues of the stability matrix are typically related to the scaling dimensions $\Delta^J$ of the operators driving the RG flow. In our case the putative relation would have the form 
$\Delta^J=6+{\cal G}\,\theta^J$. Here $6$ is the classical scaling dimension of the operator and the second term is its anomalous dimension.
In relativistic unitary theories, where scale invariance at the fixed point is embedded in the full conformal symmetry \cite{Komargodski:2011sw,Luty:2013aa}, the dimensions of the operators are known to be real and positive \cite{Mack:1975je}. This property is commonly assumed to hold in unitary field theories even without the Lorentz invariance, though we are not aware of any general proof for theories with the Lifshitz scaling.\footnote{See \cite{Nishida:2010tm,Goldberger:2014hca} for the proof in the special case of nonrelativistic {\it conformal} theories with Lifshitz exponent $z=2$.}
Thus, it is logically possible that the presence of complex anomalous dimensions is compatible with unitarity in HG due to the lack of the Lorentz symmetry. 

We believe, however, that a more plausible explanation has to do with the gauge invariance of HG. At the asymptotically free fixed point the full FDiff transformations (\ref{FDiff1}) get replaced by their linearized version. On the other hand, everywhere outside the fixed point the theory enjoys the full nonlinear FDiff invariance. Thus, the operators that deform the theory away from the fixed point are not gauge invariant under the linearized FDiffs and do not need to obey unitarity constraints of the (free) theory at the fixed point. In particular, their scaling dimensions need not be real. It would be interesting to explore this possibility further and look for other examples of unitary gauge theories featuring stability matrices with complex eigenvalues.
This task is, however, outside the scope of this paper.

All fixed points in Table~\ref{EVfull} have at least one repulsive eigendirection with ${\rm Re}\, \theta^J <0$. That means that every fixed point can serve as UV completion of the theory. As already mentioned, however, all fixed points F1--F5 lie in the left part of the unitary domain at $\l<1/3$. Since the RG trajectories cannot cross the non-unitary region $\l\in[1/3,1]$, they cannot flow from these fixed points to $\lambda\to1^+$, i.e. the GR form of the kinetic term is unattainable in the IR domain. We describe trajectories flowing out of the first two fixed points F1, F2 in Sec.~\ref{flowsfull}.

\subsection{Fixed points at $\boldsymbol{\l=\infty}$}


Here we switch to the variables $g_i=\{\varrho,u_s,v_a\}$. The fixed points we are interested in lie in the hyperplane $\varrho=1$. Since the beta-function for the coupling $\varrho$ is proportional to $(1-\varrho)$, see Eq.~(\ref{betarho}), all elements in the $\varrho$-row of the stability matrices $B_\varrho{}^j$ vanish, except for the diagonal element 
\be
B_\varrho{}^\varrho =-\frac{3(u_s-3/2)}{20\pi^2 u_s} \bigg|_{u_s=u_s^\star}\;.
\ee
The eigenvalue equation $B_i{}^jw_j=\theta w_i$ then for $i=\varrho$ becomes 
$B_\varrho{}^\varrho w_\varrho=\theta w_\varrho$. This implies existence of a single eigenvector $w^1$ with non-zero $\varrho$-component and the corresponding eigenvalue $\theta^1=B_\varrho{}^\varrho$. The eigenvectors corresponding to all other eigenvalues lie entirely in the $\varrho=1$ hyperplane. We list the eigenvalues of the stability matrices for fixed points \textnumero 1--8 in Table~\ref{EVlam}. We again observe that some eigenvalues come in complex conjugate pairs.
\begin{table*} 
\begin{center}
\begin{tabular}{@{}| c | c | c | c | c | c |@{}}
 \hline
\makecell{Fixed point\\label} & $\theta^1$ & $\theta^2$ & $\theta^3$ & $\theta^4$& $\theta^5$  \\ [0.5ex]
\hline\hline
1&1.154&-1.235& \multicolumn{2} {c|}{ -0.2734 $\pm$ 0.2828 $i$}& 0.9825 \\ [0.5ex]
\hline
 2&0.5302& \multicolumn{2}{c|}{-71.95 $\pm$ 5.134 $i$}& -0.3207& 12.35 \\ [0.5ex]  
 \hline
3&0.3970& \multicolumn{2}{c|}{-64.72 $\pm$ 0.6149 $i$}& 0.3012&10.77 \\ [0.5ex]
 \hline
4&-0.01334&-0.3436& -0.09353 &  \multicolumn{2}{c|}{0.2200 $\pm$ 0.1806 $i$}\\ [0.5ex]
 \hline
5 (A)&-0.01414 & -0.06998& 0.06569& 0.2565& 0.3204 \\ [0.5ex]
 \hline
6 (B)&-0.01515& \multicolumn{2}{c|}{0.0924 $\pm$ 0.2890 $i$}& 0.3079&0.6032 \\ [0.5ex]
 \hline
7& -0.01516&-1.722&  \multicolumn{2}{c|}{-0.3324 $\pm$ 0.3289 $i$}& 0.1328 \\ [0.5ex]
\hline
8&-0.01517& -0.3657& \multicolumn{2}{c|}{0.4340 $\pm$ 0.4849 $i$}& 1.326 \\ [0.5ex]
 \hline
  \end{tabular}
    \caption{Eigenvalues $\theta^J$ of the stability matrix for the fixed points at $\l=\infty$ ($\varrho=1$). At each fixed point, the eigenvalue $\theta^1$ corresponds to the unique eigenvector with non-zero $\varrho$-component. The remaining eigenvalues are listed in the ascending order of their real parts.} 
    \label{EVlam}
\end{center}
\end{table*}

The tangent vector to an RG trajectory emanating from the fixed point can be decomposed in the basis of eigenvectors $w^J$. In order to escape from the hyperplane $\varrho=1$, it must contain a contribution of the vector $w^1$. However, this is not sufficient --- the direction $w^1$ must also be repulsive, i.e. $\theta^1$ must be negative. We see from Table~\ref{EVlam} that this condition is not fulfilled for the first three fixed points, so the trajectories starting from them always stay in the $\varrho=1$ plane. Such trajectories will not be of interest to us. The fixed points 6 and 8 are not asymptotically free. Thus, only the fixed points 
4, 5 and 7 can give rise to consistent RG trajectories connecting regions with infinite and finite $\l$.

\section{RG trajectories}
\label{SectRGflows}

Now we can construct RG trajectories emanating from asymptotically free UV fixed points at finite $\l$ and at $\l=\infty$. We numerically solve the RG equations (\ref{newbetas}) from $\tau=0$ towards $\tau=-\infty$ with the initial conditions slightly offset from the fixed point $g_i^\star$ in the repulsive direction,
\be
\label{RGeq}
g_i (0) = g_i^\star + \varepsilon\, c_J\, w_i^{J}, 
\ee
where $\varepsilon$ is a small positive parameter, $c_J$ are constants satisfying $\sum_J(c_J)^2=1$, and $w^J_i$ are eigenvectors enumerated by the index $J$, corresponding to the eigenvalues with negative real parts, 
${\rm Re}\, \theta^J <0$. We normalize them as $\sum_i(w_i^J)^2=1$. 
The parameter $\varepsilon$ is chosen sufficiently small, so that the trajectory approaches the fixed point $g^\star_i$ for $\tau\to+\infty$. We typically take $\varepsilon=10^{-5}$.

\subsection{RG flow from fixed points at finite $\boldsymbol{\lambda}$ }
\label{flowsfull}

There are five fixed points at finite values of $\l$, see Table \ref{tabFP1}. We focus on the first two of them since the remaining ones lie at very large values of the couplings 
$u_s$, $v_1$ and it is unclear if one-loop approximation is sufficient for the construction of the RG trajectories emanating from them.

\subsubsection{Flow from fixed point F1}

According to Table~\ref{EVfull} the first fixed point has two negative eigenvalues $\theta^1$ and $\theta^2$.
The components of the corresponding eigenvectors $w^1$, $w^2$ are presented in the first two rows of Table~\ref{EVFP1} (labeled FP1$w^1$ and FP1$w^2$).
Note that these eigenvectors are almost collinear and aligned along the  
with $v_1$-direction.
\begin{table*}
\centering
\begin{tabular}{@{}|c| c || c | c | c | c | c |@{}}
 \hline
\makecell{Eigenvector\\label}&$\theta^J$& $w_\l^J$ & $w_{u_s}^J$ & $w_{v_1}^J$ & $w_{v_2}^J$ & $w_{v_3}^J$   \\ [0.5ex]
\hline\hline
F1$w^1$&-0.3416& 7.159$\times10^{-9}$& 3.411$\times 10^{-4}$&-0.9999& -2.323$\times 10^{-3}$& 4.48$\times 10^{-5}$\\ [0.5ex]
\hline
F1$w^2$&-0.06495& 8.536$\times10^{-6}$& 0.08206 &-0.9909& 0.09028& -0.05745\\ [0.5ex]  
 \hline\hline
F2$w^1$& -0.06504& 2.511$\times10^{-6}$& 0.9999&1.339$\times 10^{-3}$& 7.199$\times 10^{-3}$& -3.395$\times 10^{-4}$\\ [0.5ex] 
\hline
  \end{tabular}
    \caption{Components of the repulsive eigenvectors $w_i^J$ for fixed points F1, F2 located at finite $\l$. We also list the corresponding eigenvalues $\theta^J$.} 
    \label{EVFP1}
\end{table*}

In the initial conditions for the RG equation \eqref{RGeq} we choose constants $c_J$ on the unit circle, 
\be
\label{FP1incond}
c_1 w^1 + c_2 w^2 = \cos\varphi\, w^1 + \sin\varphi\, w^2\;, 
\ee
and construct many trajectories scanning the interval $\varphi \in [0,2\pi)$ with a small step. One expects to see a one-parameter family of RG trajectories parametrized by the angle $\varphi$. However, since the absolute value of $\theta^1$ is substantially larger than that of $\theta^2$ (see Table~\ref{EVFP1}), all trajectories are quickly attracted to the direction set by the eigenvector $w^1$ and practically coincide. The only difference arises depending on the sign of the coefficient $\cos\varphi$ in front of $w^1$. 

\begin{figure}[h]
\centering
\includegraphics[width=\columnwidth]{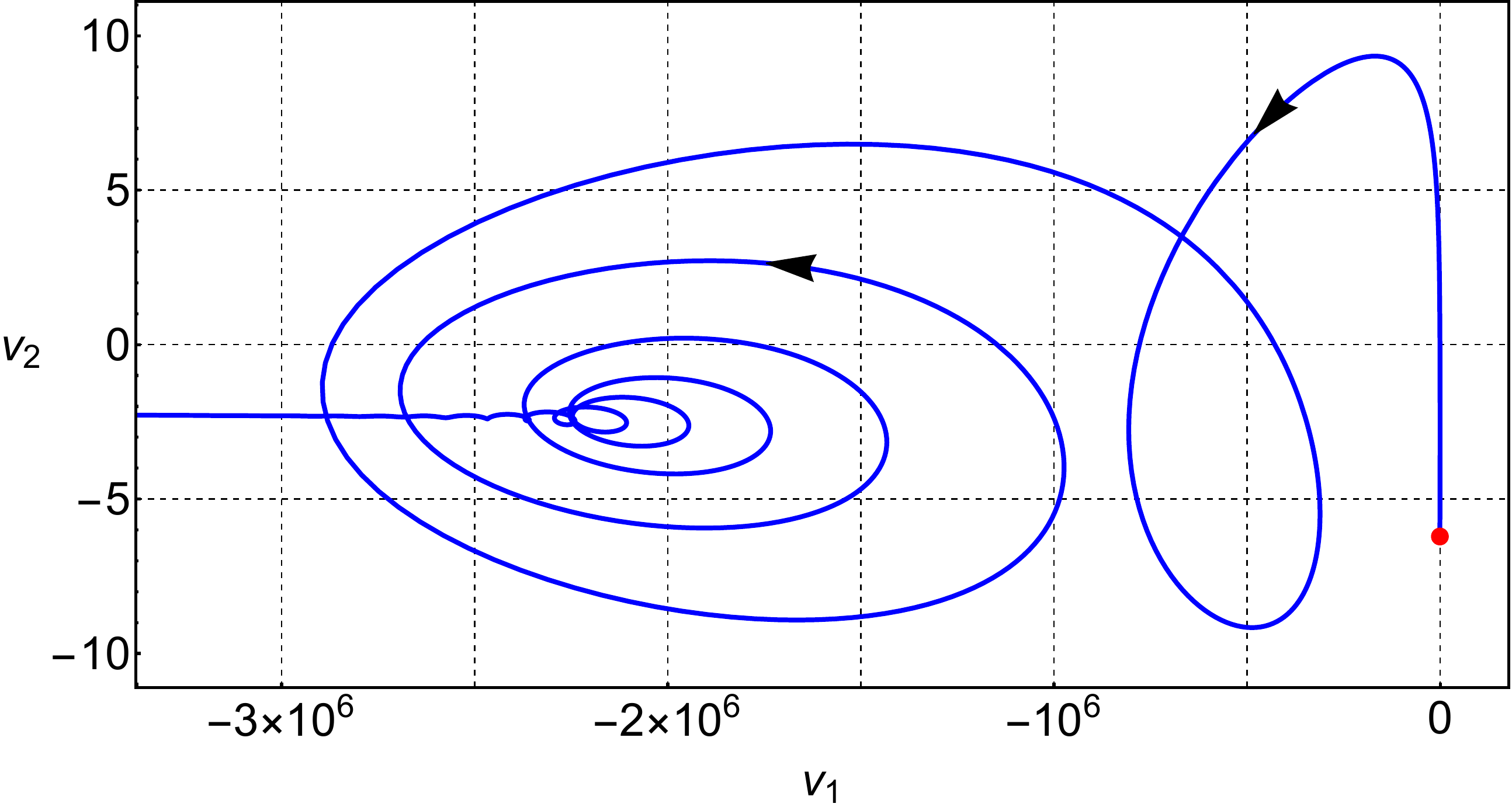}\\
~\\
\includegraphics[width=\columnwidth]{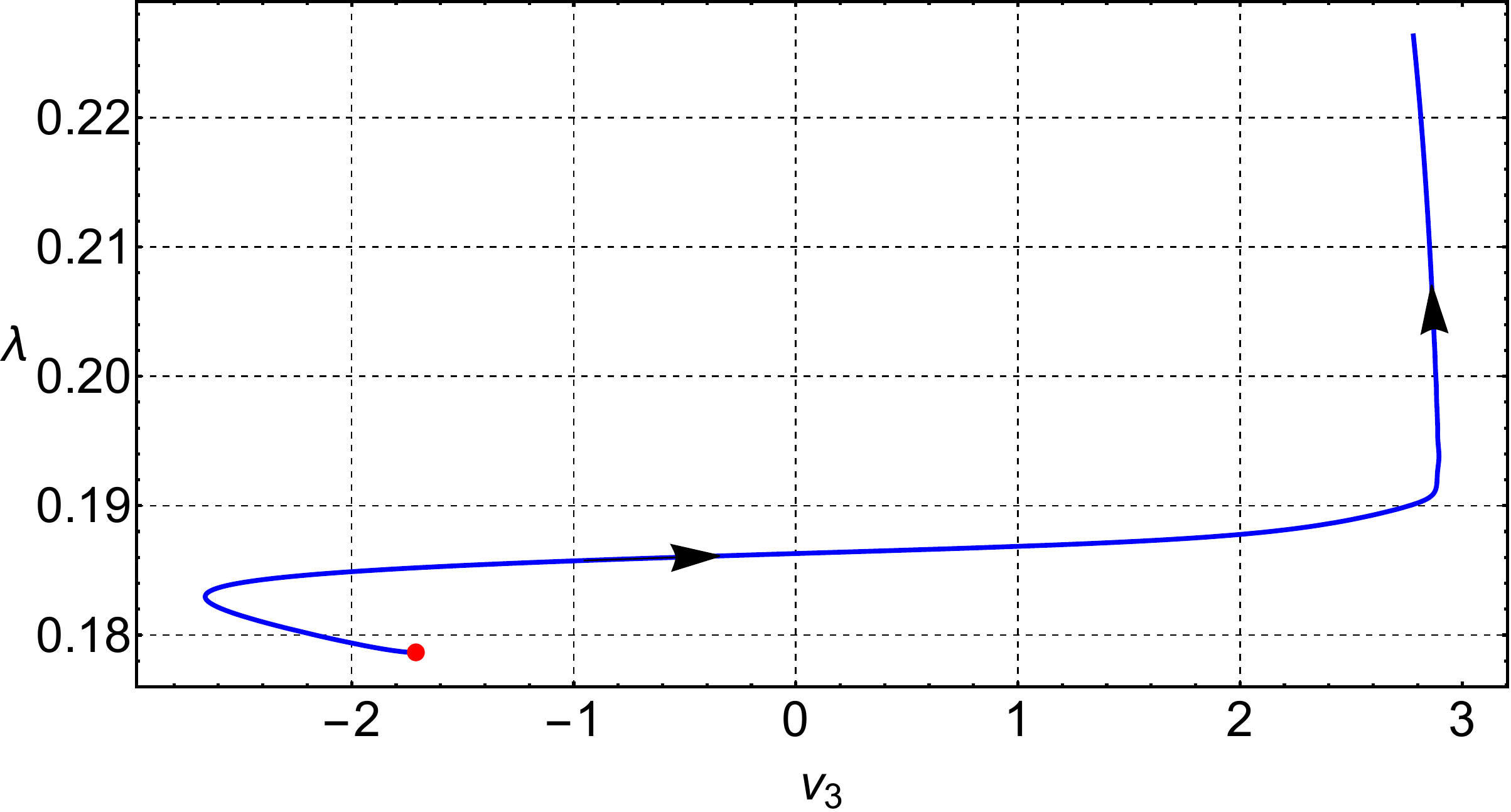}\\~\\
\includegraphics[width=\columnwidth]{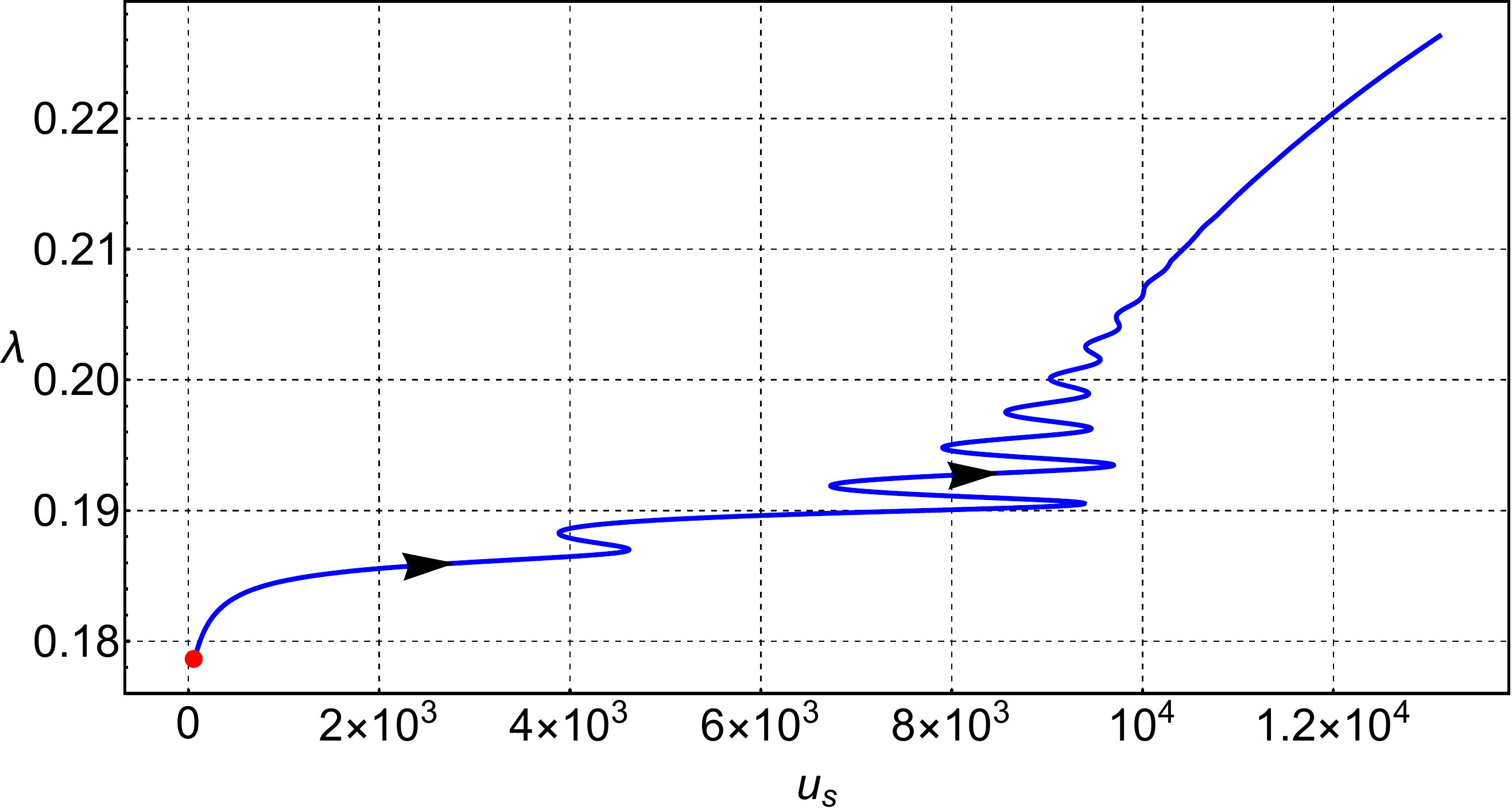}
\caption{RG flow from fixed point F1 (red dot) depicting trajectories with $\cos\varphi\leq 0$ in the initial conditions (\ref{FP1incond}). The trajectories practically coincide for all $\varphi\in[\pi/2,3\pi/2]$. Arrows indicate the flow direction from UV to IR. 
}
\label{FP1flows}
\end{figure}

When $\cos\varphi\leq 0$
we obtain a trajectory whose two-dimensional projections on various coordinate planes in the five-dimensional space of couplings are shown in Fig.~\ref{FP1flows}. 
We observe that the trajectory exhibits an intricate behavior, with $u_s$ and $v_1$ growing rapidly (in absolute magnitude). This growth leads to the loss of numerical precision, so that the trajectory cannot be continued further. We interpret this as evidence that the couplings $u_s$ and $v_1$ diverge at a finite RG scale $\tau$, signaling appearance of strong coupling. The coupling $\l$ increases towards the boundary of 
the unitary domain $\l=1/3$ but does not reach it. 

When $\cos\varphi$ is strictly positive the coupling $v_1$ grows and diverges even faster. The corresponding plots are not instructive and we do not present them.

\subsubsection{Flow from fixed point F2}

\begin{figure}[h]
\centering
\includegraphics[width=0.95\columnwidth]{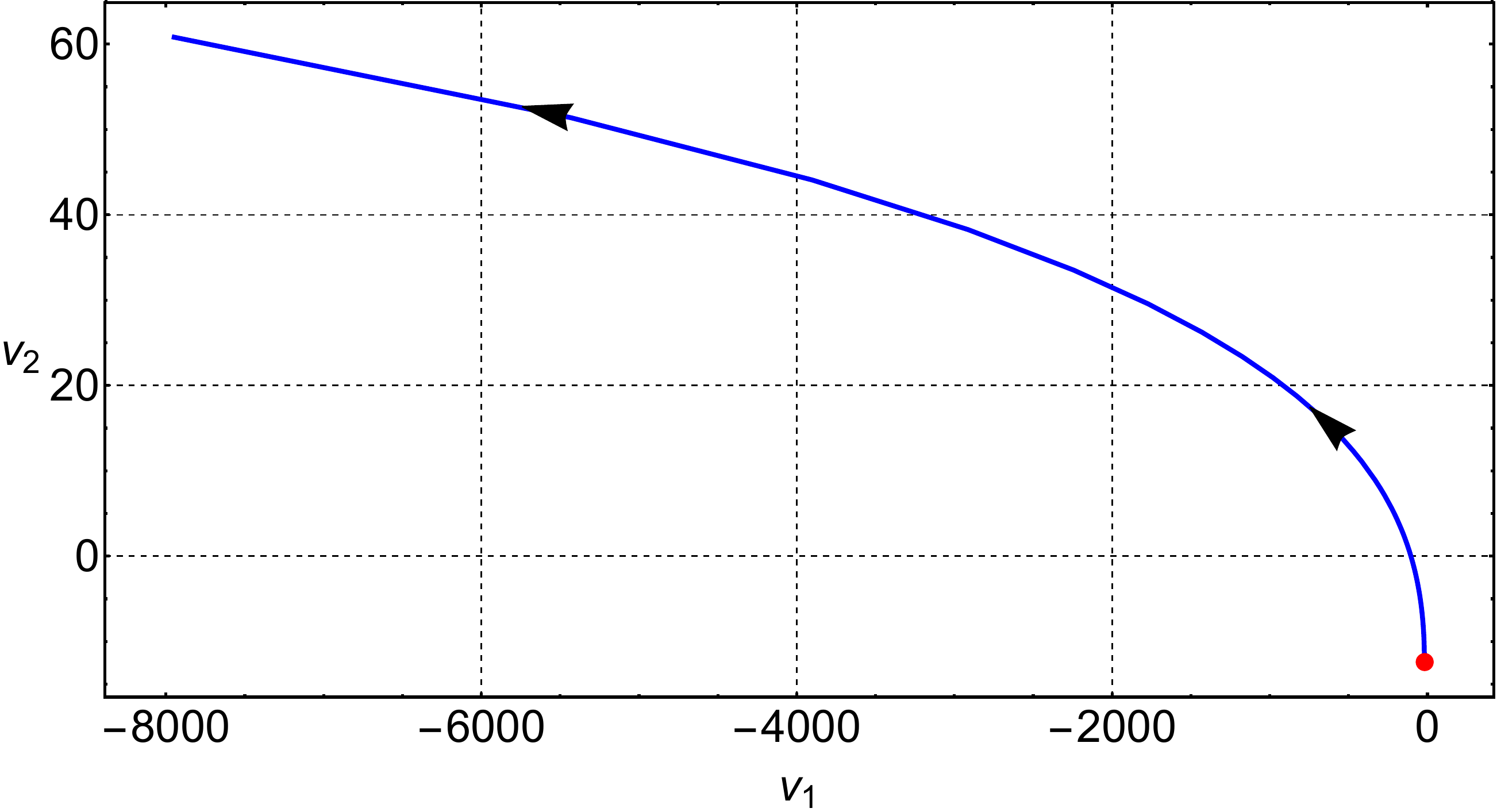}\\
~\\
\includegraphics[width=\columnwidth]{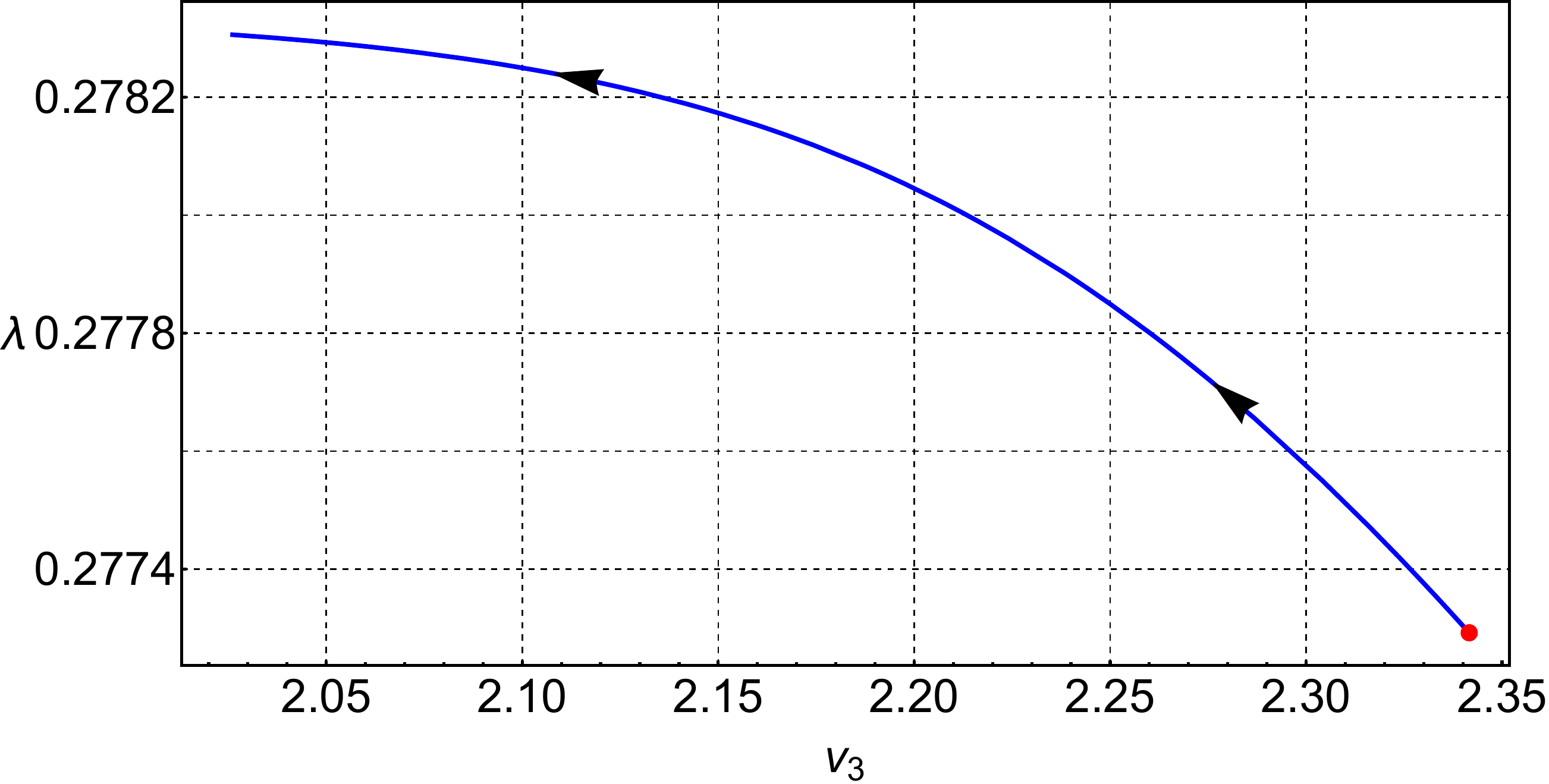}\\
~\\
\includegraphics[width=\columnwidth]{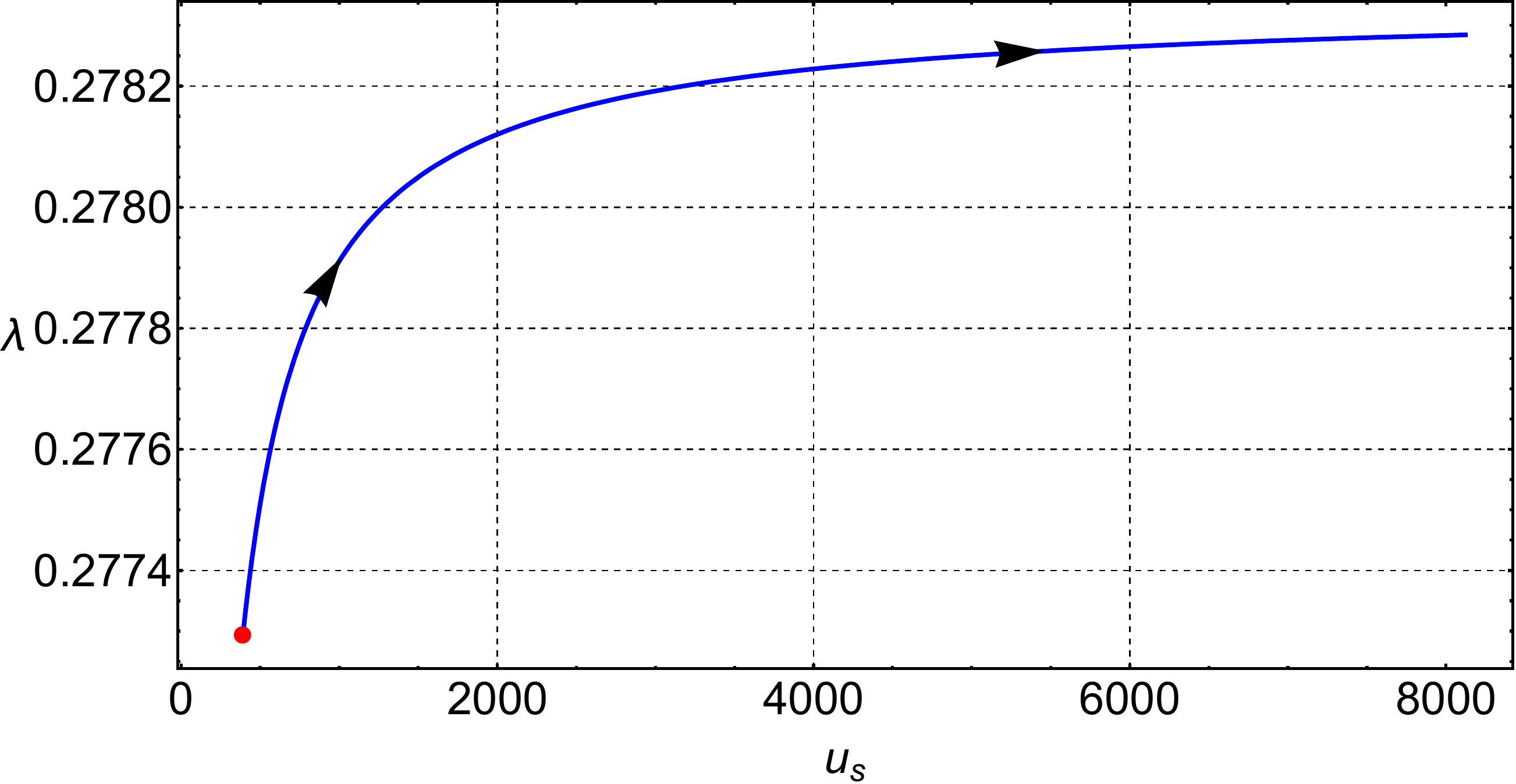}
\caption{RG flow from fixed point F2 (red dot) initiated along the positive direction of the repulsive vector. Arrows indicate the direction from UV to IR. The trajectories cannot be continued further due to loss of numerical precision. }
\label{FP2flows}
\end{figure}

According to Table \ref{EVfull}, the second fixed point has only one repulsive direction.
The components of the corresponding eigenvector $w^1$ are given in the third row of
Table~\ref{EVFP1} (labeled FP2$w^1$). 
Note that, as in the case of the first fixed point, this eigenvector nearly aligns along one of the coordinate axes --- $u_s$-direction in this case.
There are only two RG trajectories corresponding to 
$c_1=\pm 1$ in Eq.~\eqref{RGeq}. The projections of the trajectory with $c_1=1$ are shown in Fig.~\ref{FP2flows}. 
The trajectory in this case is rather monotonous. Still, as in the case of trajectories emanating from F1, the couplings $u_s$ and $v_1$ become very large and numerical integration breaks down. We again interpret this as divergence of the couplings at finite $\tau$. Note that the coupling $\l$ almost does not change along the trajectory. 

For $c_1=-1$ we have a similar loss of numerical precision due to divergence of couplings, without any appreciable change in $\l$. We do not show the corresponding plots since they are not illuminating.\\

To sum up, the RG trajectories emanating from the fixed points F1, F2 do not make any significant excursions in $\l$, quickly running into the strong coupling due to divergence of $u_s$ and $v_1$.

\subsection{RG flow from fixed points at $\boldsymbol{\l=\infty}$ ($\boldsymbol{\varrho=1}$)}
\label{RGflow_inflam}

We turn to fixed points at infinite $\l$. As discussed above, there are only three fixed points (4, 5 and 7 in Table \ref{tabFP2}) that are asymptotically free and can give rise to trajectories flowing out of the plane $\l=\infty$ ($\varrho=1$). As shown in Appendix~\ref{FPs47}, the situation for the points 4 and 7 is very similar to what we encountered for the fixed points F1 and F2. Namely, the trajectories starting from them run into strong coupling at finite `RG time' $\tau$ without any appreciable change in $\varrho$. This leaves only the point 5 as a possible origin of RG trajectories connecting $\l=\infty$ to the region where $\l$ is of order $1$. 
We presently show that such trajectories indeed exist. In what follows we conveniently refer to point 5 as point~A.

\subsubsection{Flow between fixed points A and B}
\label{ssec:AtoB}

\begin{figure}[H]
\centering
\includegraphics[width=\columnwidth]{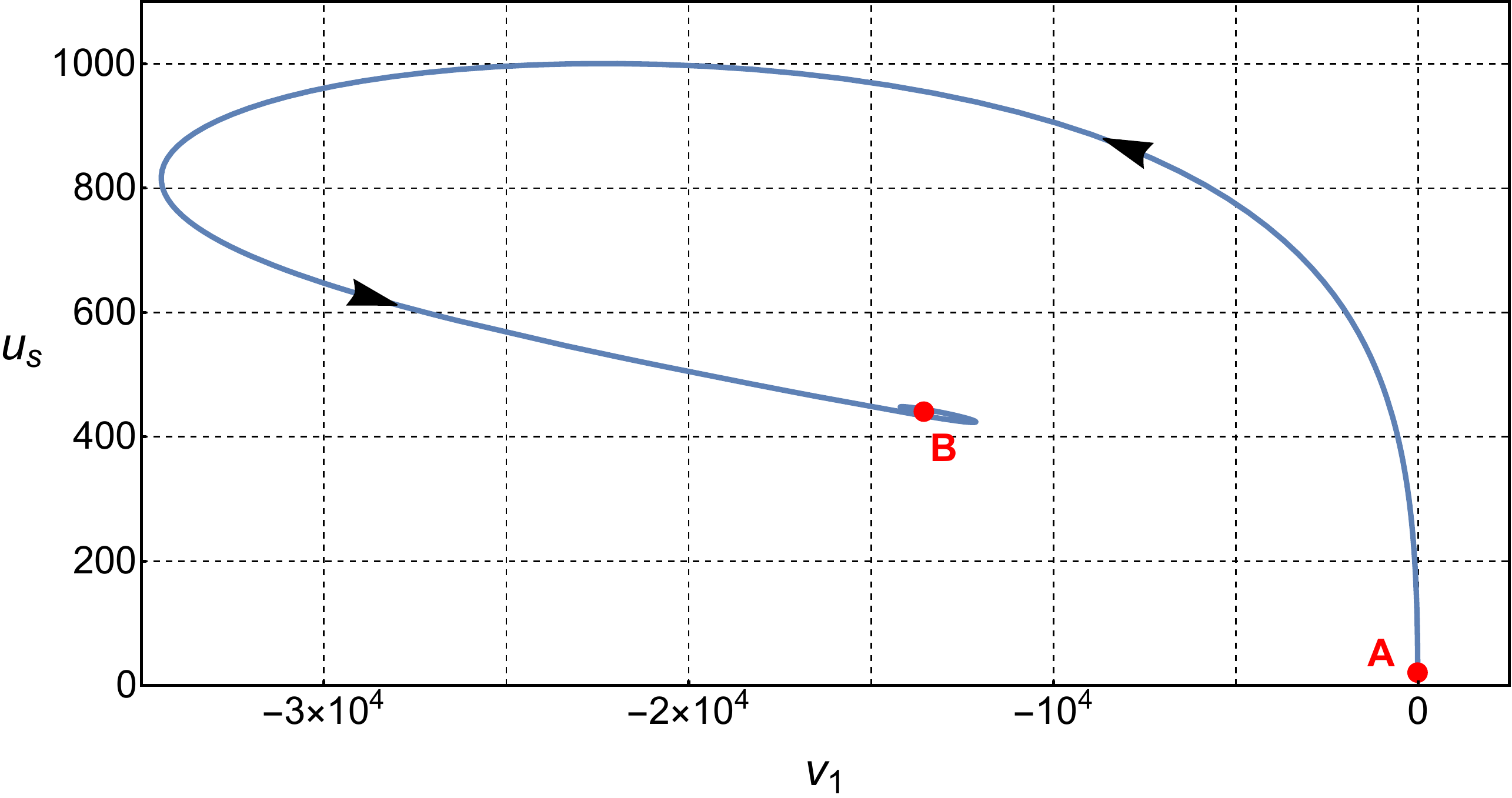}
\includegraphics[width=\columnwidth]{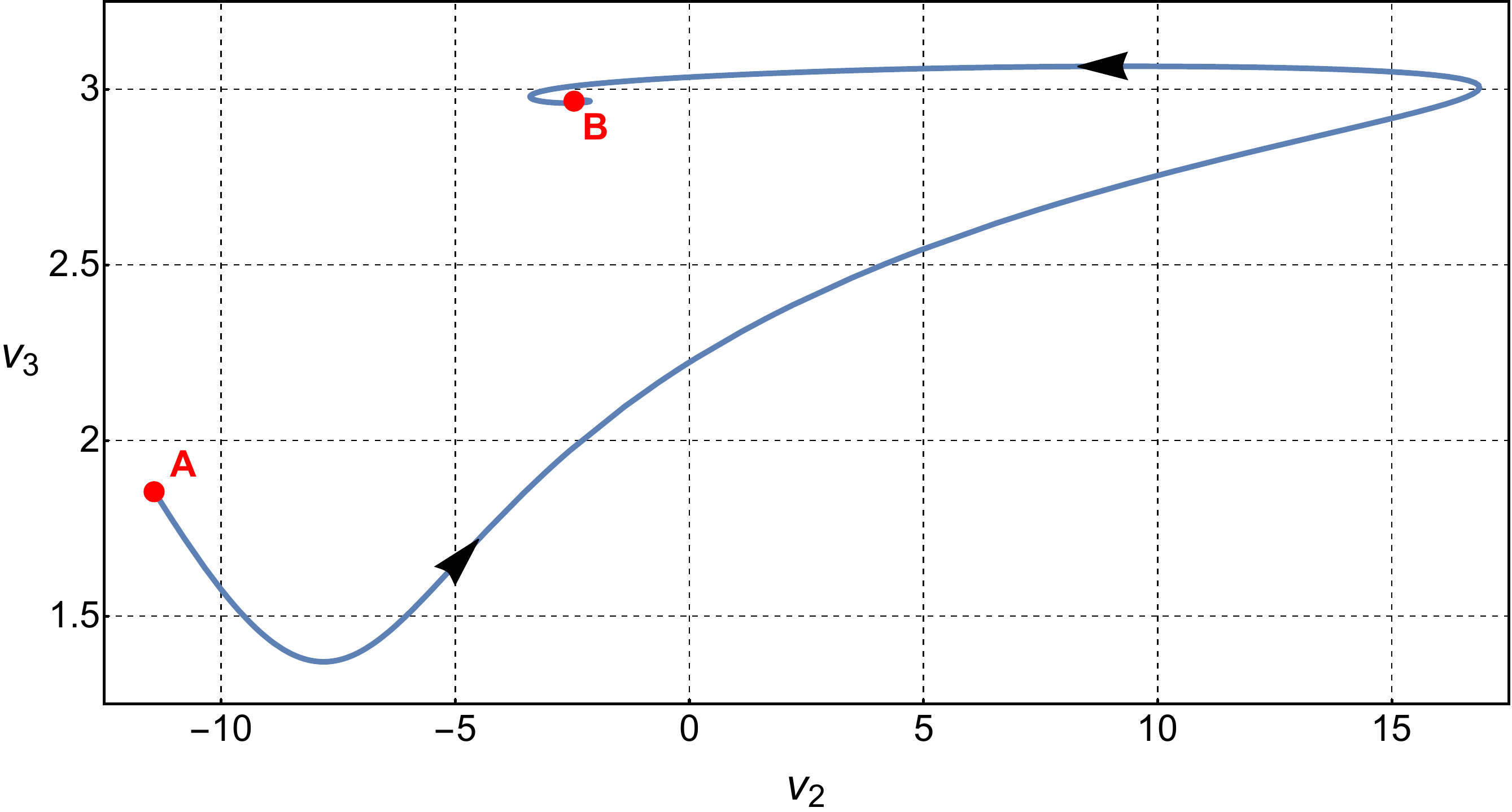}
\caption{RG trajectory connecting fixed points A and B. The trajectory lies entirely in the hyperplane $\varrho=1$. Panels show its projections on the $(u_s,v_1)$ and $(v_2,v_3)$ planes. Arrows indicate the flow from UV to IR.\label{AtoB}}
\end{figure}

The fixed point A has two negative eigenvalues, see Table~\ref{EVlam}. The components of the corresponding repulsive eigenvectors are listed in the first two rows of Table~\ref{EVlam5} (labeled $w^{A1}$ and $w^{A2}$).\footnote{Note that we define these eigenvectors with the opposite sign compared to \cite{Barvinsky:2023uir}.} 
\begin{table*}
\centering
\begin{tabular}{@{}| c | c || c | c | c | c | c |@{}}
 \hline
\makecell{Eigenvector\\label}& $\theta^J$  & $w_\varrho^J$& $w_{u_s}^J$ & $w_{v_1}^J$  & $w_{v_2}^J$ & $w_{v_3}^J$   \\ [0.5ex]
\hline\hline
$w^{A1}$&-0.01414 & -0.04229&-0.9983 & 0.03985 & -5.247$\times10^{-3}$ & -5.566$\times10^{-3}$ \\ [0.5ex]
 \hline
 $w^{A2}$ &-0.06998 &0& 0.9666 & 0.1150 & 0.2239 & -0.04799 \\ [0.5ex]
 \hline\hline
  $w^{B1}$&-0.01515 & 2.190$\times10^{-5}$&0.01622 & -0.9999 & 1.874$\times10^{-5}$ & 5.691$\times10^{-6}$ \\ [0.5ex]
  \hline
 \end{tabular}
    \caption{Components of the repulsive eigenvectors $w_i^J$ for fixed points A and B located at $\l=\infty$ ($\varrho=1$). The corresponding eigenvalues $\theta^J$ are also shown.
    \label{EVlam5}}
\end{table*}
The initial conditions for the RG trajectories are set according to Eq.~(\ref{RGeq}) with 
\be
\label{lkA}
c_{A1} w^{A1} + c_{A2} w^{A2} = \cos\varphi_A w^{A1}+ \sin\varphi_A w^{A2}\;,
\ee
where $\varphi_A \in [0,2\pi)$. 
It is instructive to first set $\varphi_A=\pi/2$ and consider the trajectory flowing out of A along the eigenvector $w^{A2}$. Though this trajectory stays in the 
hyperplane $\varrho=1$ because 
the vector $w^{A2}$ has zero $\varrho$-component, it plays an important role.
We show its projections in Fig.~\ref{AtoB}. 
Remarkably, it arrives in IR at the fixed point 6 from Table~\ref{tabFP2}, which we call point B for convenience. 

The fact that we found an RG trajectory interpolating between two fixed points may at first seem surprising.  
However, as one can see from Table \ref{EVlam}, all eigenvalues of the stability matrix at the point B, except $\theta^1$, have positive real parts.
Since $\theta^2,\ldots,\theta^5$ correspond to the eigenvectors lying in the hyperplane $\varrho=1$, we conclude that the point B is absolutely attractive within this plane. It just happens that the fixed point A belongs to (the boundary of) the basin of attraction of B, and the direction $w^{A2}$ points inside this basin.\footnote{It is worth mentioning that the trajectory with $\varphi_A=3\pi/2$ in the initial displacement (\ref{lkA}), i.e. initialized in the opposite direction of vector $w^{A2}$, does not reach the point B. Instead, it exhibits a singularity corresponding to divergence of couplings at finite $\tau$.} 

It is now clear that if $\varphi_A$ in Eq.~(\ref{lkA}) slightly deviates from $\pi/2$, i.e. the initial conditions for the RG trajectory have a small admixture of the vector $w^{A1}$ with non-zero $\varrho$-component, the flow will approach the point B but will not terminate there. Instead, it will pick up the repulsive direction $w^{B1}$ and will run along it away from the ${\varrho=1}$ plane. This motivates us to look closely at the RG trajectories starting from the point B.

\subsubsection{Flow from point B to $\lambda\to1^+$} 

The fixed point B is not asymptotically free. Thus, it cannot serve as a UV fixed point of the RG flow. However, as we saw above, it can play the role of an attractor at intermediate RG scales. The flow leaves the attractor along the unique repulsive eigenvector $w^{B1}$ whose components are listed in the third row of Table~\ref{EVlam5}. This vector points away from the $\varrho=1$ plane. Depending on the choice $c_{B1}=\pm1$ in the initial condition (\ref{RGeq}) we obtain two RG trajectories.  

\begin{figure}[H]
\centering
\includegraphics[width=\columnwidth]{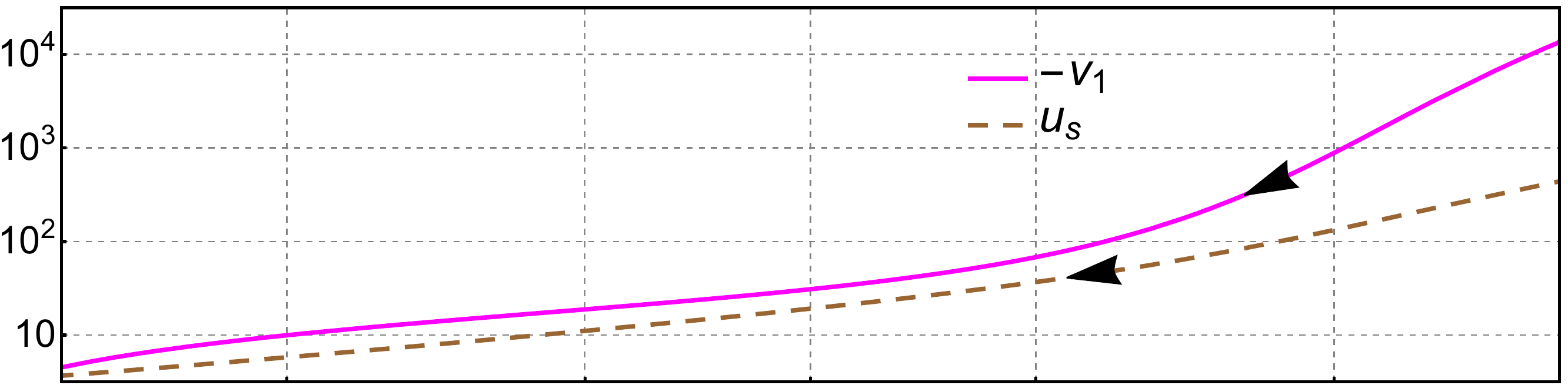}
\includegraphics[width=\columnwidth]{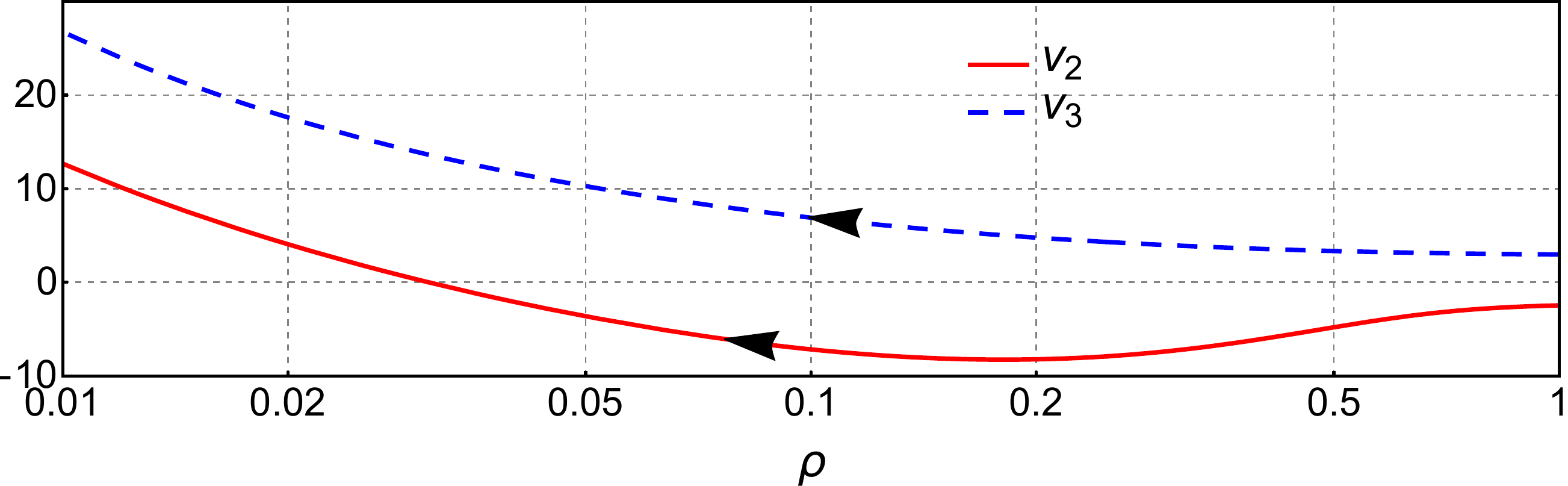}
\caption{Couplings $\{u_s,v_1,v_2,v_3\}$ as functions of $\varrho$ along the RG trajectory from the fixed point B to $\varrho\to 0$ ($\l\to1^+$). 
Arrows indicate the flow from UV to IR.}
\label{Btolam1+}
\end{figure}

On the trajectory with $c_{B1}=-1$, the coupling $\varrho$ monotonically decreases and at $\tau\to-\infty$ reaches the boundary of the unitary domain \eqref{lambdaunitary} $\varrho\to0$ ($\l\to1^+$).  The behavior of other couplings is shown in Fig.~\ref{Btolam1+}. 
The couplings $v_a$ approach some finite values of order $O(1)$ or $O(10)$ before they start rapidly growing in a small vicinity of $\varrho=0$ (not shown in the plot). This divergence can be attributed to the presence of large inverse powers of $(1-\lambda)$ in the beta functions \eqref{beta_chi}. The coupling $u_s$ tends to zero when $\varrho\to0$. More details about the behavior of the couplings at $\varrho\to 0$ ($\l\to1^+$) are given in Appendix~\ref{asympt}.

\subsubsection{Flow from point B to $\lambda\to1/3^-$} 
\label{ssec:Bto13}

\begin{figure}[H]
\centering
\includegraphics[width=\columnwidth]{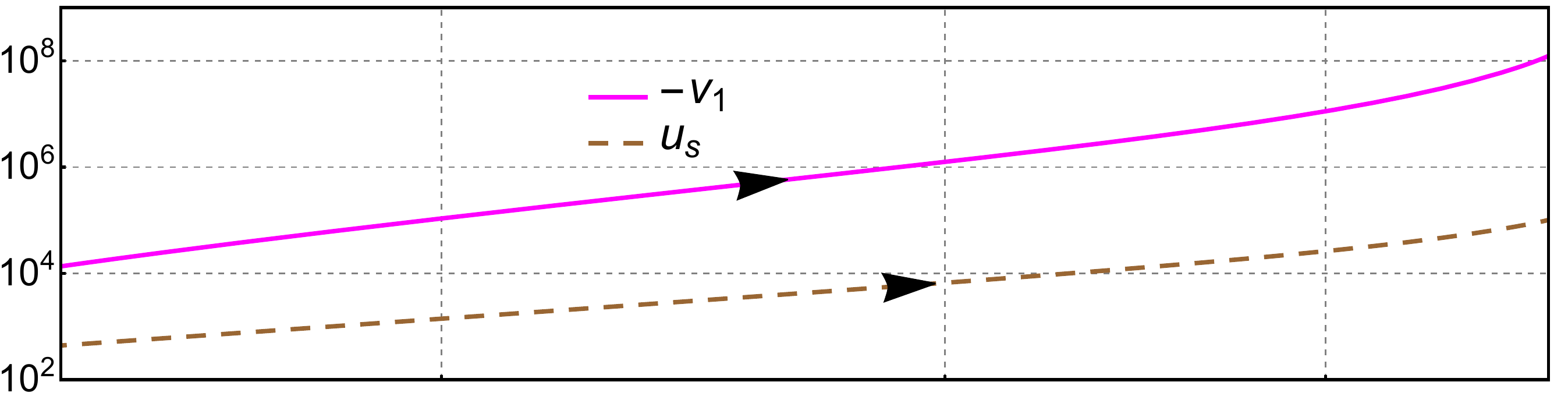}
\includegraphics[width=\columnwidth]{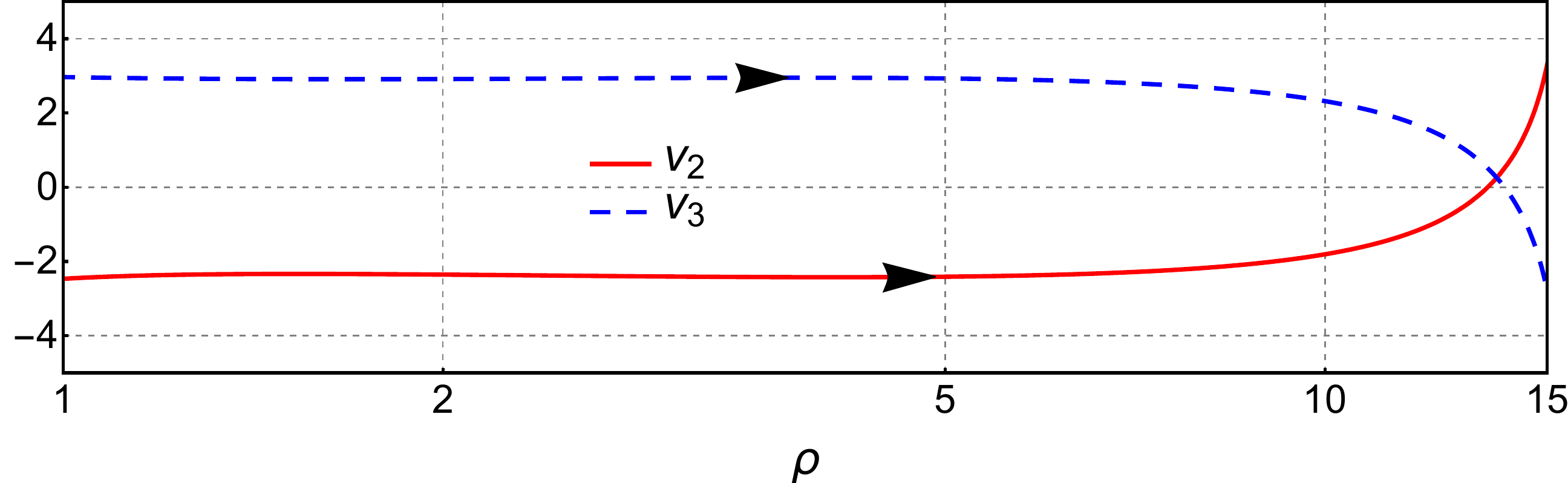}
\caption{Couplings $\{u_s,v_1,v_2,v_3\}$ as functions of $\varrho$ along the RG trajectory from the fixed point B to $\varrho\to\infty$ ($\l\to1/3^-$). 
Arrows indicate the flow from UV to IR.}
\label{Btolam13-}
\end{figure}

 On the trajectory with $c_{B1}=1$ the coupling $\varrho$ monotonically increases and at $\tau\to-\infty$ reaches another boundary of the unitarity domain \eqref{lambdaunitary} $\varrho\to\infty$ ($\l\to1/3^-$). The couplings $u_s$, $v_1$ grow rapidly in absolute value along the trajectory, while $v_2$, $v_3$ are of order 1 until the trajectory reaches vicinity of $\l=1/3$. Despite large values of $u_s$ and $v_1$, we have not encounters any numerical difficulties with integration of this trajectory until it gets quite close to $\l=1/3$. 
 The dependence of the coupling on $\varrho$ is shown in Fig.~\ref{Btolam13-}. Note that $\varrho=15$ corresponds to $\l=0.286$. 

Thus, the RG trajectories starting from B are regular and span the whole unitary domain (\ref{lambdaunitary}).  This allows us to construct a flow from the asymptotically free fixed point A which covers all admissible values of $\l$.

\subsubsection{Flow from fixed point A to $\l=O(1)$}
\label{FP5}

\begin{figure}[H]
\centering
\includegraphics[width=0.6\columnwidth]{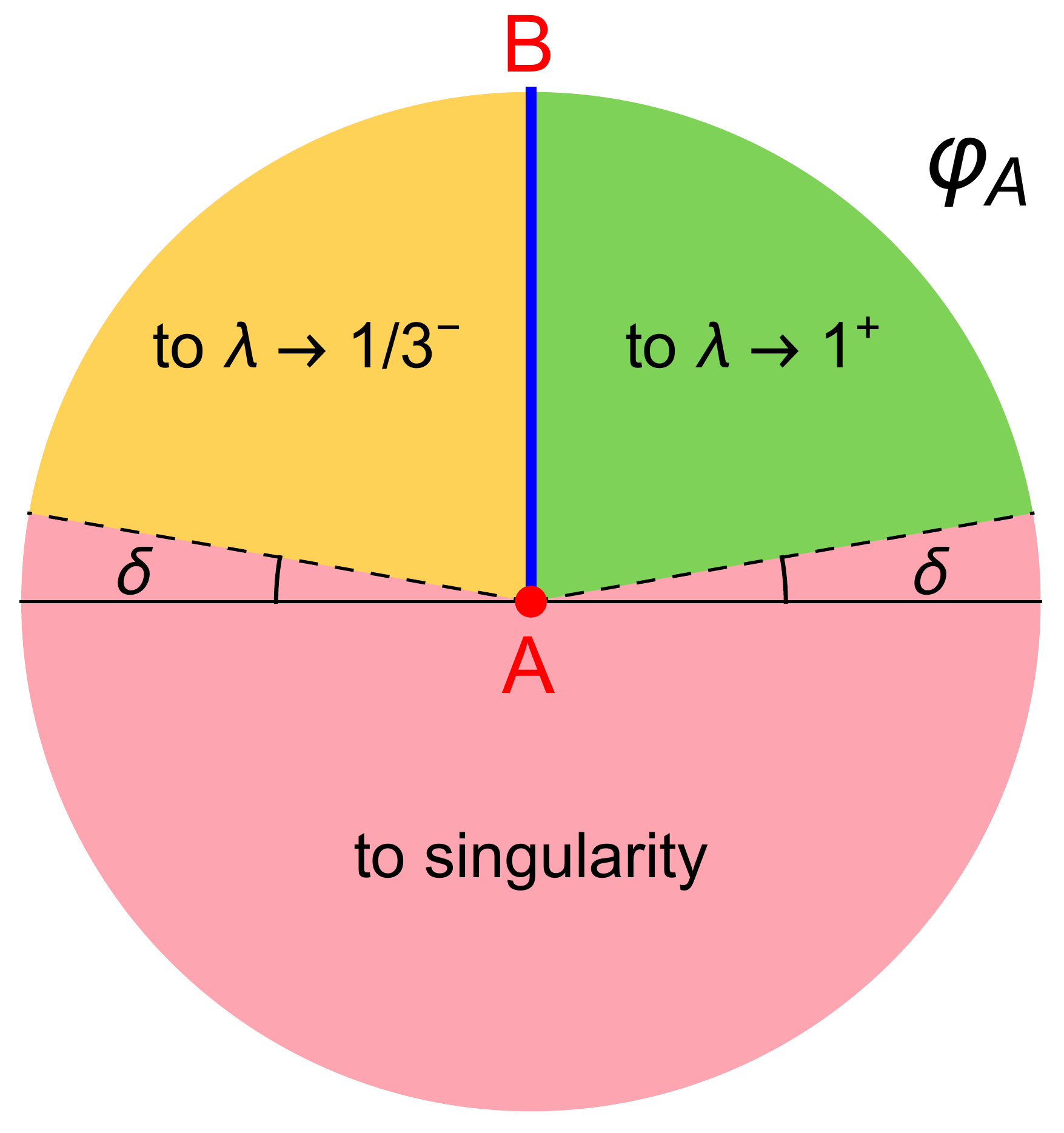}
\caption{A chart illustrating global properties of the RG trajectories flowing from the fixed point A along different directions parametrized by the angle 
$\varphi_A$, see \eqref{lkA}.}
\label{phiA}
\end{figure}

\begin{figure*}
\centering
\includegraphics[width=\columnwidth]{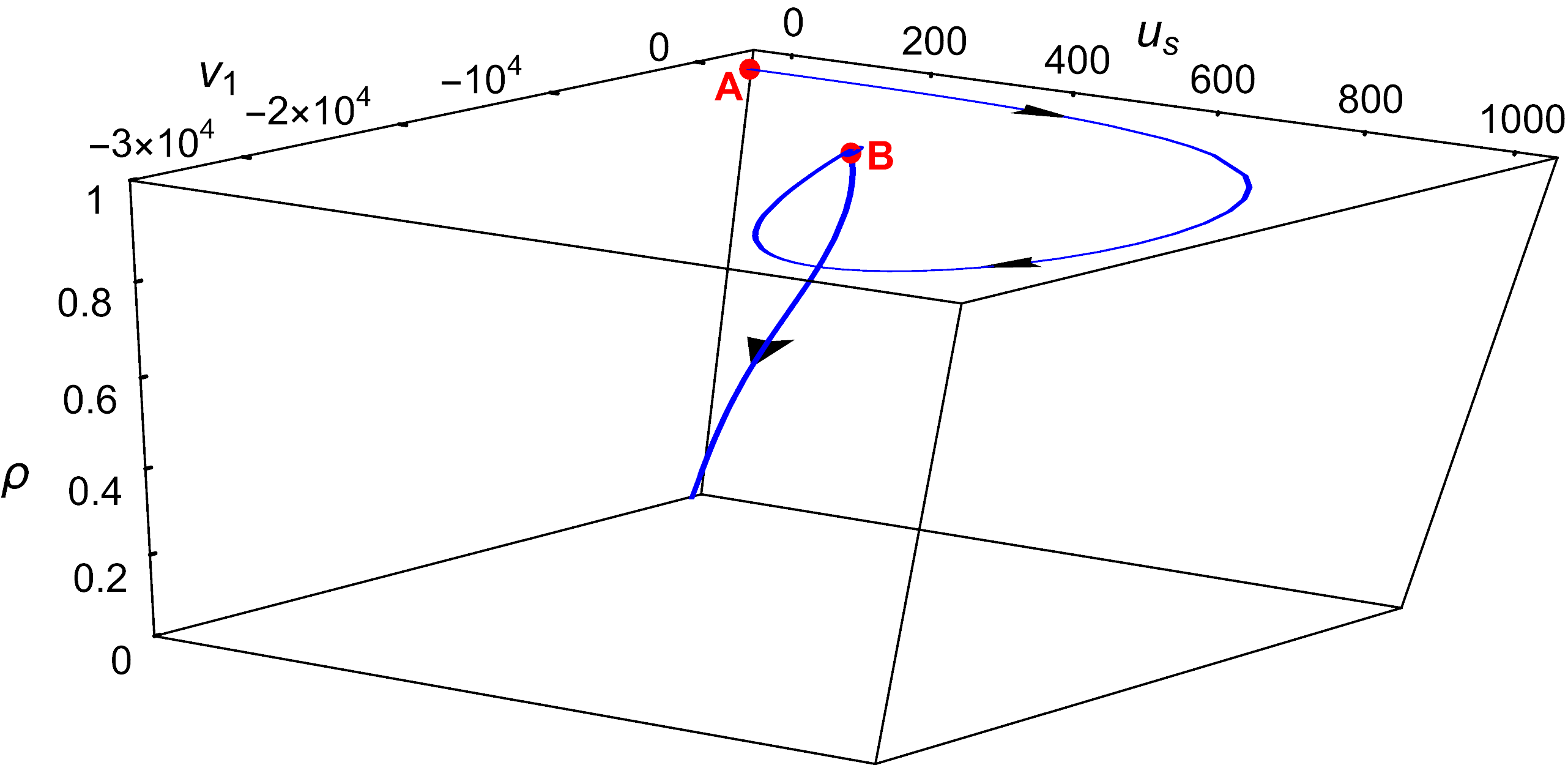}
\includegraphics[width=\columnwidth]{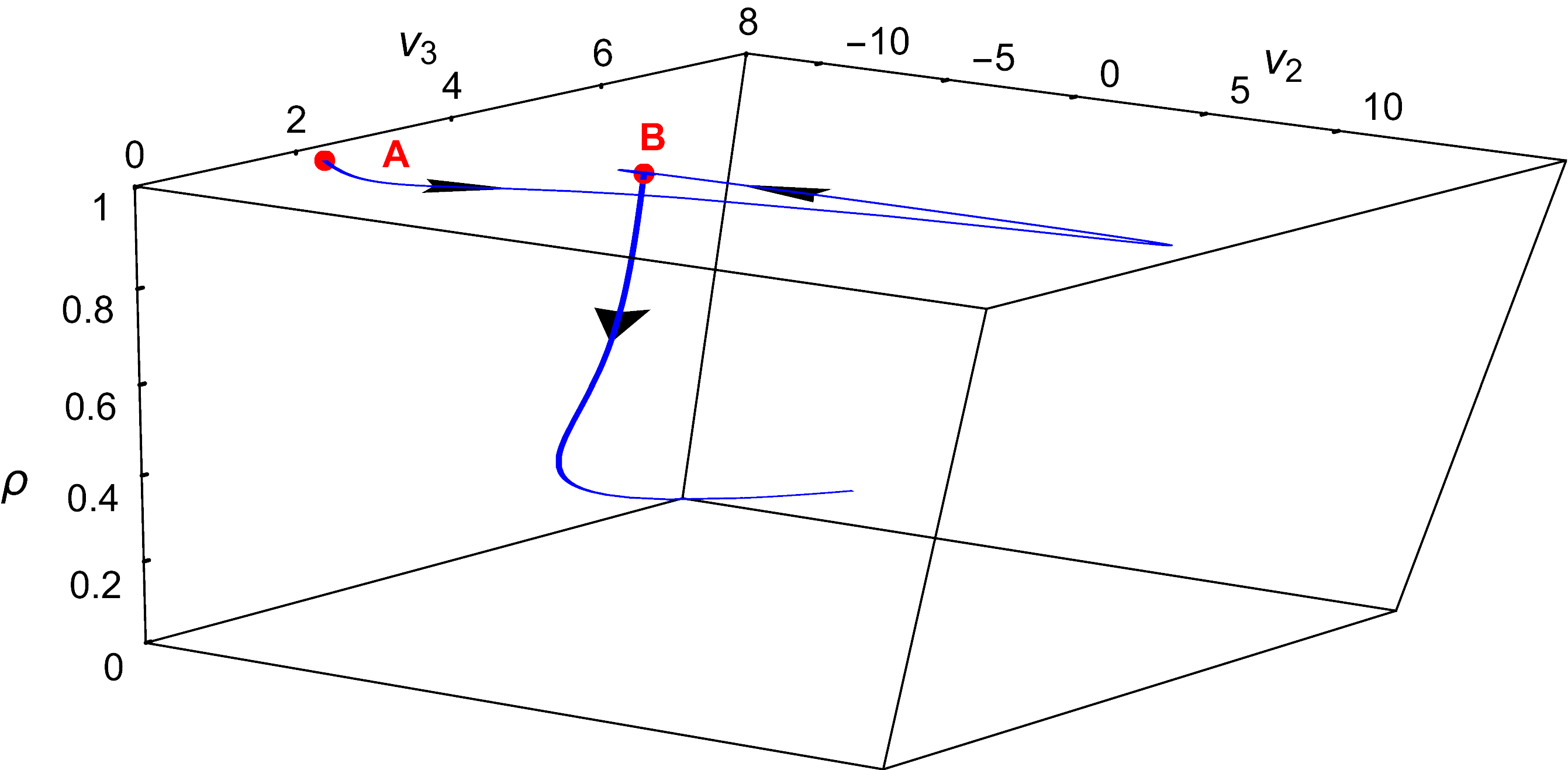}
\caption{RG trajectory from fixed point A to $\varrho\to0$ ($\l\to1^+$). Arrows indicate the flow from UV to IR. After the trajectory bypasses the neighborhood of the point B it gets attracted to the trajectory shown in Fig.~\ref{Btolam1+}. }
\label{Atolam1+}
\end{figure*}

\begin{figure*}
\centering
\includegraphics[width=\columnwidth]{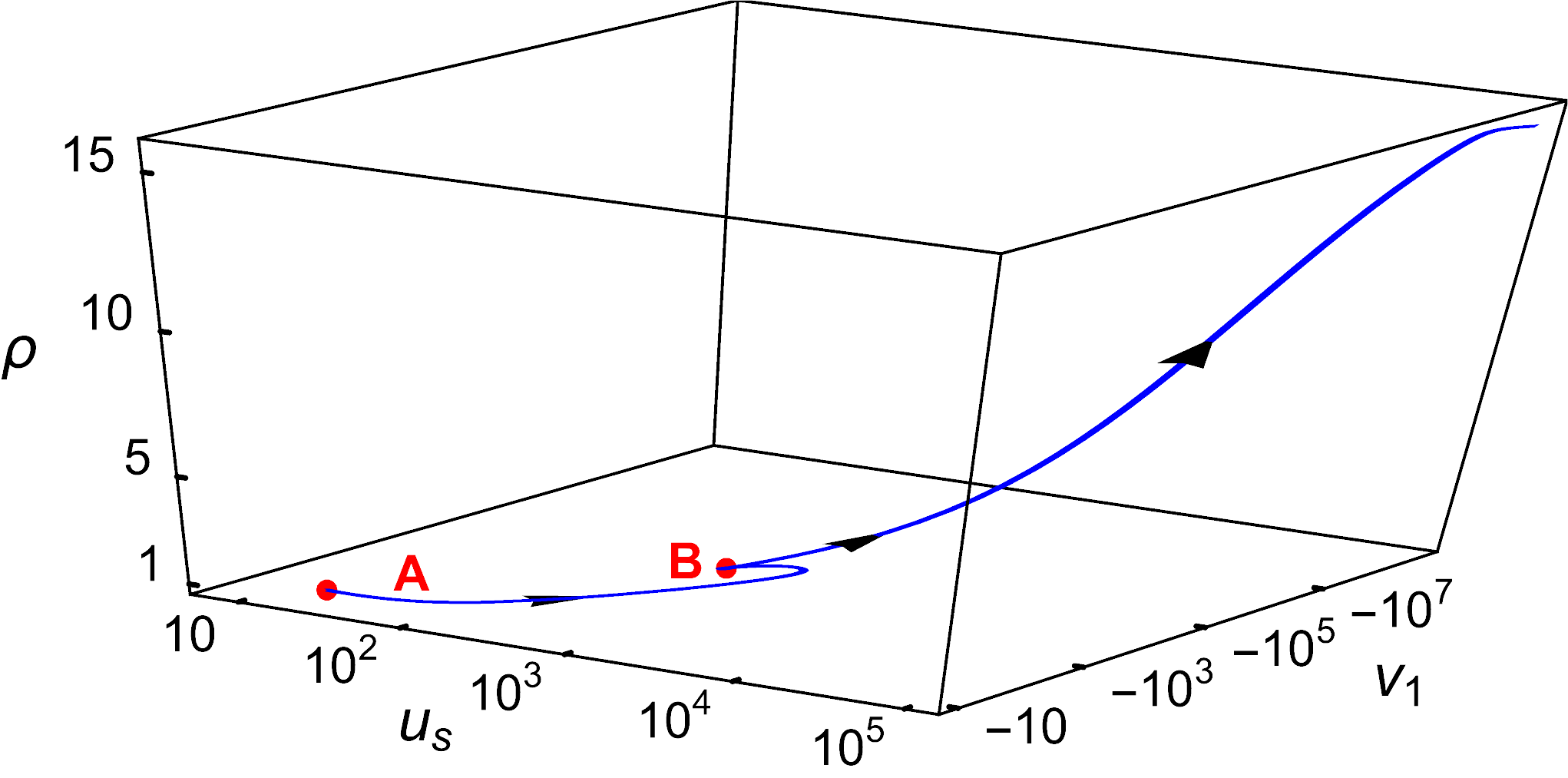}
\includegraphics[width=\columnwidth]{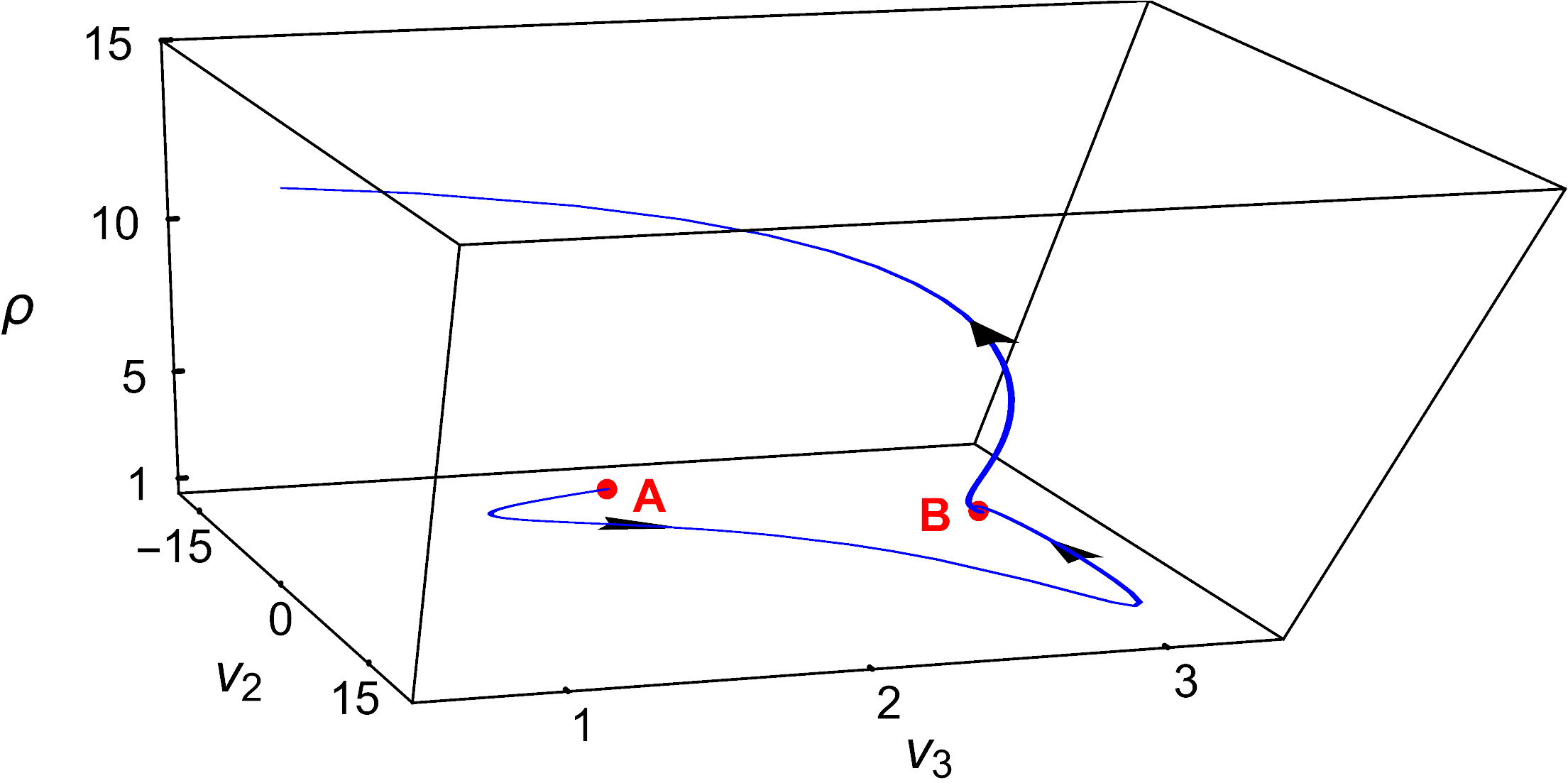}
\caption{RG trajectory from fixed point A to $\varrho\to\infty$ ($\l\to1/3^-$). Arrows indicate the flow from UV to IR. After the trajectory bypasses the neighborhood of the point B it gets attracted to the trajectory shown in Fig.~\ref{Btolam13-}. Note the logarithmic scale of $u_s$ and $v_1$ in the left plot.}
\label{Atolam13-}
\end{figure*}

We now consider a general linear combination of vectors $w^{A1}$ and $w^{A2}$ in the initial condition (\ref{lkA}) at the point A. 
For different values of $\varphi_A$ we obtain the following global behavior of the RG trajectories schematically  illustrated in Fig.~\ref{phiA}:
\begin{itemize}
\item $\varphi_A \in [\delta,\frac\pi2)$, where $\delta\ll 1$:\footnote{Precise value of $\delta$ depends on the choice of $\varepsilon$ in Eq.~(\ref{RGeq}). We find  
$\delta\simeq 2\times 10^{-8}$ for $\varepsilon=10^{-5}$.} 
the trajectory passes in the neighborhood of the point B and gets attracted to the trajectory shown in Fig.~\ref{Btolam1+}. 
This gives a family of RG trajectories starting from the asymptotically free UV fixed point A and running into the region $\varrho=0$ ($\l\to 1^+$) in IR. 
A complete trajectory for $\varphi_A=\pi/4$ is shown in Fig.~\ref{Atolam1+}. 

\item $\varphi_A=\frac{\pi}2$: the trajectory connects fixed points A and B. It is shown in  Fig. \ref{AtoB}.

\item $\varphi_A \in (\frac\pi2,\pi-\delta]$:  
the trajectory passes in the neighborhood of the point B and gets attracted to the trajectory shown in Fig.~\ref{Btolam13-}. This gives a family of trajectories starting from the asymptotically free UV fixed point A and running into the region $\varrho=\infty$ ($\l=1/3^-$) in IR. They cover the left part of the unitary domain (\ref{lambdaunitary}) $\l\in(-\infty,1/3)$.
Fig.~\ref{Atolam13-} shows such a trajectory with $\varphi_A=3\pi/4$. Absolute values of the coupling $u_s$ and $v_1$ grow very rapidly on this trajectories, while $v_2$ and $v_3$ stay of order one in magnitude.

\item $\varphi_A\in (\pi-\delta,2\pi+\delta)$: the trajectory runs into a singularity with $v_1$ diverging to negative infinity at finite value of RG parameter $\tau$.
 \end{itemize}



To sum up, the results of this subsection, together with Appendix~\ref{FPs47}, complete the investigation of RG trajectories which start from asymptotically free fixed points at $\l=\infty$ and can flow to finite $\l$. We found that only one fixed point --- point A --- gives rise to long RG trajectories reaching the region $\l=O(1)$. These trajectories cover the whole admissible range (\ref{lambdaunitary}) of the coupling $\l$ and are remarkably universal. After bypassing another (not asymptotically free) fixed point B they are attracted to one of the trajectories emanating from this point. The universal behavior of the couplings $\{u_s,v_1,v_2,v_3\}$ along these two trajectories is shown in Figs.~\ref{Btolam1+} and \ref{Btolam13-} as function of the variable $\varrho$ related to $\l$ by Eq.~(\ref{rho}). The trajectory shown in Fig.~\ref{Btolam1+} approaches in IR the phenomenologically interesting region $\l\to 1^+$ where the kinetic term of HG coincides with that of GR.

\section{Running of the gravitational coupling}
\label{sec:strength}

\subsection{Non-monotonic flow of $\boldsymbol{\cal G}$}

Until now we considered the RG flow (\ref{newbetas}) in the space of couplings $g_i=\{\l,u_s,v_a\}$ which separates from the behavior of the overall gravitational coupling ${\cal G}$. We presently discuss the running of ${\cal G}$ itself. Its RG equation has the form 
\be
\label{GRG}
\frac{d {\cal G}}{d \tau}  =  {\cal G} \hat{\beta}_{\cal G}\;,
\ee
where $\tau$ is defined in (\ref{tau}) and $\hat\beta_{\cal G}$ depends only on $g_i$. For a given RG trajectory $g_i(\tau)$ this equation can be easily integrated: 
\be
\label{Gsol}
{\cal G}(\tau) = {\cal G}_0 \exp\bigg[\int_{0}^\tau d \tau'\, \hat{\beta}_{\cal G}\big(g_i(\tau')\big)\bigg] .
\ee
Thus, the behavior of ${\cal G}$ along a trajectory is determined up to an overall normalization ${\cal G}_0$. We are primarily interested in the trajectories running from the point A towards $\l\to 1^+$ or $\l\to 1/3^-$. Along these trajectories $\l$ changes monotonically, so $\tau$ and hence ${\cal G}$ can be expressed as functions of $\l$.  

\begin{figure}[H]
\centering
\includegraphics[width=\columnwidth]{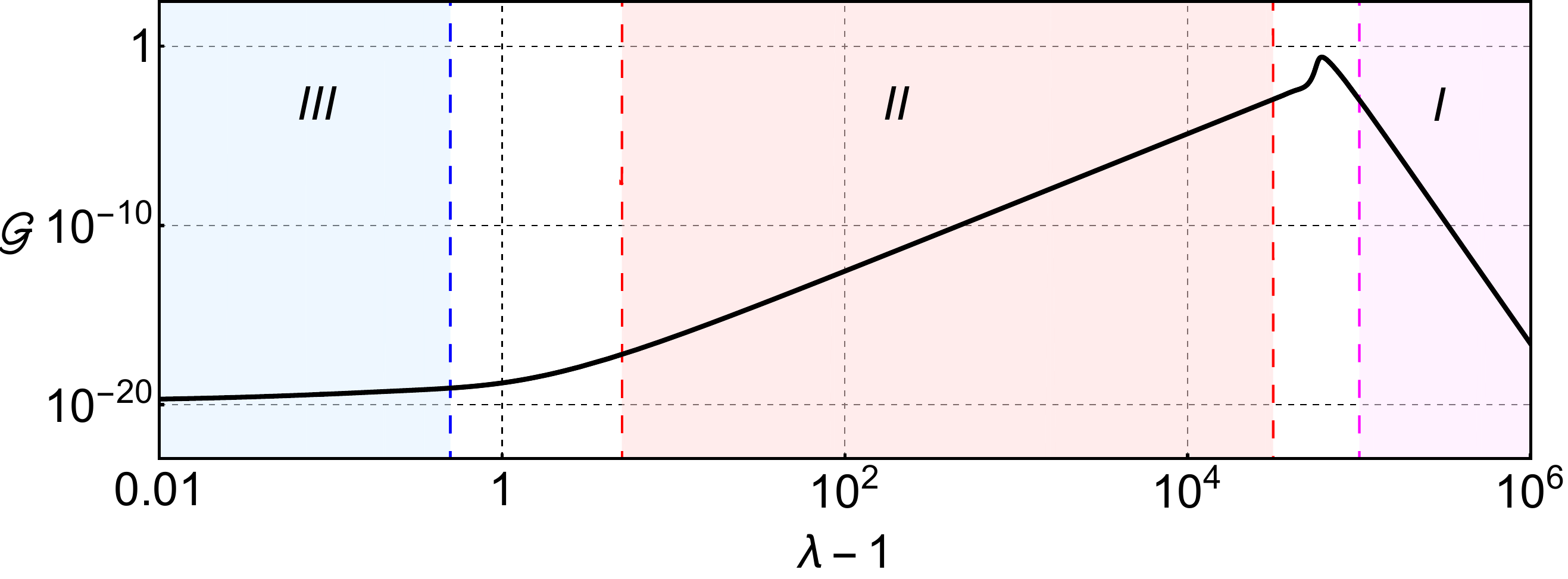}
\caption{Behavior of ${\cal G}$ as a function of $(\l-1)$ along an RG trajectory connecting the point A to $\l\to 1^+$. In regions $I$, $II$ and $III$ the dependence is well described by the power law ${\cal G}\propto (\l-1)^\kappa$ 
with $\kappa_I=-13.69$, $\kappa_{II}=3.84$, $\kappa_{III}\approx 0.37$. 
}
\label{PlotG}
\end{figure}

In Fig.~\ref{PlotG} we plot ${\cal G}(\l)$ on the RG trajectory from Fig.~\ref{Atolam1+} which belongs to the flow from A to $\lambda\to 1^+$. 
A striking feature of this flow is non-monotonic behavior of ${\cal G}$. This has a transparent explanation. When the trajectory leaves the asymptotically free fixed point A ($\l$ decreases from infinity) ${\cal G}$ grows. The trajectory is then attracted to the point B which is not asymptotically free. Somewhere in between the points A and B the beta-function $\hat\beta_{\cal G}$ changes sign and ${\cal G}$ reaches a maximum. The trajectory spends a long `RG time' in the vicinity of B where $\hat\beta_{\cal G}$ is positive, so the gravitational coupling decreases.\footnote{Recall that $\tau\to-\infty$ along the flow.} 
Eventually, the trajectory escapes from the points B and gets into the region $\l=O(1)$. 

The dependence of ${\cal G}$ on $\l$ in regions $I$ and $II$ on the plot is dominated by the points A and B, respectively. It is well described by the power law
\be
\label{Gpl}
{\cal G}\propto \l^\kappa\;,
\ee
where the exponent $\kappa$ can be found as follows. Consider
\be
\frac{d {\cal G}}{d\l} = \frac{\beta_{\cal G}}{\beta_\l}= \frac{{\cal G}}{\l} \cdot\frac{\hat\beta_{\cal G}}{\hat\beta_\l}\;,
\ee 
where both $\hat\beta_{\cal G}$ and
$\hat\beta_\l\equiv\beta_\l/\l$ are finite at $\l\to\infty$ (see Eqs.~(\ref{beta_lam}), (\ref{betaG})).
The ratio $\hat\beta_{\cal G}/\hat\beta_\l$ is approximately constant in the vicinity of each fixed point, giving rise to the power law (\ref{Gpl}) with  
\bseq
\label{kappa12}
\begin{align}
\label{kappa1}
&\kappa_I=(\hat\beta_{\cal G}/\hat\beta_\l)\big|_A=-13.69\;,\\
\label{kappa2}
&\kappa_{II}=(\hat\beta_{\cal G}/\hat\beta_\l)\big|_B=3.84\;.
\end{align}
\eseq

The steep dependence of ${\cal G}$ on $\l$ in the region $II$ has an important consequence. For the validity of the perturbative expansion, ${\cal G}$ must be less than unity everywhere along the flow. In particular, ${\cal G}<1$ must be satisfied when the trajectory bypasses the fixed point B. Let us denote the corresponding value of $\l$ by $\l_B$. Since the point B lies at $\l=\infty$, we have $\l_B\gg 1$. Then the value of he gravitational coupling at $\l\sim 1$ is ${\cal G}<\l_B^{-\kappa_{II}}\ll 1$. In other words, the IR value of ${\cal G}$ is necessarily very small. This is indeed observed in Fig.~\ref{PlotG}. We will discuss this property in more detail below.

At $(\l-1)\ll 1$ the dependence ${\cal G}(\l)$ flattens out. We find numerically that at $0.01\lesssim (\l-1)\lesssim 1$ it also approximately follows a power law, now of the form ${\cal G}\propto (\l-1)^{0.37}$. Finally, in the region $(\l-1)\ll 10^{-3}$ (not shown in Fig.~\ref{PlotG}) the curves flattens further and asymptotically approaches the form 
\be
\label{GIRscaling}
{\cal G}\,\big|_{\,\l\to1} \propto (\l-1)^{17/448}\;.
\ee
Derivation of this asymptotics is given in Appendix~\ref{asympt}.

The fact that ${\cal G}$ decreases towards IR may suggest that in the IR limit $\l\to 1^+$
the theory becomes free. 
This is unlikely. The inverse powers of $(\l-1)$ in the beta-functions (\ref{betafun}) lead to growth of other couplings which jeopardizes the perturbative expansion. 
In Appendix~\ref{asympt} we derive the behavior of $u_s$ and $v_a$ in deep IR,
\be
\label{IRscalings}
u_s\,\big|_{\,\l\to1} \propto (\l-1)^{241/448}\;,\qquad
v_a\,\big|_{\,\l\to1} \propto (\l-1)^{-1}\:.
\ee
The rapid growth of $v_a$ is not compensated by the slow decrease of ${\cal G}$ (\ref{GIRscaling}) in the physical observables, such as e.g. differential scattering cross section of tt-gravitons \cite{Radkovski:2023cew}. This indicates that the theory gets strongly coupled if the RG reaches too close to $\l=1$. 

It is worth noting in this connection that the properties of the RG flow in deep IR will be altered by the relevant operators in the HG action, which we have neglected in our analysis. Depending on the scale of these operators, the change can happen before the theory reaches into strong coupling. While studying the RG flow in the presence of relevant operators is beyond the scope of this paper, we note that they are expected to stabilize the strength of the gravitational interaction from increasing towards IR since HG at low energies is equivalent to GR coupled to a scalar field with derivative self-interactions~\cite{Blas:2010hb}. 

\begin{figure}[H]
\centering
\includegraphics[width=\columnwidth]{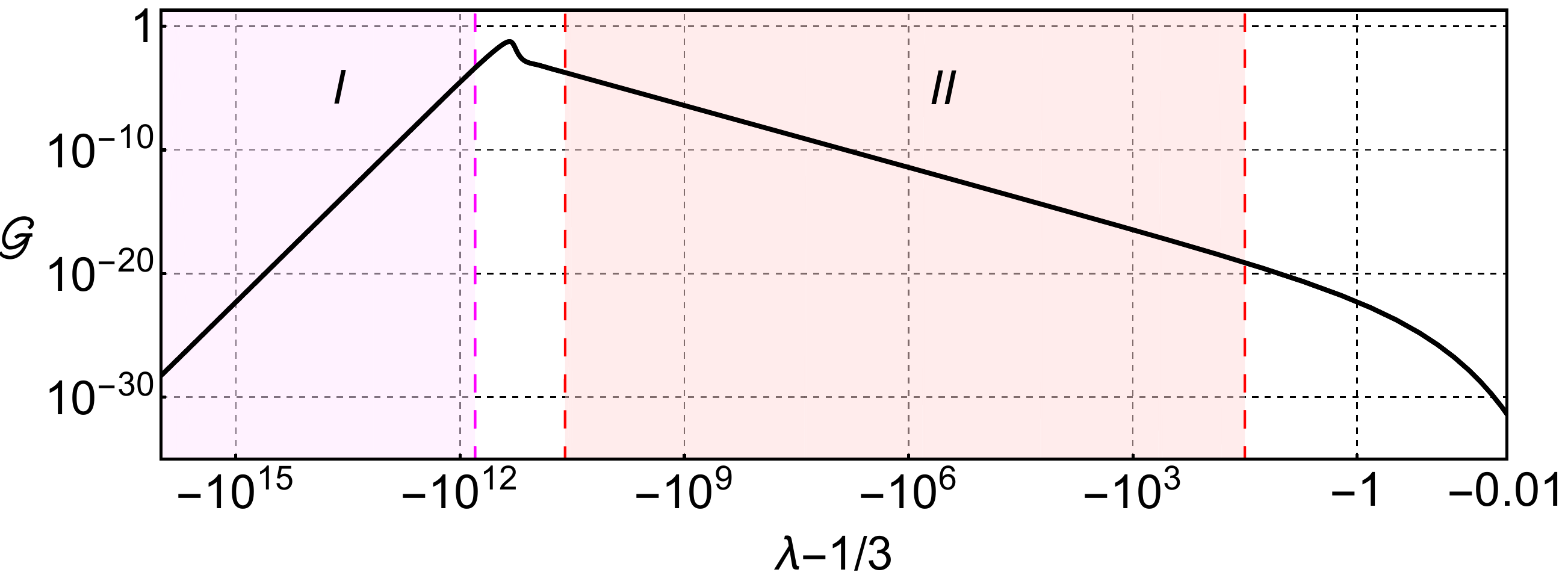}
\caption{Behavior of ${\cal G}$ as a function of $(\l-1/3)$ along an RG trajectory connecting the point A to $\l\to 1/3^-$. In regions $I$ and $II$  the dependence is well described by the power law ${\cal G}\propto |\l|^\kappa$ 
with $\kappa_I=-13.69$, $\kappa_{II}=3.84$. 
}
\label{PlotG13}
\end{figure}

For completeness, we show in Fig.~\ref{PlotG13} the running of ${\cal G}$ along the RG trajectory from Fig.~\ref{Atolam13-} 
interpolating between the fixed point A and the region $\l\to1/3^-$.
In regions $I$ and $II$ of the plot the trajectory is again dominated by the fixed points A and B, respectively, leading to non-monotonic variation of ${\cal G}$. The dependence ${\cal G}(\l)$ in these regions is captured by the power law (\ref{Gpl}) with the exponents (\ref{kappa12}). When $\l$ gets close to $1/3$ the coupling ${\cal G}$  decreases abruptly. We do not know if this implies weak coupling along this trajectory in IR since other couplings $\{u_s,v_a\}$ rapidly grow, see Sec.~\ref{ssec:Bto13}. 

\subsection{Hierarchy between Planck mass and Lorentz violating scale}

In this subsection we focus on the RG trajectories interpolating between the asymptotically free fixed point A and the phenomenologically interesting region $\l\to1^+$. We have seen that the applicability of the perurbaation theory along the trajectory implies a very small IR value of the gravitational coupling. Here we explore this property and its consequences in more detail. 

We define the IR gravitational coupling as the running coupling ${\cal G}$ evaluated at a fixed value of $\l$ close to $\l=1$,
\be
\label{GIRdef}
{\cal G}_{IR}={\cal G}_{IR}(\l)\big|_{\l=1.01}.
\ee
Let us analyze how ${\cal G}_{IR}$ depends on the initial direction of the RG trajectory at the point A. This direction is controlled by the angle $\varphi_A$ in the linear combination (\ref{lkA}) of the repulsive vectors $w^{A1}$, $w^{A2}$. The projection of the trajectory on the subspace of couplings $\{u_s,v_a\}$ is practically independent of the angle, as long as it lies in the interval $\varphi_A\in (\delta,\pi/2)$, see Sec.~\ref{FP5}. However, this angle determines how closely the trajectory bypasses the fixed point B, which has a major effect on ${\cal G}_{IR}$.  

If $\varphi_A\sim \pi/2$, the flow initially runs very close to the plane $\varrho=1$ ($\l=\infty$) and almost hits the point B, before being deflected towards $\l\to 1^+$. 
The trajectory spends a long `RG time' in the vicinity of the point B, during which the gravitational coupling decreases from its maximal value ${\cal G}_{\rm max}$ down to ${\cal G}_{IR}$. Consistency of the perturbation theory requires\footnote{This can be achieved by adjusting the initial value ${\cal G}_0$ near the asymptotically free point A.} ${\cal G}_{\rm max}<1$
which leads to ${\cal G}_{IR}$ being very small. 

By tuning $\varphi_A$ close to $\delta$ --- the boundary of the flow attracted to B --- we can decrease the amount of the `RG time' spent by the trajectory in the neighborhood of B. This increases the value of ${\cal G}_{IR}$. In Appendix~\ref{app:D} we derive its scaling at small $(\varphi_A-\delta)$,
\be
\label{GIRscalingphi}
{\cal G}_{IR}={\cal G}_{IR}^{(0)}\,(\varphi_A-\delta)^\alpha,~~~~
\alpha=-\kappa_{II}\cdot\frac{\theta^1}{\theta^2}=-0.776\;,
\ee
where $\kappa_{II}$ is defined in Eq.~(\ref{kappa2}), and $\theta^1$, $\theta^2$ are the eigenvalues corresponding to the repulsive vectors $w^{A1}$, $w^{A2}$ at the point A, see Table~\ref{EVlam5}. The coefficient ${\cal G}_{IR}^{(0)}$ depends on the maximal value of ${\cal G}$ along the flow.

\begin{figure}[h]
\centering
\includegraphics[width=0.85\columnwidth]{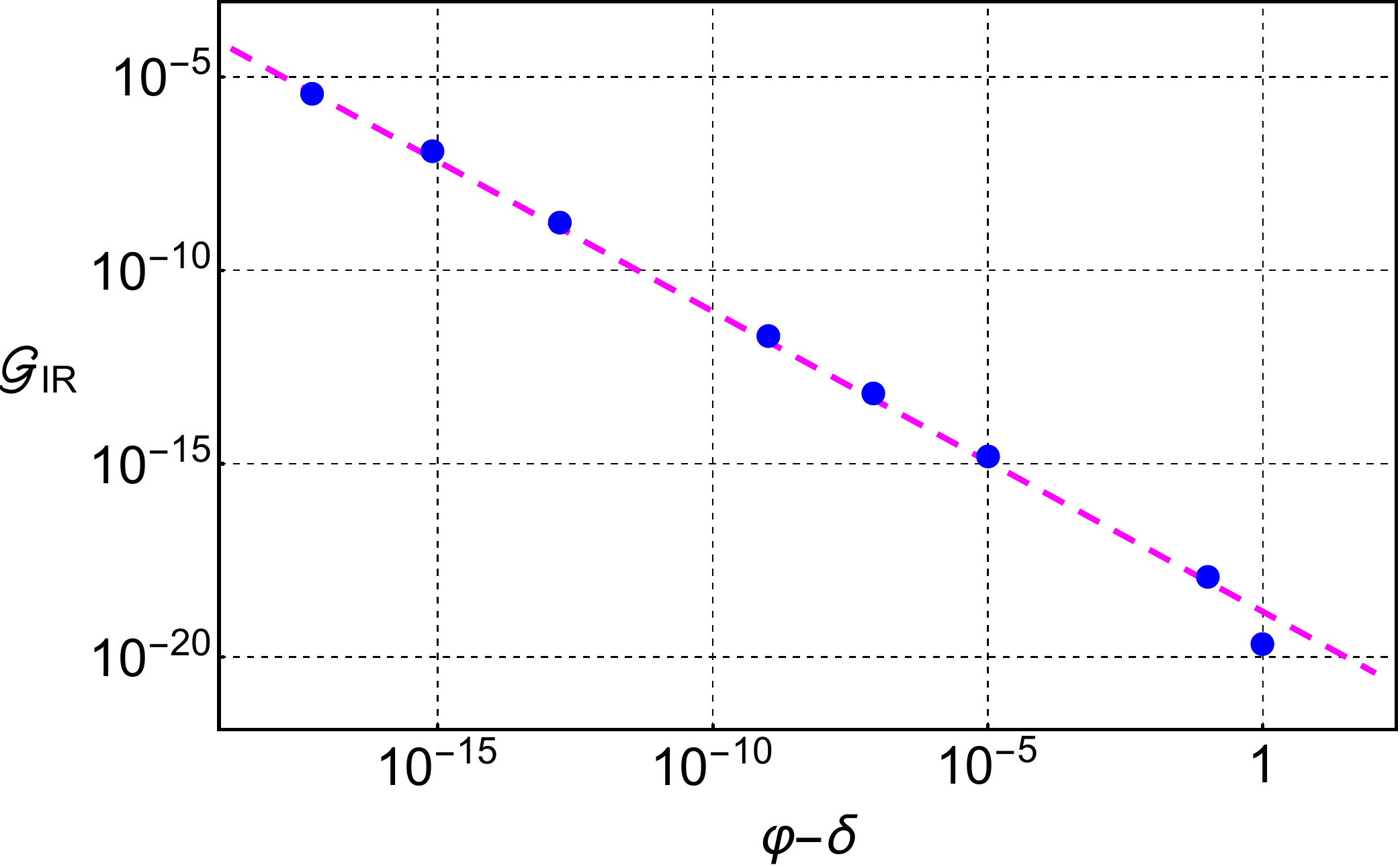}
\caption{Dependence of the IR value of the gravitational coupling on the initial direction of the trajectory at the point~A. Dots show the result of numerical integration of RG equations while dashed line is the fit by Eq.~(\ref{GIRscalingphi}).} 
\label{plotGphi}
\end{figure}

Fig.~\ref{plotGphi} shows the dependence of ${\cal G}_{IR}$ on $(\varphi_A-\delta)$ found numerically\footnote{To explore very small deviations $(\varphi_A-\delta)\sim 10^{-17}$ we have determined the critical angle $\delta$ with the accuracy $10^{-19}$.} with ${\cal G}_{\rm max}=0.25$. We observe that even extreme fine-tuning at the level $(\varphi_A-\delta)\sim 10^{-17}$ gives at most ${\cal G}_{IR}\sim 10^{-5}$, whereas typical values 
$(\varphi_A-\delta)\sim 1$ correspond to ${{\cal G}_{IR}\sim 10^{-20}}$. The dependence indeed obeys the scaling (\ref{GIRscalingphi}) with ${\cal G}_{IR}^{(0)}\simeq 1.5\cdot 10^{-19}$. 

The smallness of ${\cal G}_{IR}$ has an important implication. Imagine that the RG flow is cut off by the relevant operators at some scale where $\l\simeq 1$. At this scale all couplings must be matched to their values in the low-energy theory. Recall that the dimensionless gravitational coupling is defined as the ratio ${\cal G} = G/\sqrt{\nu_5}$. As discussed in Sec.~\ref{sec:projectableHG}, the parameter $\nu_5^{-1/4}$ is identified within the low-energy theory with the Lorentz breaking scale $M_{LV}$, see Eq.~(\ref{MLV}). On the other hand, the dimensionful coupling $G$ becomes essentially the Newton's constant.\footnote{Barring the issues related to the Minkowski space instability.}  
Expressing it conventionally through the Planck mass, $G=M_{Pl}^{-2}$, we obtain  a vast hierarchy between the latter and the Lorentz violation scale,
\be
\label{hier}
\frac{M_{LV}}{M_{Pl}}=\sqrt{{\cal G}_{IR}}\ll 1\;.
\ee
It is remarkable that this hierarchy is not only technically natural, but actually enforced by the consistency of the RG flow. It would be interesting to see if a similar hierarchy (\ref{hier}) naturally arises in other versions of HG, in particular, the non-projectable model. In that case it could lead to strong suppression of Lorentz violating effects in gravity at low energies \cite{Blas:2010hb,Blas:2014aca,EmirGumrukcuoglu:2017cfa}.

\section{Summary}
\label{sec:conclusions}

We performed a comprehensive study of the RG flow of marginal couplings in $(3+1)$-dimensional projectable HG. We first identified all fixed points of the flow: five points at finite value of the kinetic coupling $\l$ and eight points at $\l=\infty$. We presented strong numerical evidence that no other fixed points exist. All fixed points at finite $\l$ are asymptotically free. However, unlike the case of HG in $(2+1)$-dimensions \cite{Barvinsky:2017kob}, they all lie at $\l<1/3$, in the left part of the domain (\ref{lambdaunitary}) allowed by unitarity. Thus, they cannot be connected by RG trajectories to the phenomenologically interesting region $\l\to1^+$. Three fixed points out of five 
are very close to the boundary of the unitary domain $\l=1/3$ and feature very large values of couplings $\{u_s,v_1\}$.
Out of the eight fixed points at infinity only three points are asymptotically free and can give rise to trajectories flowing towards $\l\to1^+$.

We next analyzed the local properties of the RG flow around the fixed points by studying eigenvalues and eigenvectors of their stability matrices. This allowed us to identify the attractive and repulsive directions of the fixed points. Quite unexpectedly, we found that some fixed points are characterized by complex conjugate pairs of eigenvalues. This appears at clash with the intuition that the eigenvalues of the stability matrix are associated with the anomalous dimensions of the operators generating the RG flow. While complex anomalous dimensions are quite common in non-unitary theories (see e.g. \cite{Hogervorst:2015akt,Gorbenko:2018ncu,Gromov:2017cja,Jepsen:2020czw}), they are believed to be incompatible with unitarity. 

Still, we do not think that the presence of complex eigenvalues of the stability matrix signals violation of unitarity in HG. Rather, the relation between properties of the fixed points and unitarity can be subtle because HG is a gauge theory. In non-Abelian theories the gauge group degenerates into its linearized version at asymptotically free fixed points. The operators describing interactions and generating the RG flow are not invariant under this linearized gauge group. In other words, they do not belong to the spectrum of gauge invariant operators at the (free) fixed point and unitarity constraints do not apply to them. 

Besides, it should be reminded that in the case of HG the dimensions of operators are defined with respect to the anisotropic (Lifshitz) scaling. We are not aware of any rigorous reality conditions in general Lifshitz theories. Exploring unitarity constraints on the dimensions of operators and RG flows in such theories will be an interesting task. To the best of our knowledge, HG is the first example of a unitary theory with complex eigenvalues of the stability matrix. It would be interesting to search for more theories with this property. 

With the insight from the local analysis we moved to the global properties of RG trajectories emanating from asymptotically free fixed points. We considered two fixed points at finite $\l$ and moderate $\{u_s,v_1\}$, initiating a family of trajectories along their repulsive directions. We found that all such trajectories run into singularity, with some couplings diverging in a finite `RG time'. Similar behavior is exhibited by trajectories starting from two of the three fixed points at $\l=\infty$ which we previously identified as potentially interesting. The parameter $\l$ along these trajectories remains very large all the way from the fixed point to the singularity. In particular, the trajectories from these points never reach into the phenomenologically interesting region $\l\to1^+$. 

However, one fixed point at $\l=\infty$, which we called point A, does give rise to long trajectories interpolating to $\l\to1^+$ or $\l\to1/3^-$, depending on the initial conditions. The RG flow emanating from the point A thus covers the whole range (\ref{lambdaunitary}) allowed by unitarity. The RG trajectories have a peculiar structure: upon leaving the point A they get attracted to another, not asymptotically free, fixed point B also at $\l=\infty$. Coming close to the point B, they `scatter' on it and continue along its single repulsive direction towards $\l=O(1)$. The trajectories from a family with one of the two asymptotics ($\l\to1^+$ or $\l\to1/3^-$) are very close to each other in the space of couplings $\{\l,u_s,v_1,v_2,v_3\}$. What distinguishes them is the running of the overall gravitational coupling~${\cal G}$. Remarkably, this running is not monotonic, with ${\cal G}$ becoming very small both in UV and IR. 

We argued that, despite vanishing of ${\cal G}$, the theory becomes strongly coupled in IR if the RG flow gets too close to the boundary of the unitary domain $\l=1$. This running into strong coupling can be prevented by the relevant operators which will cut the flow at some finite IR values $\l_{IR}$ and ${\cal G}_{IR}$. We showed that if $\l_{IR}$ is order-one, the value ${\cal G}_{IR}$ is naturally tiny. This implies that the Planck mass inferred from the strength of gravitational interactions in the low-energy theory is hierarchically larger --- by as many as 10 orders of magnitude --- than the scale of Lorentz symmetry breaking. If such natural strong hierarchy arises also in the non-projectable HG, it can strongly suppress its deviations from GR at low energies \cite{Blas:2010hb,Blas:2014aca,EmirGumrukcuoglu:2017cfa}, as well as leakage of Lorentz violating effects from gravity to the matter sector \cite{Pospelov:2010mp}. The study of this intriguing possibility is left for future.

\vspace{0.5cm}

{\bf Acknowledgements} - We are grateful to Cliff Burgess, Davide Gaiotto, Ted Jacobson, Vladimir Kazakov, Shinji Mukohyama, Jury Radkovski and Marc Schiffer for valuable discussions.
The work of A.B. and A.K. was supported by the Russian Science Foundation Grant No.~23-12-00051,  https://rscf.ru/en/project/23-12-00051. 
The work of S.S. is supported by the Natural Sciences and Engineering Research Council (NSERC) of Canada. Research at Perimeter Institute is supported in part by the Government of Canada through the Department of Innovation, Science and Economic Development Canada and by the Province of Ontario through the Ministry of Colleges and Universities.

\appendix

\section{Numerical search for fixed points}
\label{AppA}

\subsection{Finite $\l$}

As described in Sec.~\ref{sec:FPfinl}, we eliminate $\l$ from the system of equations for the fixed points using Eq.~(\ref{lamus}) and introduce the variable $u_t$ according to Eq.~(\ref{ut}). We scan over the values $u_t^*$ in the interval from $\sqrt{10}$ to $10^8$ with different steps $\epsilon$ in different intervals:
\begin{itemize}
\item $u_t^*\in (\sqrt{10},5]$: step $\epsilon=10^{-5}$;
\item $u_t^*\in [5,100]$: step $\epsilon=10^{-4}$;
\item $u_t^*\in [100, 1000]$: step $\epsilon=10^{-3}$;
\item $u_t^*\in [10^{m+3}, 10^{m+4}]$, $m = 0,\ldots,4$: step ${\epsilon=10^{m}}$.
\end{itemize}

\begin{figure}[H]
\centering
\quad\:\includegraphics[width=0.8\columnwidth]{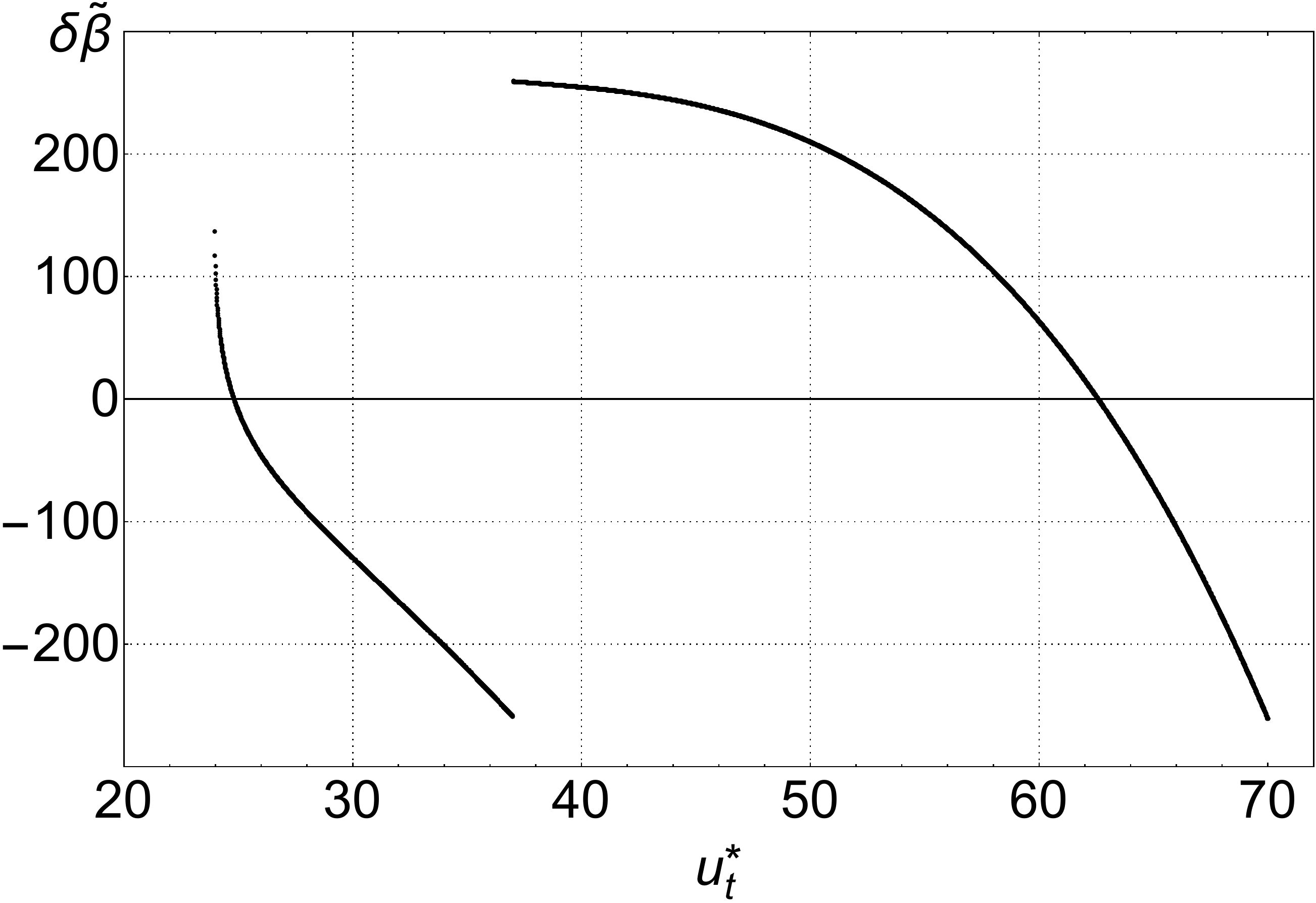}\\
~\\
\includegraphics[width=0.85\columnwidth]{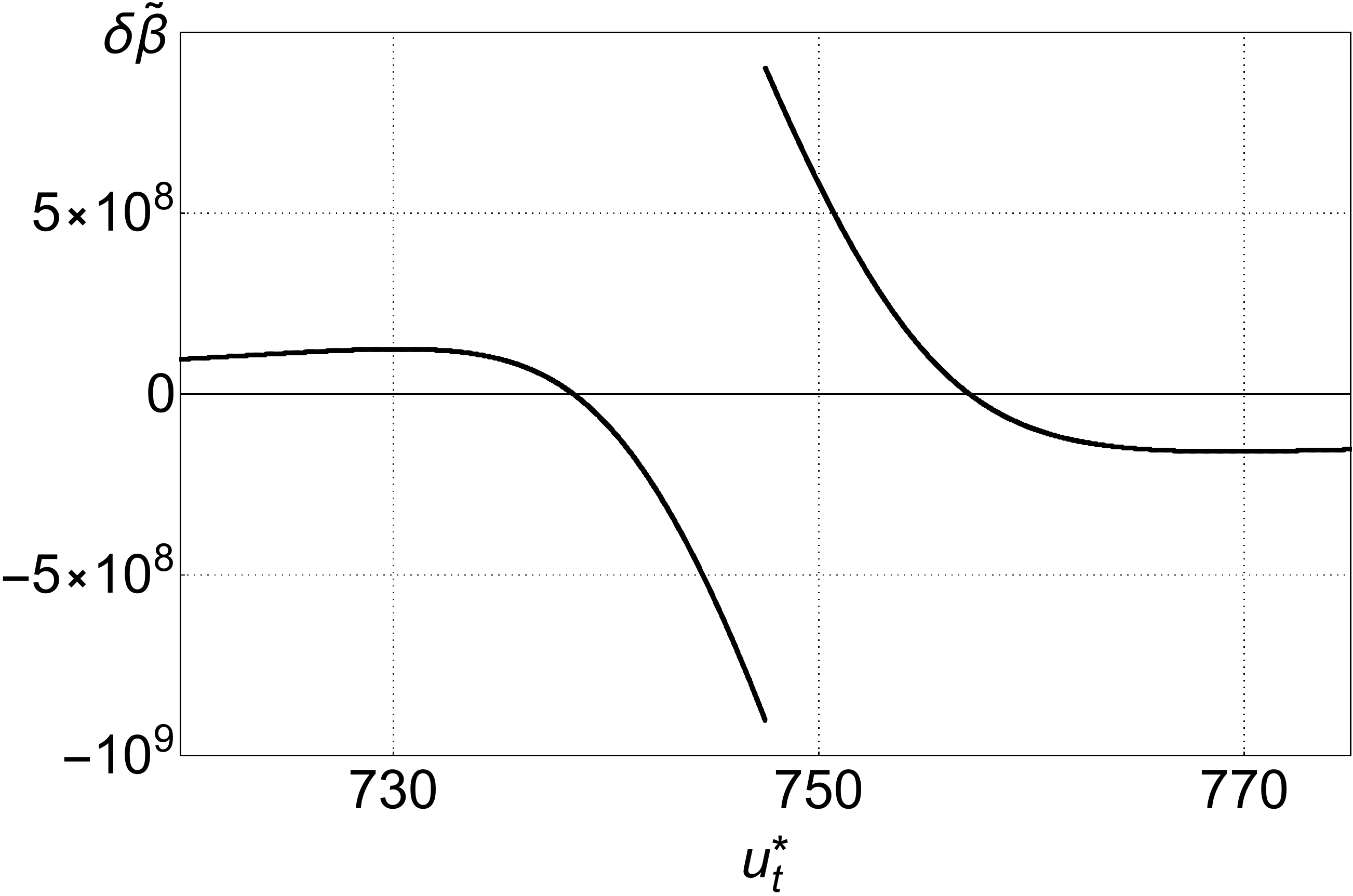}\\
\vspace{5mm}
\includegraphics[width=0.85\columnwidth]{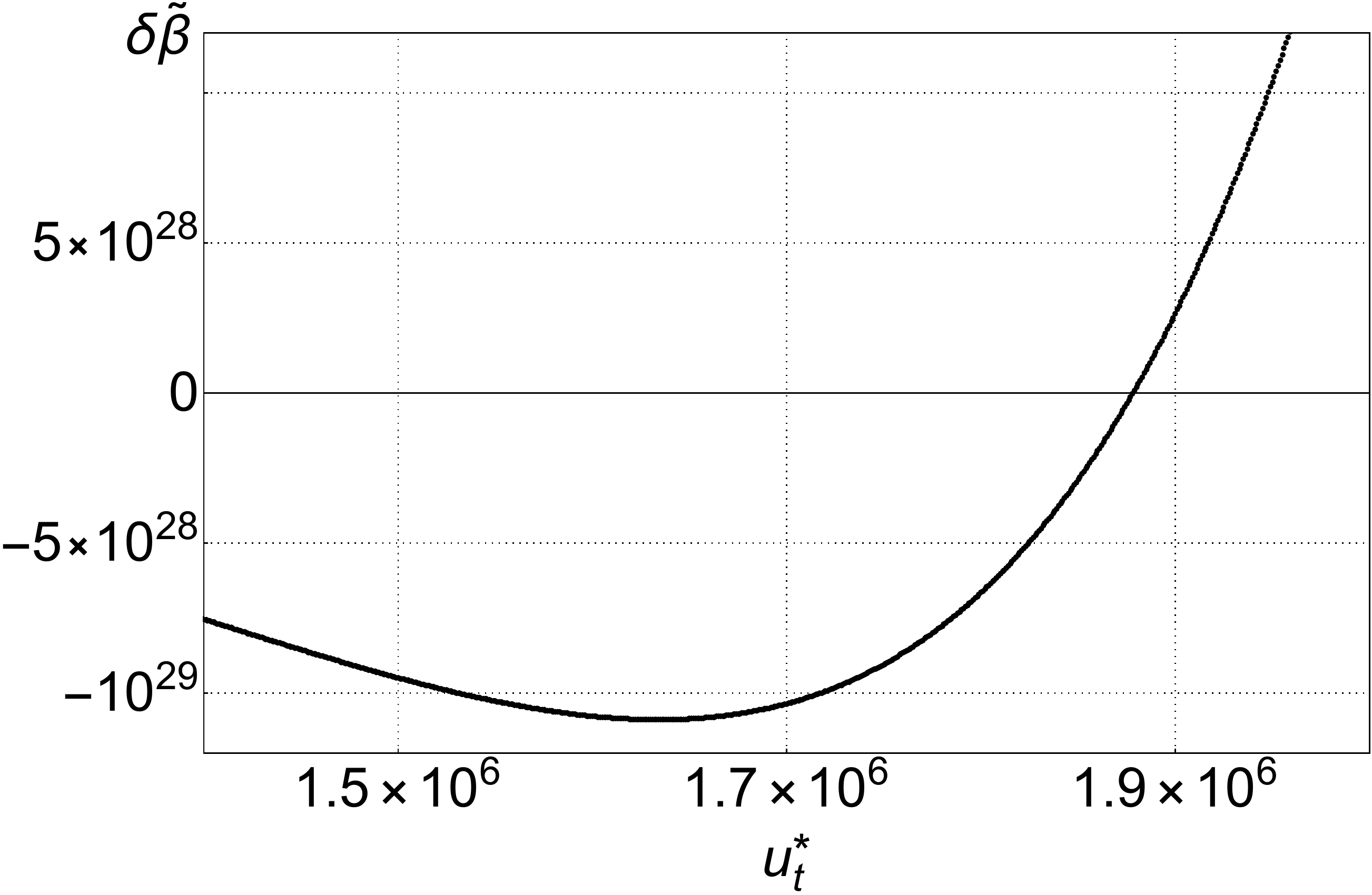}
\caption{Dependence of the residual $\delta\tilde\beta$ defined in Eq.~(\ref{minbeta}) on $u_t^*$ in the intervals where $\delta\tilde\beta$ crosses zero. For other $u_t^*$ values $\delta\tilde\beta$ is strictly positive or negative. The discontinuities in $\delta\tilde\beta$ occur when the partial root minimizing the residual switches to a different branch. }
\label{plot_ut}
\end{figure}
For each value of $u_t^*$ we construct the residual $\delta\tilde\beta$ as described in Sec.~\ref{sec:FPfinl}. Figure~\ref{plot_ut} shows the plots $\delta\tilde\beta(u_t^*)$ in several intervals where they cross zero. We observe five crossings which correspond to five solutions of the system (\ref{fullsystem}). Discontinuities of $\delta\tilde\beta(u_t^*)$ occur when the partial root minimizing the residual switches from one branch of roots to another. 

\begin{figure}[H]
\centering
\includegraphics[width=\columnwidth]{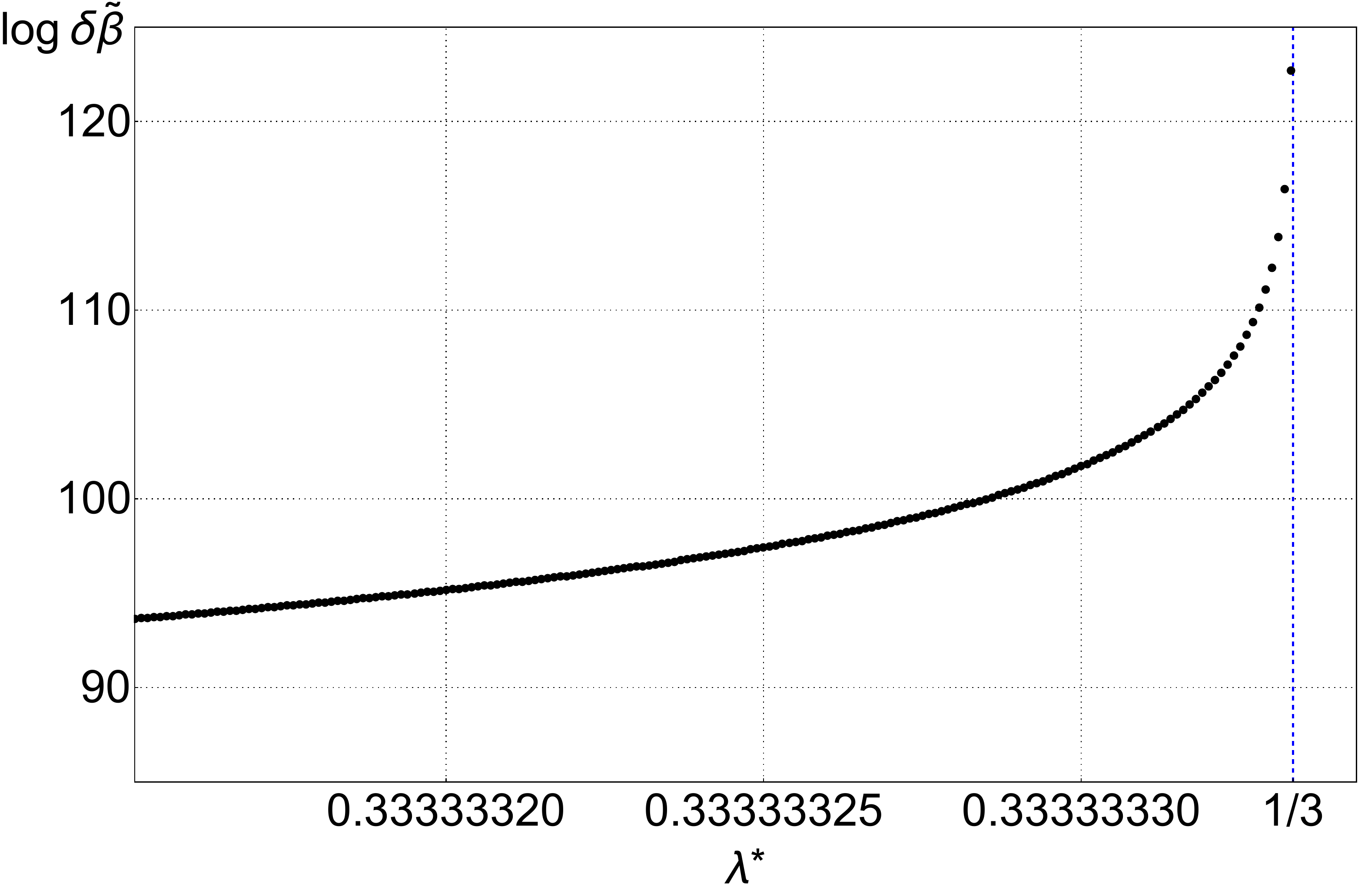}
\caption{
Behavior of the residual $\delta\tilde\beta(\l^*)$ in the small neighborhood of the point $\l^*=1/3$. Scanning step in $\l^*$ is $\epsilon=10^{-9}$.}
\label{lam13}
\end{figure}

The fixed points listed in Table~\ref{tabFP1} accumulate towards $\l=1/3$, so one may wonder if there are more fixed points with $\l$ even closer to $1/3$. These would correspond to yet larger values of $u_s$ (and $u_t$) than used in the $u_t$-scan. We rule out this possibility by eliminating $u_s$ from the fixed-point equations in favor of $\l$ and performing the search for roots by scanning over $\l^*$. We focus on the region $0<1/3-\l^*\ll 1$, with a step $\epsilon=10^{-9}$. We again remove the equation $\tilde\beta_{v_1}=0$ from the system. The remaining equations are solved explicitly and their solutions are used to derive the residual $\delta\tilde\beta(\l^*)$. The result is shown in Fig.~\ref{lam13}. We observe that the residual grows monotonically at $\l^*\to 1/3^-$ implying absence of any further solutions in this limit.



\subsection{Infinite $\l$}

Beta-functions for the variables $\chi=\{u_s,v_a\}$ in the limit $\l\to\infty$ ($\varrho=1$) have the form, 
\be\label{beta_lam1}
\begin{split}
\tilde\beta_\chi \Big|_{\varrho=1} &=\frac{\bar A_\chi}{26880\pi^2(1+u_s)^3 u_s^5} \sum_{n=0}^9 u_s^n {\cal S}^\chi_n [v_a] \;,\\
&\bar A_{u_s}= u_s\;,\quad \bar A_{v_1}=1\;,\quad \bar A_{v_2}=\bar A_{v_3}=2\;,
\end{split}
\ee
where the polynomials ${\cal S}^\chi_n[v_a]$ are obtained from the polynomials ${\cal P}^\chi_n[\l,v_a]$ in Eq.~(\ref{beta_chi}) as
\be\label{Spoly}
\begin{split}
{\cal S}^\chi_n [v_a] = &\lim_{\l\to\infty}\frac{ B_\chi {\cal P}^\chi_n [\l,v_a]}{(1-\l)^3(1-3\l)^3}\;,\\
& B_{u_s} = (1-\l)\;,\quad B_{v_a} = 1\;, 
\end{split}
\ee
To solve the system (\ref{lmsystem}), we fix the value $u_s=u_s^*$ and remove the equation $\tilde\beta_{v_1}\big|_{\varrho=1}=0$. Then we solve the system of three remaining equations for the variables $v_a$ using the NSolve Mathematica routine. We make sure that we recover all complex solutions predicted by the B\'ezout's theorem and pick the real roots. After that we compute the residual of the first equation and scan over the values of $u_s^*$. 
We take the following scanning steps in different $u_s^*$-intervals: 
\begin{itemize}
\item $u_s^* \in (0,1]$: step $\epsilon=10^{-5}$;
\item $u_s^*\in [1,100]$: step $\epsilon=10^{-3}$;
\item $u_s^*\in [100, 1000]$: step $\epsilon=10^{-2}$;
\item $u_s^*\in [10^{m+3}, 10^{m+4}]$, $m = 0, \ldots, 11$:
step $\epsilon=10^{m}$. 
\end{itemize}
The points where the residual vanishes are identified as the roots of the full system (\ref{lmsystem}). The obtained solutions are listed in Table~\ref{tabFP2}.

We have run the numerical scan till large but finite $u_s^*$. The following argument shows that there are no more solutions at a higher value of $u_s$. The beta-functions (\ref{beta_lam1}) at large $u_s$ are dominated by the terms with the highest powers of this variable. The coefficient ${\cal S}_9^{u_s}$ happens to be a pure non-zero number, so restricting to the terms with $u_s^9$ in the fixed-point equations does not give any solutions. Keeping the first subleading terms we have, up to non-zero factors:
\be
\label{big_us}
\tilde\beta_\chi \Big|_{\substack{\varrho=1\quad\\u_s\to\infty}} 
\propto u_s \,{\cal S}^\chi_9[v_a] +{\cal S}^\chi_8[v_a]\;.
\ee
The system 
\be
u_s {\cal S}^\chi_9[v_a] +{\cal S}^\chi_8[v_a]=0\;,
\ee
is easily solved and has two real solutions which approximate the fixed points 7 and 8 from the Table~\ref{tabFP2}. We conclude that there are no more solutions of (\ref{lmsystem}) with $u_s$ bigger than its value at the point 8.

Finally, we have solved the system (\ref{lmsystem}) directly with the NSolve command of Mathematica and obtained exactly the same real roots at $u_s>0$ as with the scanning procedure. This provides an additional verification that the list of fixed points in Table~\ref{tabFP2} is complete.

\section{RG flows from fixed points \textnumero 4 and \textnumero 7}
\label{FPs47}

This Appendix complements the analysis of Sec.~\ref{RGflow_inflam} and studies the 
RG flows starting from asymptotically free fixed points 4 and 7 at $\l=\infty$, see  Table~\ref{tabFP2}. We work with the coordinate $\varrho$ defined in (\ref{rho}) and mapping the hyperplane $\l=\infty$ to $\varrho=1$.

\subsection{Flow from point \textnumero 4}
\label{FP4}

\begin{table*}
\centering
\begin{tabular}{@{}|c | c || c | c | c | c | c |@{}}
 \hline
\makecell{Eigenvector\\label}&$\theta^J$& $w_\varrho^J$ & $w_{u_s}^J$  & $w_{v_1}^J$ & $w_{v_2}^J$ & $w_{v_3}^J$   \\ [0.5ex]
\hline\hline
$w^{4|1}$&-0.01334& 2.53$\times10^{-3}$& 0.0807&-0.997& -6.24$\times 10^{-4}$& 5.16$\times 10^{-3}$\\ [0.5ex]
\hline
$w^{4|2}$&-0.3436&0& 1.22$\times 10^{-3}$ &-0.999& -0.0117& 1.07$\times 10^{-3}$\\ [0.5ex]  
 \hline
  $w^{4|3}$&-0.09353&0& -9.25$\times 10^{-3}$ &-0.941& -0.279& 0.190\\ [0.5ex]
\hline\hline
$w^{7|1}$&-0.01516&5.92$\times10^{-4}$& 0.999&-7.77$\times10^{-3}$& 5.51$\times10^{-3}$& -7.43$\times10^{-3}$ \\ [0.5ex]
\hline
$w^{7|2}$& -1.722&0& 0.509 &-0.127& 0.625& 0.578 \\ [0.5ex]  
 \hline
$w^{7|3}$&  \multirow{2}{*}{ -0.3324 $\pm$ 0.3289$i$ }&0& -0.999&0.0218& -0.0310& 0.0213\\ [0.5ex]
\cline{1-1}\cline{3-7}
$w^{7|4}$&&0& 0&-6.11$\times10^{-3}$& 0.0178& -5.07$\times10^{-3}$\\ [0.5ex]
 \hline
  \end{tabular}
  \caption{Components of the repulsive vectors $w_i^J$ for fixed points 4 and 7 located at $\l=\infty$ ($\varrho=1$) with their corresponding eigenvalues $\theta^J$. Vectors $w^{7|3}$, $w^{7|4}$ are real and imaginary parts of a pair of complex eigenvectors corresponding to two complex conjugate eigenvalues. 
  We set $w^{7|4}_{u_s}=0$ using the freedom to rotate the phase of a complex vector and impose the normalization $\sum_{i}\big[\big(w^{7|3}_i\big)^2+\big(w^{7|4}_i\big)^2\big]=1$.} 
    \label{EVlam4}
\end{table*}

Stability matrix at the point 4 
has three negative eigenvalues $\theta^J$. The components of the corresponding eigenvectors $w^J$ are collected in three upper rows of Table \ref{EVlam4} (labeled $w^{4|1}$, $w^{4|2}$, $w^{4|3}$). Note that the eigenvectors are almost collinear and nearly coincide with $v_1$-direction.
In the initial conditions of RG equation \eqref{RGeq}, we choose constants $c_J$   on the unit sphere 
\be
\label{FP4init}
\begin{split}
&c_1 w^{4|1} +c_2 w^{4|2} + c_3 w^{4|3}\\ 
&=\sin\vartheta \cos\varphi\, w^{4|1} + \cos\vartheta\, w^{4|2} 
+ \sin\vartheta\sin\varphi\, w^{4|3}\:, 
\end{split}
\ee
where $\vartheta\in [0,\pi]$ and $\varphi \in [0,2\pi)$.

\begin{figure}[h]
\centering
\includegraphics[width=0.85\columnwidth]{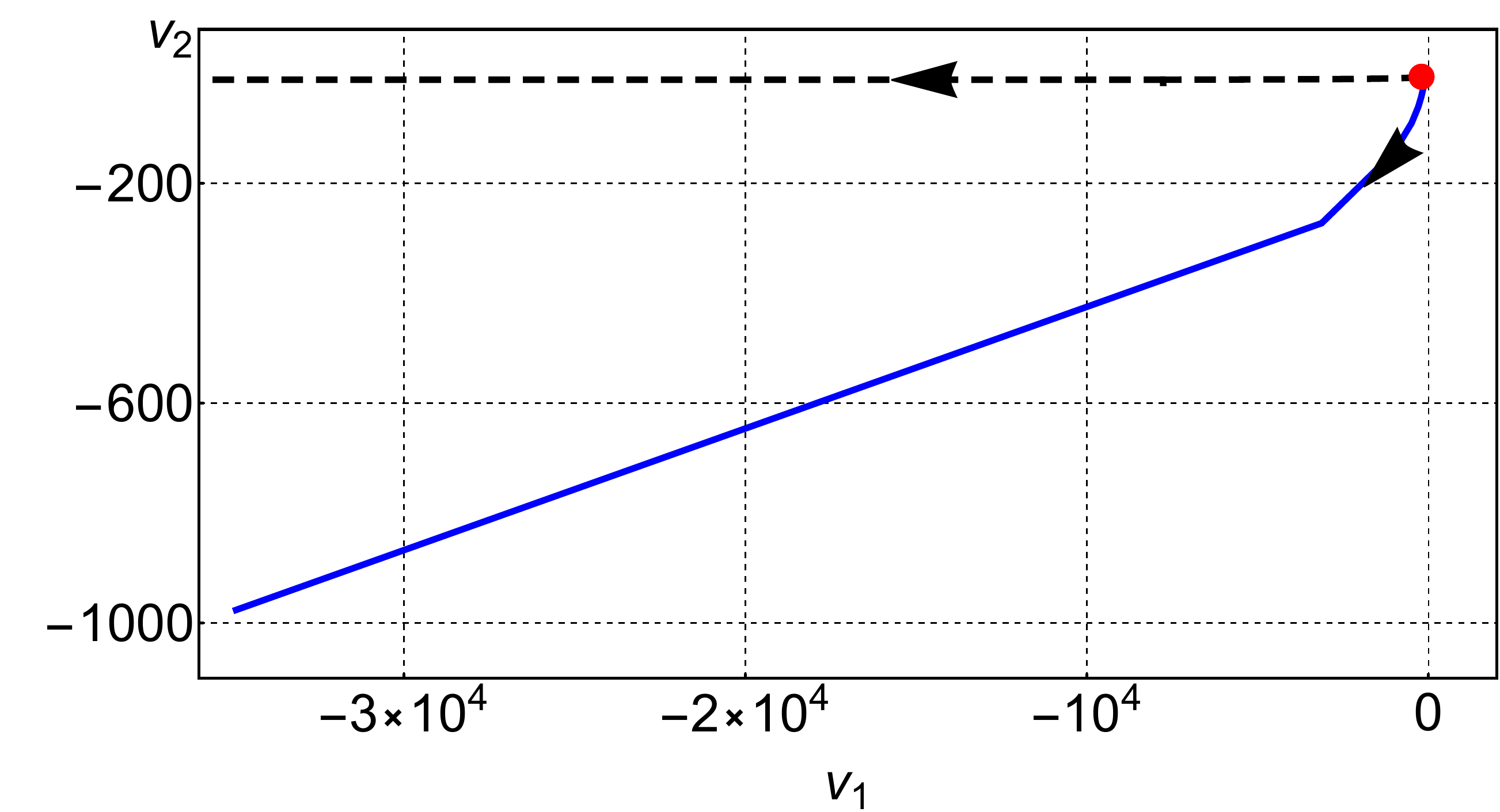}
\\~\\
\includegraphics[width=0.85\columnwidth]{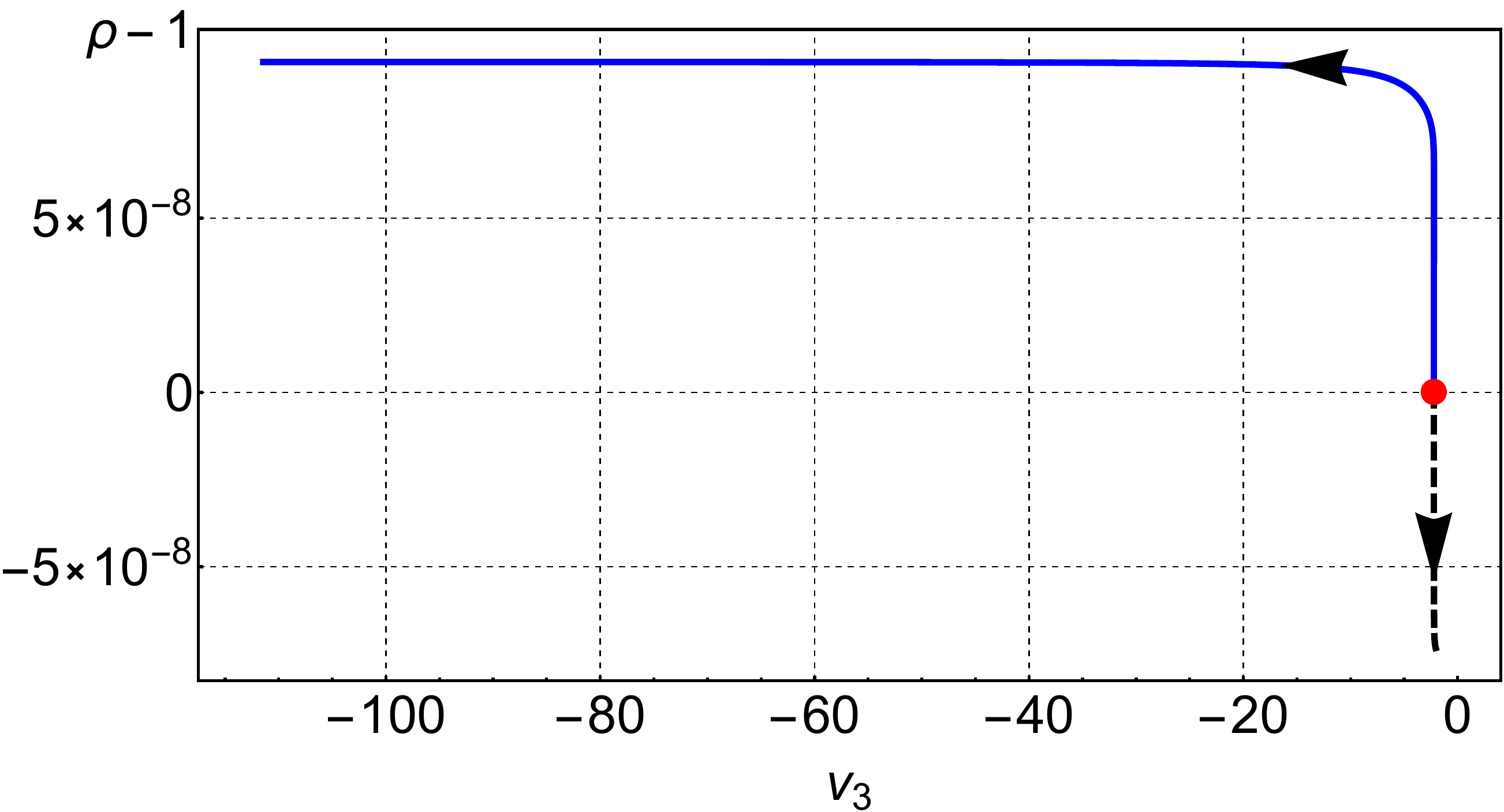}
\\~\\
\includegraphics[width=0.85\columnwidth]{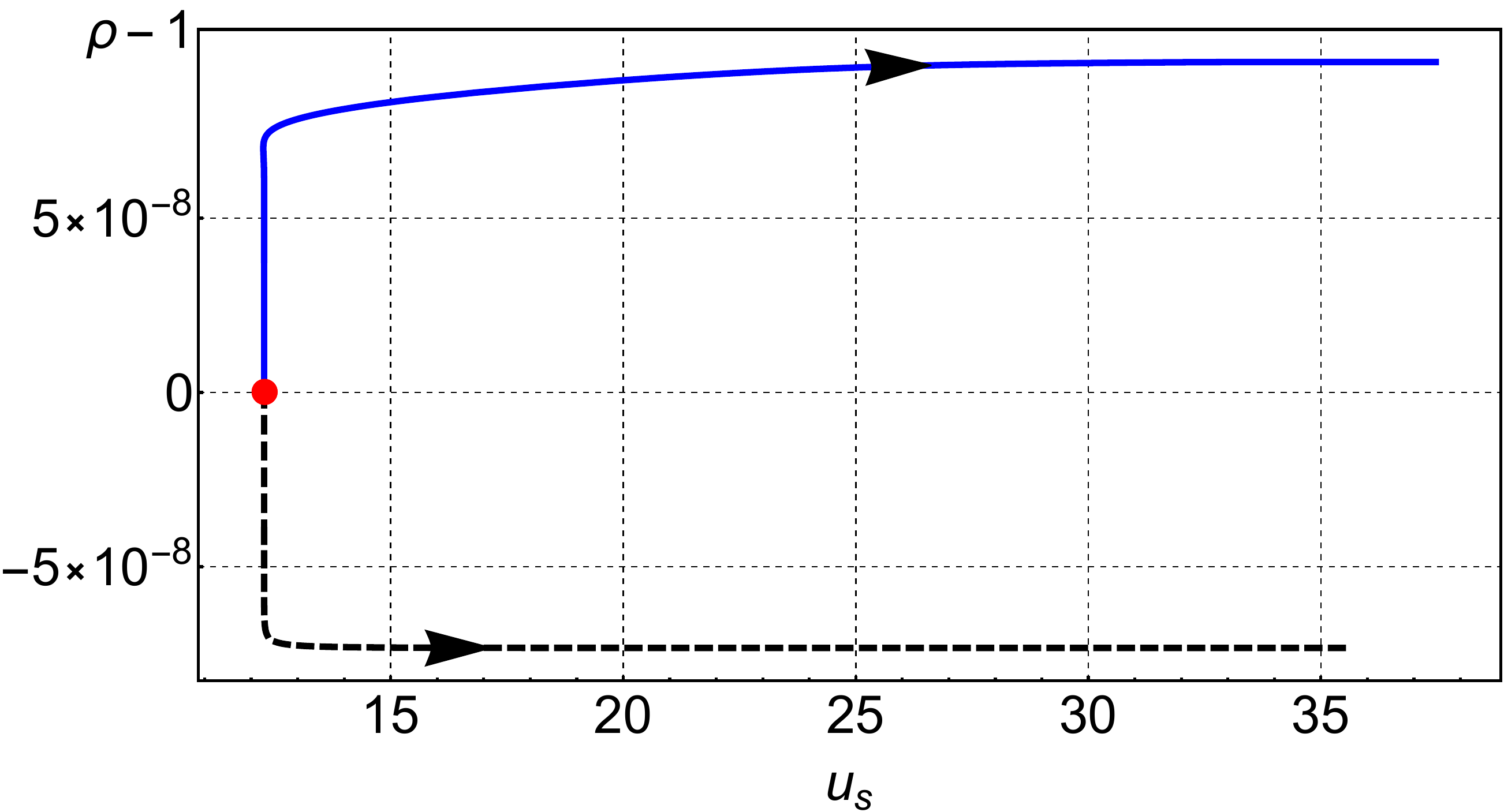}
\caption{RG flow from fixed point 4 (red dot). Dashed black (solid blue) curves show the projections of the trajectories with ${c_2>0}$ ($c_2\leq0$).  
In both cases ${\rm sign}\, c_1 = -{\rm sign}\, c_2$. The opposite choice of the $c_1$-sign will reflect the trajectory with respect to the fixed point  in $\varrho$-axis.
The trajectories stay very close to the initial value $\varrho=1$.  
Arrows indicate the direction from UV to IR.
The break in the blue curve in the upper panel signals the loss of numerical precision.
}
\label{FP4flows}
\end{figure}

We construct a two-parameter family of trajectories scanning the space $(\vartheta, \varphi)$ with steps $\sim 0.1$ in each direction. We find that all trajectories practically coincide with one of the two trajectories corresponding to $c_2=\pm 1$. This is explained by the dominance of the eigenvalue $\theta^2$ (see Table~\ref{EVlam4}) which implies that all trajectories are quickly attracted to the direction set by the vector $w^{4|2}$. More in detail, we have: 
\begin{enumerate}
\item $\vartheta\in [0,\pi/2)$, $\varphi$-any: the projections of the trajectory on different planes are shown by black dashed curves in Fig.~\ref{FP4flows};
\item $\vartheta\in[\pi/2,\pi]$, $\varphi$-any: the trajectory is shown by the blue solid curves in Fig.~\ref{FP4flows}.
\end{enumerate}

The trajectories cannot be continued any further because of the loss of numerical precision due to a rapid growth of $v_1$. In other words, the trajectories run into singularity in a finite `RG time' $\tau$. Note that at the same time $u_s$, $v_2$, $v_3$ remain relatively small. The coupling $\varrho$ increases if $c_1>0$ in Eq.~(\ref{FP4init}) or decreases if $c_1<0$. In both cases $\varrho$ changes very little over the whole span of the trajectory, see middle and lower panels in Fig.~\ref{FP4flows}, so the flow effectively stays in the plane $\varrho=1$.

\subsection{Flow from point \textnumero 7}
\label{FP7}

Fixed point 7 has four negative eigenvalues $\theta^J$. 
Two of them $(\theta^1,\theta^2)$ 
are real and two $(\theta^3,\theta^4)$ are complex conjugate.  
The repulsive vectors are listed 
in the four lower rows of Table \ref{EVlam4} (labeled $w^{7|1}$ through $w^{7|4}$). The vectors $w^{7|1}$ and $w^{7|2}$ are eigenvectors corresponding to $\theta^1$, $\theta^2$, whereas the vectors $w^{7|3}$, $w^{7|4}$ are real and imaginary parts of the complex eigenvectors corresponding to $\theta^3$, $\theta^4$. Using the freedom in the choice of the phase of a complex eigenvector, we set the component $w_{u_s}^{7|4}$ to zero. Note that $w^{7|1}$ and $w^{7|3}$ happen to be almost (anti-)collinear and aligned along the $u_s$-direction. On the other hand, the vectors $w^{7|2}$, $w^{7|4}$ have comparable components along other directions.

\begin{figure}[h]
\centering
\includegraphics[width=0.85\columnwidth]{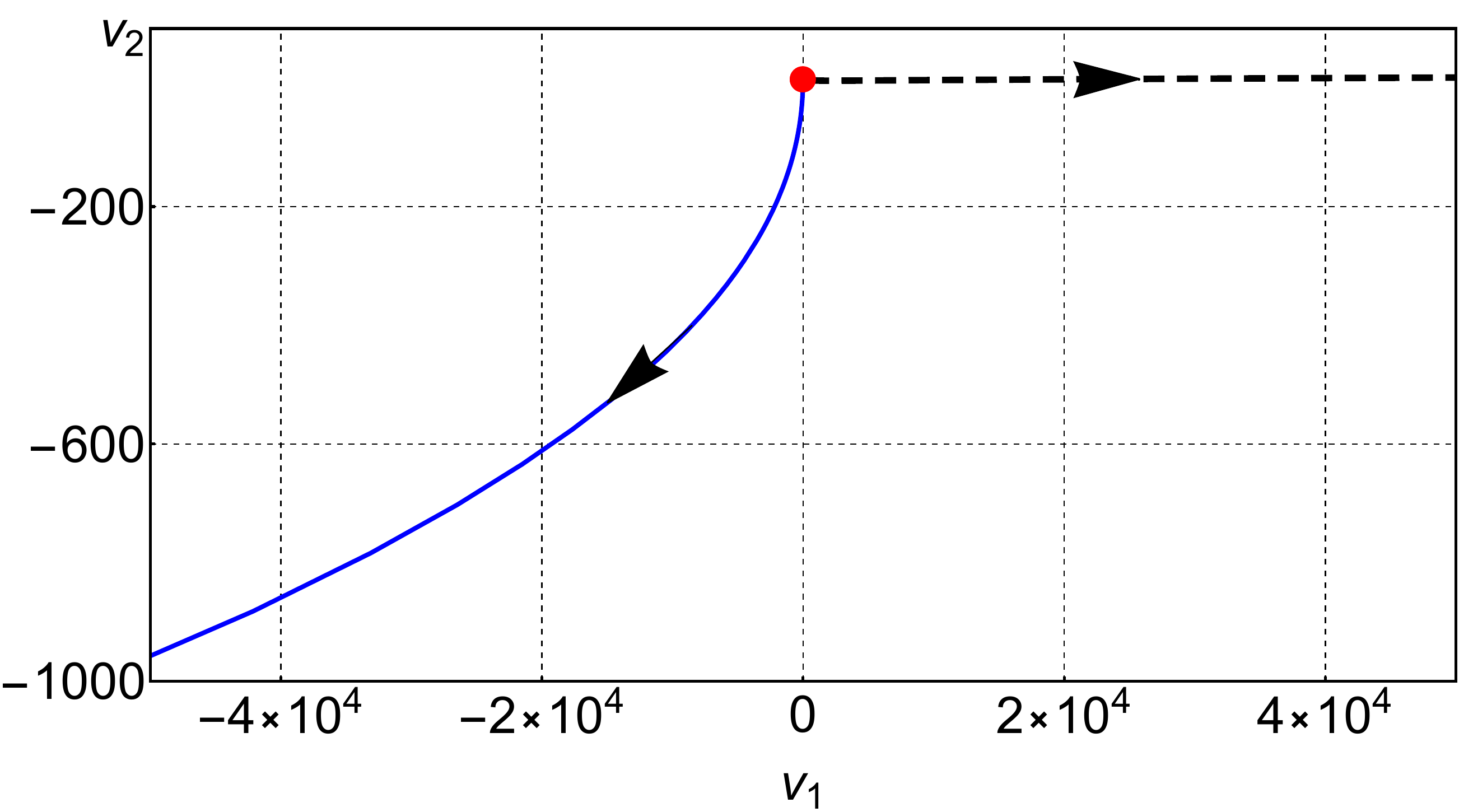}\\~\\
\includegraphics[width=0.85\columnwidth]{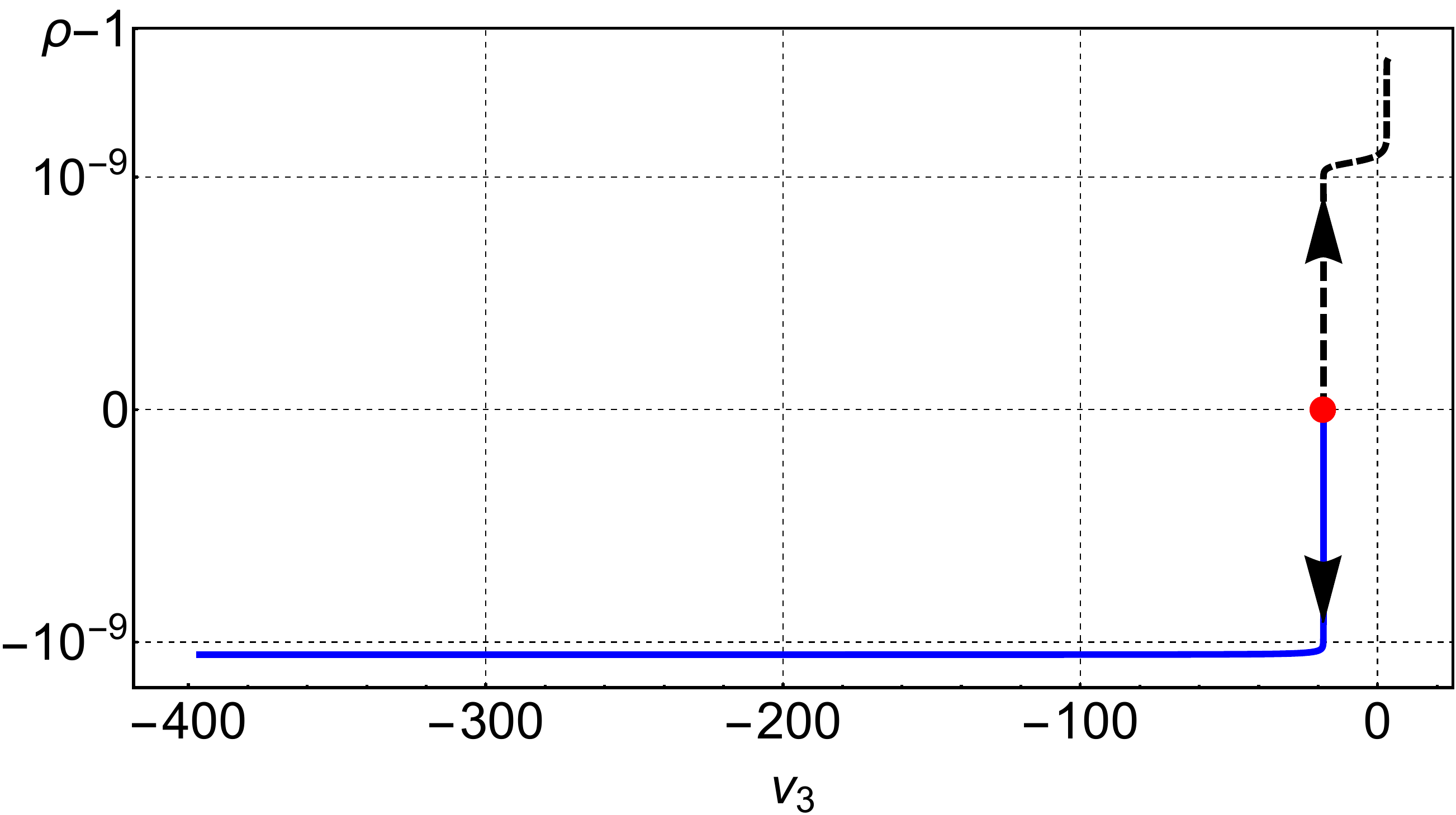}\\~\\
\includegraphics[width=0.85\columnwidth]{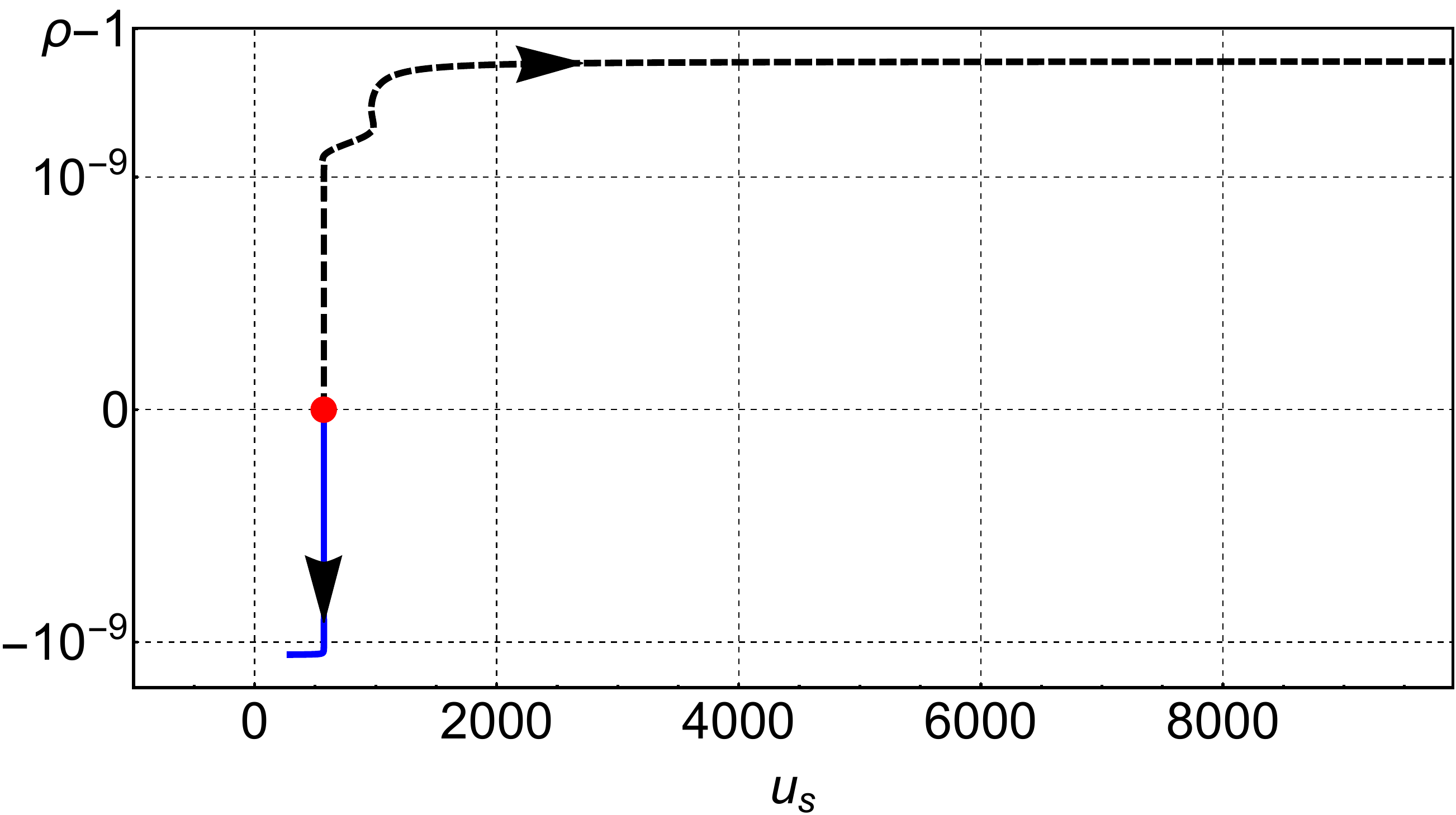}
\caption{RG flow from fixed point 7 (red dot). 
Dashed black (solid blue) curves show the projections of the trajectories with ${c_2>0}$ ($c_2\leq0$).  
In both cases ${\rm sign}\, c_1 = {\rm sign}\, c_2$. The opposite choice of the $c_1$-sign will reflect the trajectory with respect to the fixed point  in $\varrho$-axis.
The trajectories stay very close to the initial value $\varrho=1$.  
Arrows indicate the direction from UV to IR.
The breaks in the black dashed curves in the middle and lower panels signal the loss of numerical precision.} 
\label{FP7flows}
\end{figure}

In the initial conditions of RG equation \eqref{RGeq}, we choose constants $c_J$ on the unit 3-sphere, parametrized by angles $\vartheta$, $\psi$ and $\varphi$,
\be
\begin{split}
&c_1w^{7|1}+c_2w^{7|2}+c_3w^{7|3}+c_4w^{7|4}\\
&= \sin\vartheta\cos\psi\, w^{7|1}
+\cos\vartheta\, w^{7|2}\\
&~~~~+\sin\vartheta\sin\psi\cos\varphi\, w^{7|3}
+ \sin\vartheta\sin\psi\sin\varphi\, w^{7|4}\;.
\end{split}
\ee
As in the previous sections, we construct a family trajectories by scanning the parameters $\vartheta\in [0,\pi]$, $\psi\in [0,\pi]$, $\varphi \in [0,2\pi)$ with a small step $\sim 0.1$ in each direction. 
All trajectories again practically coincide with one of two trajectories whose 
projections on various coordinate planes are shown in
Fig.~\ref{FP7flows}.
This is again explained by the dominance of the eigenvalue $\theta^2$ (see Table~\ref{EVlam4}) which implies that all trajectories are quickly attracted to the direction set by the vector $w^{7|2}$ and the only relevant characteristics is the sign of the coefficient $c_2$.
More in detail, we have:
\begin{enumerate}
\item $\vartheta\in [0,\pi/2)$, $\psi$-any, $\varphi$-any: the projections of the trajectory on different planes are shown by black dashed curves in Fig.~\ref{FP7flows};
\item $\vartheta\in[\pi/2,\pi]$,  $\psi$-any, $\varphi$-any: the trajectory is shown by the blue solid curves in Fig.~\ref{FP7flows}.
\end{enumerate}

The trajectories cannot be continued any further because of the loss of precision due to a rapid growth of the couplings in absolute values. We conclude that the trajectories run into singularity in a finite `RG time'. The behavior of $\varrho$ depends only on the sign of the coefficient $c_1$: the coupling increases when $c_1>0$  and decreases when $c_1<0$. The change of $\varrho$ on all trajectories is negligible and the flow effectively stays in the $\varrho=1$ plane.

\section[Asymptotics of RG flow at $\l\to1^+$]{Asymptotics of RG flow at $\boldsymbol{\lambda\to1^+}$}
\label{asympt}

In the main text we have obtained numerically the behavior of essential couplings as functions of $(\l-1)$ along RG trajectories connecting the point A to the domain of $\l\to 1^+$, see Figs. \ref{Atolam1+} and \ref{PlotG}. 
However, it is problematic to continue the numerical solutions down to very small values of $(\l-1)$ due to the accumulation of numerical errors. In this Appendix we derive approximate analytic expressions for the running couplings at $\l\to1^+$ exploiting the simplification of beta-functions (\ref{betafun}) in this region.  

\begin{figure}[h]
\centering
\includegraphics[width=0.9\columnwidth]{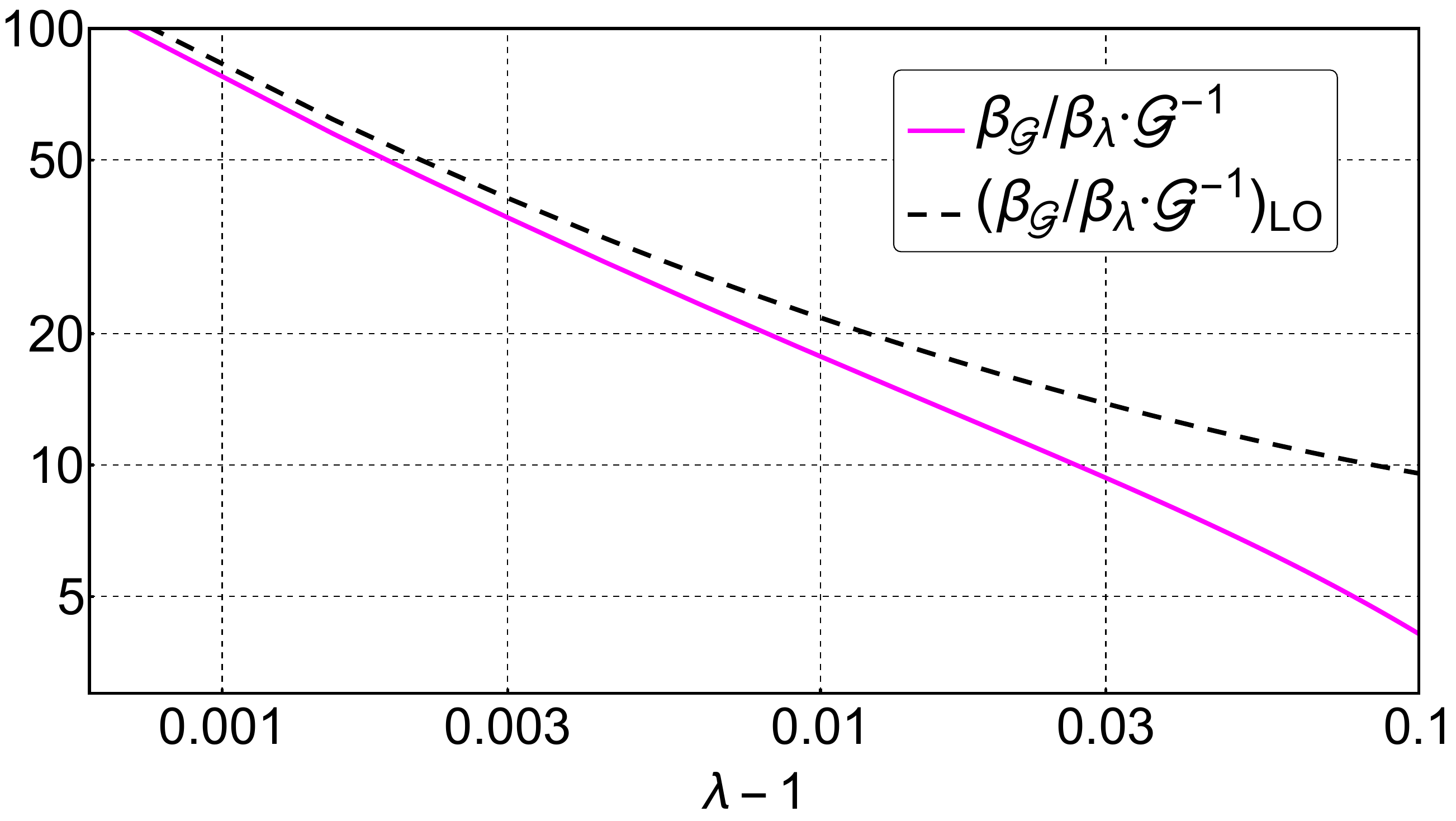}
\caption{Beta-function of the coupling ${\cal G}$ as a function of $(\l-1)$ on a typical RG trajectory from the fixed point A to $\l\to 1^+$ (solid), compared to the analytic approximation (\ref{asymptG}) (dashed). The initial angle of the trajectory at point A is $\varphi_A=\pi/4$. }
\label{beta_G_asympt}
\end{figure}

\begin{figure*}
\centering
\includegraphics[width=0.9\columnwidth]{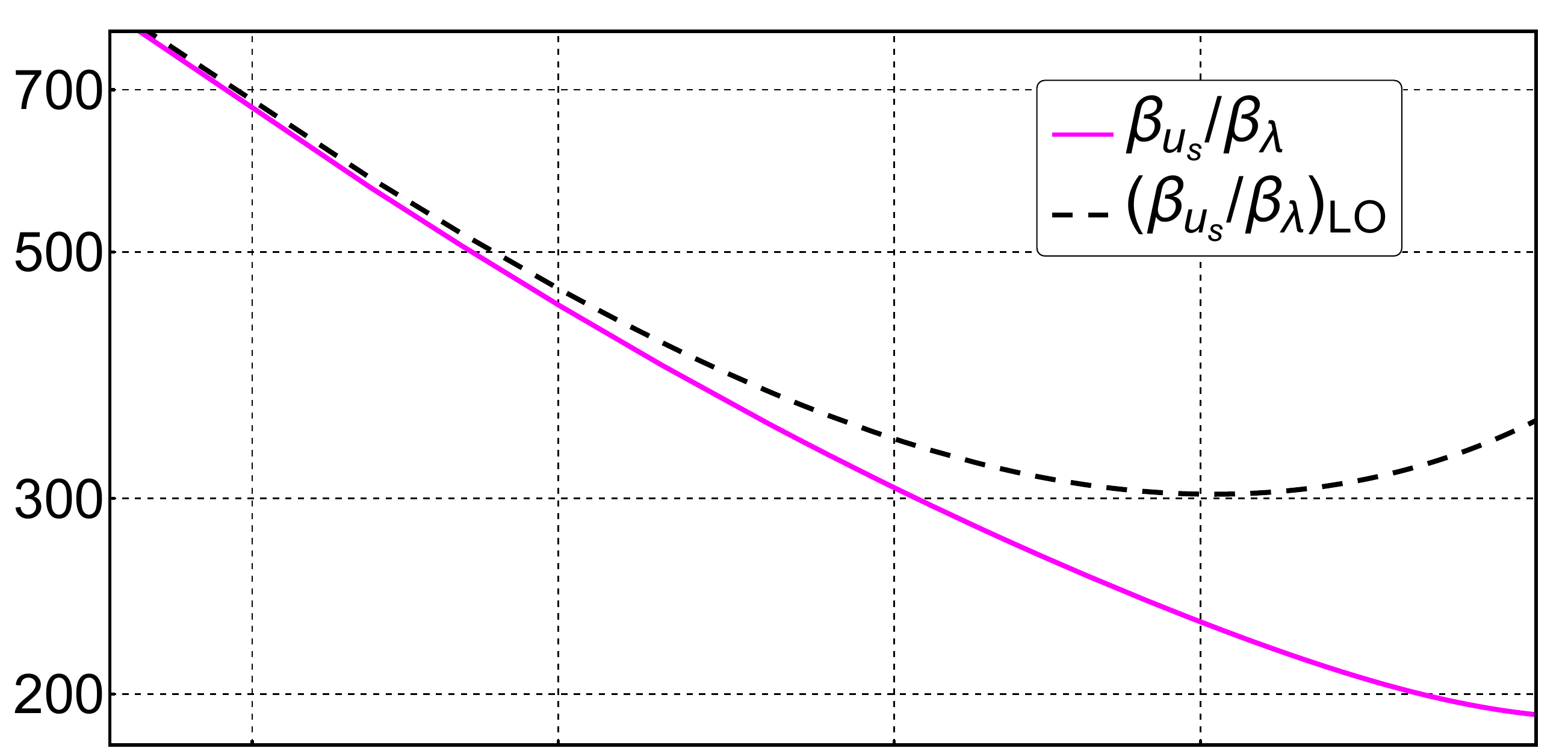}\qquad
\includegraphics[width=0.9\columnwidth]{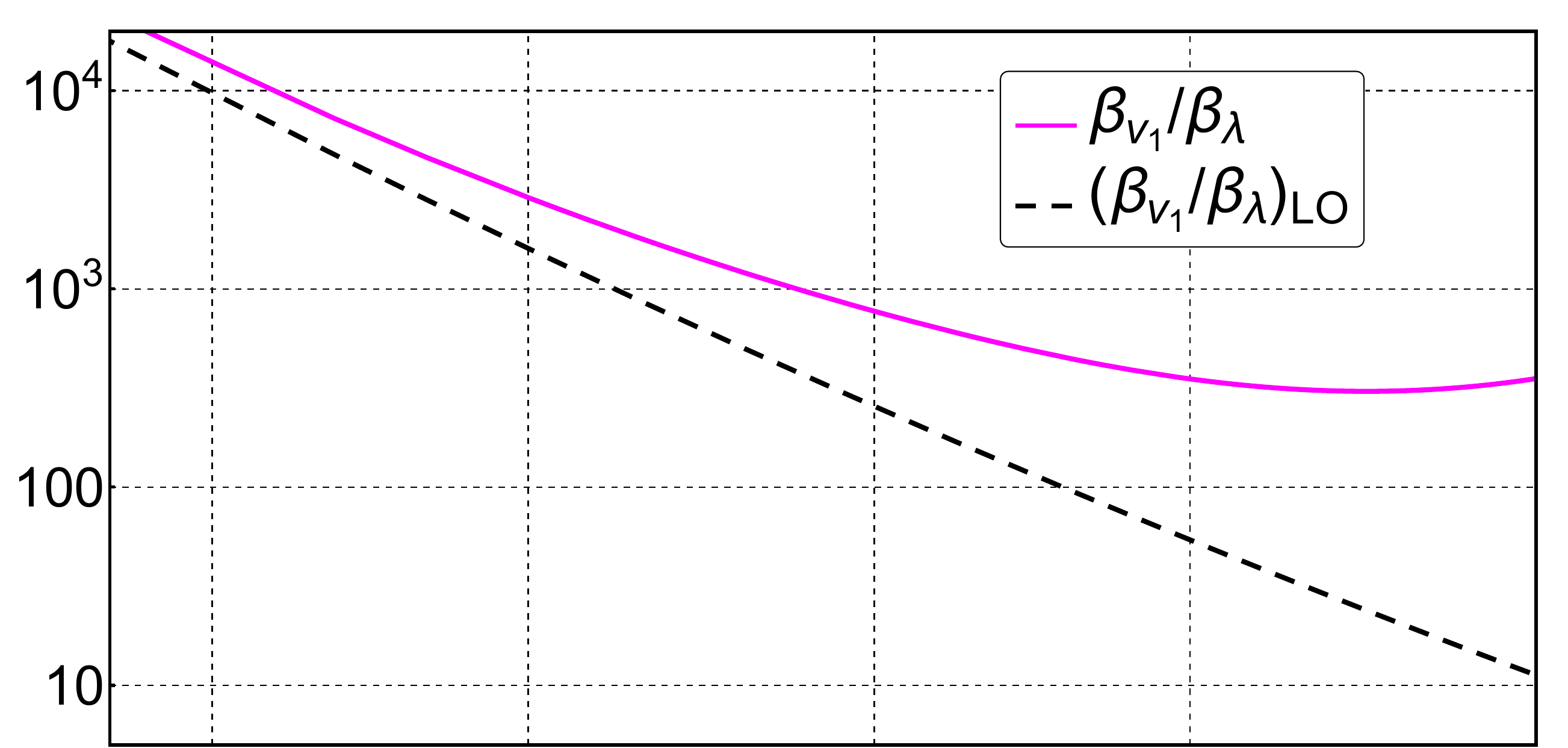}
\includegraphics[width=0.9\columnwidth]{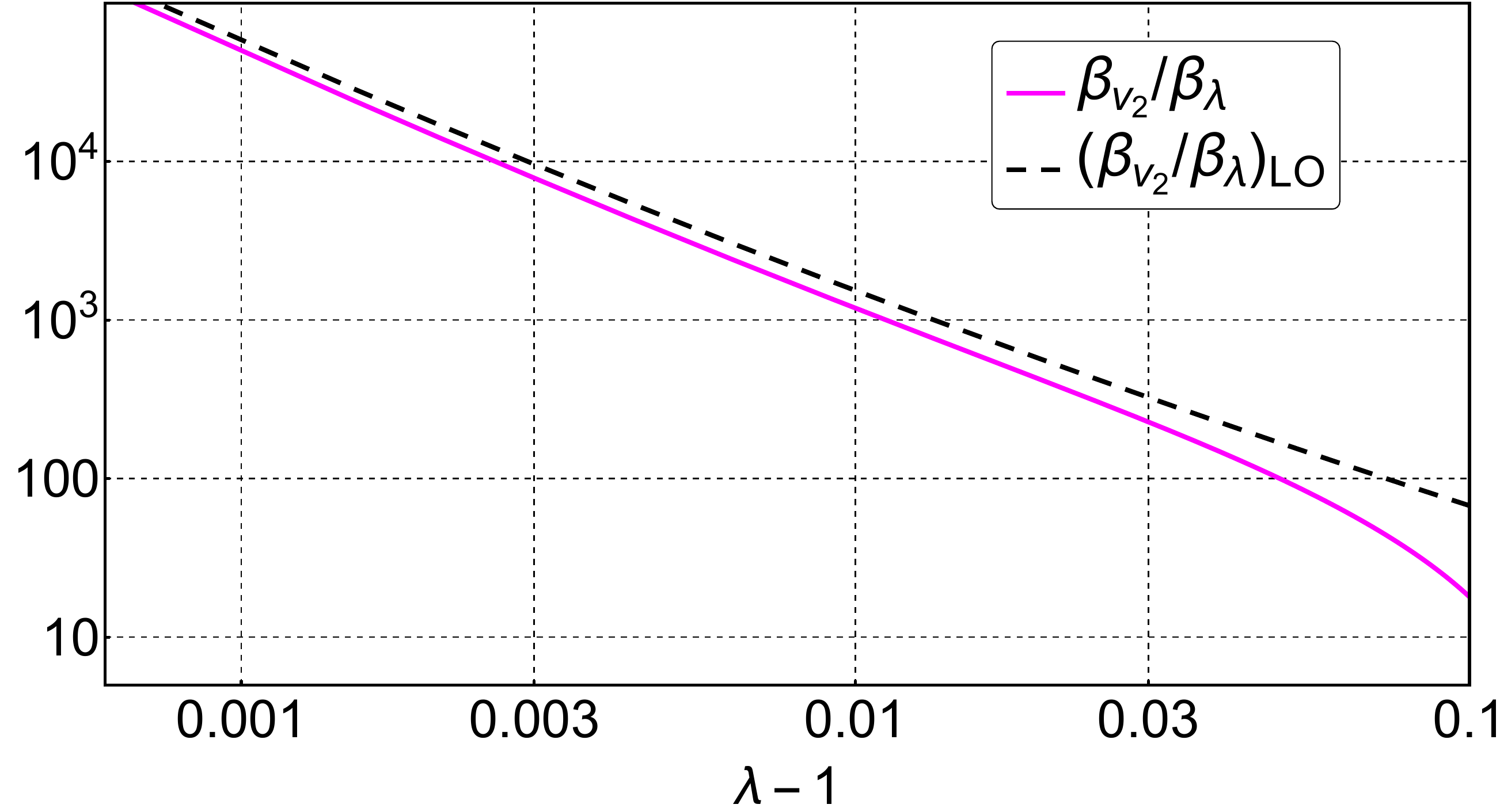}\qquad
\includegraphics[width=0.9\columnwidth]{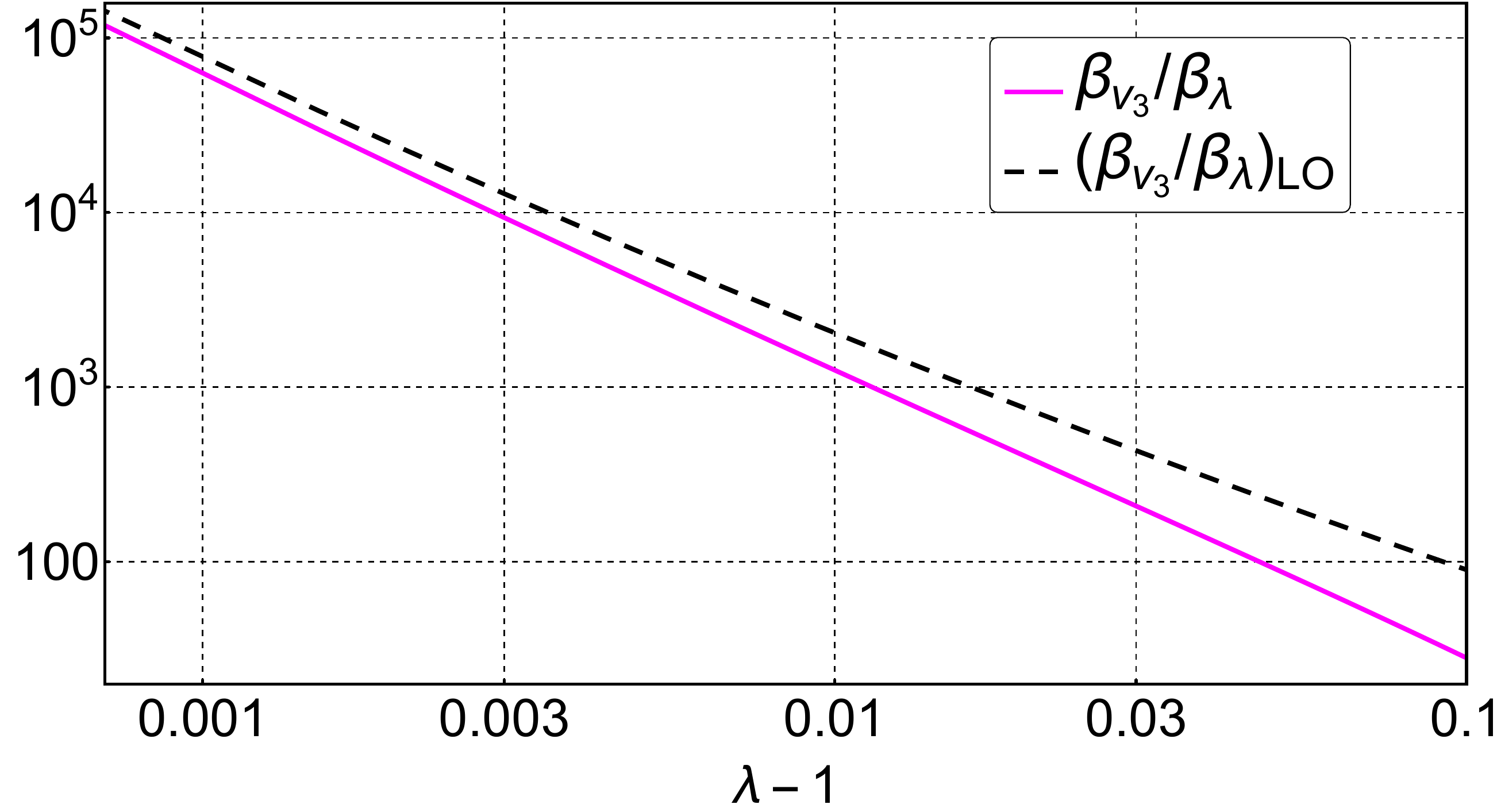}
\caption{Same as Fig.~\ref{beta_G_asympt} but for the beta-functions of the couplings $u_s$, $v_a$. }
\label{beta_chi_asympt}
\end{figure*}

We are interested in the ratios of beta-functions $\beta_{\cal G}/\beta_\l$ and $\beta_{\chi}/\beta_\l$, $\chi=\{u_s,v_a\}$, which determine the running of all couplings with respect to $\l$. It is natural to assume that at $\l\to 1^+$ these ratios will be dominated by the leading terms in their Laurent series at $\l=1$.\footnote{This is not guaranteed, though, since a priori the subleading terms in the series can be enhanced by the other couplings entering the beta-functions.}
Using the expressions for the polynomials ${\cal P}^{\cal G}_n$, ${\cal P}^\chi_n$ from \cite{3+1} we obtain,
\bseq
\label{betaIR}
\begin{align}
\label{asymptG}
\left(\frac{\beta_{\cal G}}{\beta_\l}\right)_{\rm LO} &=
{\cal G}\frac{17}{448}\frac{1+u_s}{\l-1}\;,\\
\left(\frac{\beta_{u_s}}{\beta_\l}\right)_{\rm LO} &= \frac{u_s(241+17u_s)}{448(\l-1)}\;,\label{asymptus}\\
\left(\frac{\beta_{v_a}}{\beta_\l}\right)_{\rm LO} &= -\alpha_{v_a} \frac{1+u_s}{(\l-1)^2} \;,\label{asymptva}
\end{align}
\eseq
where $\alpha_{v_1} = \frac1{224}$, $\alpha_{v_2} = \frac3{112}$, $\alpha_{v_3} = \frac1{28}$. 
Figs.~\ref{beta_G_asympt}, \ref{beta_chi_asympt} show comparison of these expressions with the exact beta-function ratios found numerically along a typical RG trajectory connecting the point A to $\l\to 1^+$. We see that the approximation works well at $(\l-1)\lesssim 0.01$.   

The system \eqref{betaIR} is easy to study analytically. First, we solve the equation (\ref{asymptus}) for $u_s$,
\be\label{us_sol_lam}
\frac{u_s}{u_s+241/17} = C_{u_s}(\l-1)^{241/448}\:, 
\ee
where $C_{u_s}$ is an integration constant. Numerically we find 
$C_{u_s}\approx 2.98$ for all trajectories from the fixed point A to $\l\to 1^+$. 
Expressing $u_s$ through $(\l-1)$ and keeping only two terms of the expansion\footnote{Retaining two terms, rather than one, significantly improves the agreement with the numerical data.} 
we obtain,
\be
\label{us_sol}
u_s\simeq \frac{241}{17} C_{u_s} (\l-1)^{241/448}\left(1+C_{u_s}  (\l-1)^{241/448}\right)\;.
\ee 
Substituting this into \eqref{asymptG} we obtains the asymptotic behavior
\be\label{GIR}
{\cal G}
\simeq C_{\cal G}\left((\l-1)^{17/448}+C_{u_s}(\l-1)^{129/224}\right)\;,
\ee
with a new integration constant $C_{\cal G}$. This depends on the initial direction $\varphi_A$ of the trajectory at the point A. For $\varphi_A=\pi/4$ we find $C_{\cal G}\approx 1.95\times 10^{-20}$. 
Finally, substituting (\ref{us_sol}) into (\ref{asymptva}) we find,
\be\label{va_sol}
v_a\simeq \frac{\alpha_{v_a}}{\l-1}\bigg(1 + \frac{107968 C_{u_s}}{3519} (\l-1)^{241/448}+C_{v_a}\bigg) \:,
\ee
where $C_{v_a}$ are further constants. We observe that the couplings $v_a$ diverge at $\l\to1^+$. Expressions (\ref{us_sol}), (\ref{GIR}), (\ref{va_sol}) yield the scalings (\ref{GIRscaling}), (\ref{IRscalings}) used in the main text.

\section[Dependence of ${\cal G}_{IR}$ on the initial flow direction]{Dependence of $\boldsymbol{{\cal G}_{IR}}$ on the initial flow direction}
\label{app:D}

Here we derive Eq.~(\ref{GIRscalingphi}) describing the scaling of the gravitational coupling in the IR region $\l\to1^+$ on the initial direction of the RG trajectory at the UV fixed point A located at $\l\to\infty$. We replace $\l$ with the coupling $\varrho$ defined in Eq.~(\ref{rho}), which brings the position of the fixed point to the finite value $\varrho=1$. 

When the RG trajectory leaves the UV fixed point, it first closely follows the trajectory $\chi^{AB}(\tau)$ (here $\chi=\{u_s,v_a\}$) in the hyperplane $\varrho=1$ which connects the point A to the point B, see Sec.~\ref{ssec:AtoB}. The gravitational coupling then evolves according to Eq.~(\ref{Gsol}), where $\hat\beta_{\cal G}$ can be considered as function of $\chi^{AB}(\tau)$. The coupling ${\cal G}$ grows along the trajectory and reaches a maximum ${\cal G}_{\rm max}$ at a point $\chi_{\rm max}$ where $\hat\beta_{\cal G}$ changes sign. Let us denote the corresponding `RG time' by $\tau_{\rm max}$. 
The RG evolution from $\tau=0$ to $\tau_{\rm max}$ corresponds to the region $I$ in Fig.~\ref{PlotG}.  

Though the deviation $(1-\varrho)_{\rm max}$ of the trajectory from the hyperplane $\varrho=1$ at $\tau_{\rm max}$ is small, it is non-zero. After the passage through $\tau_{\rm max}$ the trajectory scatters on the point B and $(1-\varrho)$ starts rapidly increasing. This corresponds to the region $II$ in Fig.~\ref{PlotG}. The power law scaling (\ref{Gpl}) in this region implies 
\be
\label{Gplrho}
{\cal G}_{IR}\propto {\cal G}_{\rm max}\big[(1-\varrho)_{\rm max}\big]^{\kappa_{II}}\;,
\ee
where $\kappa_{II}$ is given in Eq.~(\ref{kappa2}). We see that for fixed ${\cal G}_{\rm max}<1$, the IR coupling ${\cal G}_{IR}$ scales as a power of the distance between the trajectory and the hyperplane $\varrho=1$ at the `RG time' $\tau_{\rm max}$. Thus, our task is to derive the dependence of this distance on the initial angle $\varphi_A$ at $\tau=0$.

Let us show that when the angle $\varphi_A$ is close to the critical value $\delta$ the trajectory spends a long `RG time' in the neighborhood of the point A. Indeed, the RG flow in this neighborhood is controlled by the stability matrix~(\ref{stabilityB}). Expanding the deviations of the couplings from their fixed-point values in the eigenvector basis, $g_i(\tau)-g^\star_i=a_J(\tau)\,w^J$, we have
\be
a_J(\tau)=a_{J,0}\,\e^{\theta^J\tau}\;.
\ee
The couplings $\chi$ evolve mostly along the dominant repulsive direction $w^{A2}$, see Table~\ref{EVlam5}. The initial condition for the coefficient of this eigenvector is small when $\varphi_A$ is close to its critical value $\delta$,
\be
a_{2,0}\propto \varphi_A-\delta\;,
\ee
thus the `RG time' $\tau_\star$ it takes the trajectory to escape from A is logarithmically large,
\be
\label{taustar}
\tau_\star\propto -\frac{1}{\theta^2}\log(\varphi_A-\delta)\;.
\ee

On the other hand, since the $\varrho$-component of $w^{A2}$ vanishes, the evolution of $\varrho$ is determined by the subdominant eigenvalue $\theta^1$. 
Using the beta-function (\ref{betarho}) we obtain,
\be
\label{rhoRGsol}
(1-\varrho)_{\rm max}=(1-\varrho)_{0}\,\e^{\theta^1\tau_\star}
\exp\bigg[\int_{\tau_\star}^{\tau_{\rm max}}d\tau'\,\hat\beta_{\cal \varrho}\big(u_s(\tau')\bigg],
\ee
where
\be
\hat \beta_\varrho=\frac{\tilde\beta_\varrho}{1-\varrho}\bigg|_{\varrho=1}\;.
\ee
The integral in the second exponent is finite in the limit $(\varphi_A-\delta)\to 0$ since it is taken over a finite portion of the trajectory $\chi^{AB}(\tau)$ outside the neighborhoods of the fixed points. The initial value $(1-\varrho)_0=\ve\cos\varphi_A\, w^{A1}_\varrho$ is also finite when $\varphi_A\to \delta$. Thus, we get 
\be
(1-\varrho)_{\rm max}\propto \e^{\theta^1\tau_\star}\;.
\ee
Combining with Eq.~(\ref{taustar}) we obtain
\be
(1-\varrho)_{\rm max}\propto (\varphi_A-\delta)^{-\theta^1/\theta^2}\;.
\ee
Finally, inserting this into (\ref{Gplrho}) yields the 
scaling~(\ref{GIRscalingphi}). 

Note that in the opposite limit $\varphi_A\to \pi/2$ the `escape time' $\tau_\star$ remains finite, whereas the initial condition $(1-\varrho)_0$ is proportional to 
$(\pi/2-\varphi_A)$. Thus, the IR gravitational coupling simply obeys the scaling 
\be
{\cal G}_{IR}\propto (\pi/2-\varphi_A)^{\kappa_{II}}
\ee
and quickly vanishes when $\varphi_A$ approaches $\pi/2$.

\bibliography{FPlong}

\begin{thebibliography}{55}%
\makeatletter
\providecommand \@ifxundefined [1]{%
 \@ifx{#1\undefined}
}%
\providecommand \@ifnum [1]{%
 \ifnum #1\expandafter \@firstoftwo
 \else \expandafter \@secondoftwo
 \fi
}%
\providecommand \@ifx [1]{%
 \ifx #1\expandafter \@firstoftwo
 \else \expandafter \@secondoftwo
 \fi
}%
\providecommand \natexlab [1]{#1}%
\providecommand \enquote  [1]{``#1''}%
\providecommand \bibnamefont  [1]{#1}%
\providecommand \bibfnamefont [1]{#1}%
\providecommand \citenamefont [1]{#1}%
\providecommand \href@noop [0]{\@secondoftwo}%
\providecommand \href [0]{\begingroup \@sanitize@url \@href}%
\providecommand \@href[1]{\@@startlink{#1}\@@href}%
\providecommand \@@href[1]{\endgroup#1\@@endlink}%
\providecommand \@sanitize@url [0]{\catcode `\\12\catcode `\$12\catcode
  `\&12\catcode `\#12\catcode `\^12\catcode `\_12\catcode `\%12\relax}%
\providecommand \@@startlink[1]{}%
\providecommand \@@endlink[0]{}%
\providecommand \url  [0]{\begingroup\@sanitize@url \@url }%
\providecommand \@url [1]{\endgroup\@href {#1}{\urlprefix }}%
\providecommand \urlprefix  [0]{URL }%
\providecommand \Eprint [0]{\href }%
\providecommand \doibase [0]{https://doi.org/}%
\providecommand \selectlanguage [0]{\@gobble}%
\providecommand \bibinfo  [0]{\@secondoftwo}%
\providecommand \bibfield  [0]{\@secondoftwo}%
\providecommand \translation [1]{[#1]}%
\providecommand \BibitemOpen [0]{}%
\providecommand \bibitemStop [0]{}%
\providecommand \bibitemNoStop [0]{.\EOS\space}%
\providecommand \EOS [0]{\spacefactor3000\relax}%
\providecommand \BibitemShut  [1]{\csname bibitem#1\endcsname}%
\let\auto@bib@innerbib\@empty
\bibitem [{\citenamefont {Ho{\v r}ava}(2009)}]{Horava:2009uw}%
  \BibitemOpen
  \bibfield  {author} {\bibinfo {author} {\bibfnamefont {P.}~\bibnamefont
  {Ho{\v r}ava}},\ }\bibfield  {title} {\bibinfo {title} {{Quantum Gravity at a
  Lifshitz Point}},\ }\href {https://doi.org/10.1103/PhysRevD.79.084008}
  {\bibfield  {journal} {\bibinfo  {journal} {Phys. Rev.}\ }\textbf {\bibinfo
  {volume} {D79}},\ \bibinfo {pages} {084008} (\bibinfo {year} {2009})},\
  \Eprint {https://arxiv.org/abs/0901.3775} {arXiv:0901.3775 [hep-th]}
  \BibitemShut {NoStop}%
\bibitem [{\citenamefont {Mukohyama}(2010)}]{Mukohyama:2010xz}%
  \BibitemOpen
  \bibfield  {author} {\bibinfo {author} {\bibfnamefont {S.}~\bibnamefont
  {Mukohyama}},\ }\bibfield  {title} {\bibinfo {title} {{Ho\v{r}ava-Lifshitz
  Cosmology: A Review}},\ }\href
  {https://doi.org/10.1088/0264-9381/27/22/223101} {\bibfield  {journal}
  {\bibinfo  {journal} {Class. Quant. Grav.}\ }\textbf {\bibinfo {volume}
  {27}},\ \bibinfo {pages} {223101} (\bibinfo {year} {2010})},\ \Eprint
  {https://arxiv.org/abs/1007.5199} {arXiv:1007.5199 [hep-th]} \BibitemShut
  {NoStop}%
\bibitem [{\citenamefont {Sotiriou}(2011)}]{Sotiriou:2010wn}%
  \BibitemOpen
  \bibfield  {author} {\bibinfo {author} {\bibfnamefont {T.~P.}\ \bibnamefont
  {Sotiriou}},\ }\bibfield  {title} {\bibinfo {title} {{Ho\v{r}ava-Lifshitz
  gravity: a status report}},\ }\href
  {https://doi.org/10.1088/1742-6596/283/1/012034} {\bibfield  {journal}
  {\bibinfo  {journal} {J. Phys. Conf. Ser.}\ }\textbf {\bibinfo {volume}
  {283}},\ \bibinfo {pages} {012034} (\bibinfo {year} {2011})},\ \Eprint
  {https://arxiv.org/abs/1010.3218} {arXiv:1010.3218 [hep-th]} \BibitemShut
  {NoStop}%
\bibitem [{\citenamefont {Barvinsky}(2023)}]{Barvinsky:2023mrv}%
  \BibitemOpen
  \bibfield  {author} {\bibinfo {author} {\bibfnamefont {A.~O.}\ \bibnamefont
  {Barvinsky}},\ }\bibinfo {title} {{Ho\v{r}ava Models as Palladium of
  Unitarity and Renormalizability in Quantum Gravity}}\ (\bibinfo {year}
  {2023})\ \Eprint {https://arxiv.org/abs/2301.13580} {arXiv:2301.13580
  [hep-th]} \BibitemShut {NoStop}%
\bibitem [{\citenamefont {Herrero-Valea}(2023)}]{Herrero-Valea:2023zex}%
  \BibitemOpen
  \bibfield  {author} {\bibinfo {author} {\bibfnamefont {M.}~\bibnamefont
  {Herrero-Valea}},\ }\bibfield  {title} {\bibinfo {title} {{The status of
  Ho\v{r}ava gravity}},\ }\href
  {https://doi.org/10.1140/epjp/s13360-023-04593-y} {\bibfield  {journal}
  {\bibinfo  {journal} {Eur. Phys. J. Plus}\ }\textbf {\bibinfo {volume}
  {138}},\ \bibinfo {pages} {968} (\bibinfo {year} {2023})},\ \Eprint
  {https://arxiv.org/abs/2307.13039} {arXiv:2307.13039 [gr-qc]} \BibitemShut
  {NoStop}%
\bibitem [{\citenamefont {Stelle}(1977)}]{Stelle:1976gc}%
  \BibitemOpen
  \bibfield  {author} {\bibinfo {author} {\bibfnamefont {K.~S.}\ \bibnamefont
  {Stelle}},\ }\bibfield  {title} {\bibinfo {title} {{Renormalization of Higher
  Derivative Quantum Gravity}},\ }\href
  {https://doi.org/10.1103/PhysRevD.16.953} {\bibfield  {journal} {\bibinfo
  {journal} {Phys. Rev.}\ }\textbf {\bibinfo {volume} {D16}},\ \bibinfo {pages}
  {953} (\bibinfo {year} {1977})}\BibitemShut {NoStop}%
\bibitem [{\citenamefont {Stelle}(1978)}]{Stelle:1977ry}%
  \BibitemOpen
  \bibfield  {author} {\bibinfo {author} {\bibfnamefont {K.~S.}\ \bibnamefont
  {Stelle}},\ }\bibfield  {title} {\bibinfo {title} {{Classical Gravity with
  Higher Derivatives}},\ }\href {https://doi.org/10.1007/BF00760427} {\bibfield
   {journal} {\bibinfo  {journal} {Gen. Rel. Grav.}\ }\textbf {\bibinfo
  {volume} {9}},\ \bibinfo {pages} {353} (\bibinfo {year} {1978})}\BibitemShut
  {NoStop}%
\bibitem [{\citenamefont {Kostelecky}\ and\ \citenamefont
  {Russell}(2011)}]{Kostelecky:2008ts}%
  \BibitemOpen
  \bibfield  {author} {\bibinfo {author} {\bibfnamefont {V.~A.}\ \bibnamefont
  {Kostelecky}}\ and\ \bibinfo {author} {\bibfnamefont {N.}~\bibnamefont
  {Russell}},\ }\bibfield  {title} {\bibinfo {title} {{Data Tables for Lorentz
  and CPT Violation}},\ }\href {https://doi.org/10.1103/RevModPhys.83.11}
  {\bibfield  {journal} {\bibinfo  {journal} {Rev. Mod. Phys.}\ }\textbf
  {\bibinfo {volume} {83}},\ \bibinfo {pages} {11} (\bibinfo {year} {2011})},\
  \Eprint {https://arxiv.org/abs/0801.0287} {arXiv:0801.0287 [hep-ph]}
  \BibitemShut {NoStop}%
\bibitem [{\citenamefont {Liberati}(2013)}]{Liberati:2013xla}%
  \BibitemOpen
  \bibfield  {author} {\bibinfo {author} {\bibfnamefont {S.}~\bibnamefont
  {Liberati}},\ }\bibfield  {title} {\bibinfo {title} {{Tests of Lorentz
  invariance: a 2013 update}},\ }\href
  {https://doi.org/10.1088/0264-9381/30/13/133001} {\bibfield  {journal}
  {\bibinfo  {journal} {Class. Quant. Grav.}\ }\textbf {\bibinfo {volume}
  {30}},\ \bibinfo {pages} {133001} (\bibinfo {year} {2013})},\ \Eprint
  {https://arxiv.org/abs/1304.5795} {arXiv:1304.5795 [gr-qc]} \BibitemShut
  {NoStop}%
\bibitem [{\citenamefont {Groot~Nibbelink}\ and\ \citenamefont
  {Pospelov}(2005)}]{GrootNibbelink:2004za}%
  \BibitemOpen
  \bibfield  {author} {\bibinfo {author} {\bibfnamefont {S.}~\bibnamefont
  {Groot~Nibbelink}}\ and\ \bibinfo {author} {\bibfnamefont {M.}~\bibnamefont
  {Pospelov}},\ }\bibfield  {title} {\bibinfo {title} {{Lorentz violation in
  supersymmetric field theories}},\ }\href
  {https://doi.org/10.1103/PhysRevLett.94.081601} {\bibfield  {journal}
  {\bibinfo  {journal} {Phys. Rev. Lett.}\ }\textbf {\bibinfo {volume} {94}},\
  \bibinfo {pages} {081601} (\bibinfo {year} {2005})},\ \Eprint
  {https://arxiv.org/abs/hep-ph/0404271} {arXiv:hep-ph/0404271 [hep-ph]}
  \BibitemShut {NoStop}%
\bibitem [{\citenamefont {Pospelov}\ and\ \citenamefont
  {Shang}(2012)}]{Pospelov:2010mp}%
  \BibitemOpen
  \bibfield  {author} {\bibinfo {author} {\bibfnamefont {M.}~\bibnamefont
  {Pospelov}}\ and\ \bibinfo {author} {\bibfnamefont {Y.}~\bibnamefont
  {Shang}},\ }\bibfield  {title} {\bibinfo {title} {{On Lorentz violation in
  Horava-Lifshitz type theories}},\ }\href
  {https://doi.org/10.1103/PhysRevD.85.105001} {\bibfield  {journal} {\bibinfo
  {journal} {Phys. Rev. D}\ }\textbf {\bibinfo {volume} {85}},\ \bibinfo
  {pages} {105001} (\bibinfo {year} {2012})},\ \Eprint
  {https://arxiv.org/abs/1010.5249} {arXiv:1010.5249 [hep-th]} \BibitemShut
  {NoStop}%
\bibitem [{\citenamefont {Pujolas}\ and\ \citenamefont
  {Sibiryakov}(2012)}]{Pujolas:2011sk}%
  \BibitemOpen
  \bibfield  {author} {\bibinfo {author} {\bibfnamefont {O.}~\bibnamefont
  {Pujolas}}\ and\ \bibinfo {author} {\bibfnamefont {S.}~\bibnamefont
  {Sibiryakov}},\ }\bibfield  {title} {\bibinfo {title} {{Supersymmetric
  Aether}},\ }\href {https://doi.org/10.1007/JHEP01(2012)062} {\bibfield
  {journal} {\bibinfo  {journal} {JHEP}\ }\textbf {\bibinfo {volume} {01}},\
  \bibinfo {pages} {062}},\ \Eprint {https://arxiv.org/abs/1109.4495}
  {arXiv:1109.4495 [hep-th]} \BibitemShut {NoStop}%
\bibitem [{\citenamefont {Anber}\ and\ \citenamefont
  {Donoghue}(2011)}]{Anber:2011xf}%
  \BibitemOpen
  \bibfield  {author} {\bibinfo {author} {\bibfnamefont {M.~M.}\ \bibnamefont
  {Anber}}\ and\ \bibinfo {author} {\bibfnamefont {J.~F.}\ \bibnamefont
  {Donoghue}},\ }\bibfield  {title} {\bibinfo {title} {{The Emergence of a
  universal limiting speed}},\ }\href
  {https://doi.org/10.1103/PhysRevD.83.105027} {\bibfield  {journal} {\bibinfo
  {journal} {Phys. Rev.}\ }\textbf {\bibinfo {volume} {D83}},\ \bibinfo {pages}
  {105027} (\bibinfo {year} {2011})},\ \Eprint
  {https://arxiv.org/abs/1102.0789} {arXiv:1102.0789 [hep-th]} \BibitemShut
  {NoStop}%
\bibitem [{\citenamefont {Bednik}\ \emph {et~al.}(2013)\citenamefont {Bednik},
  \citenamefont {Pujolas},\ and\ \citenamefont {Sibiryakov}}]{Bednik:2013nxa}%
  \BibitemOpen
  \bibfield  {author} {\bibinfo {author} {\bibfnamefont {G.}~\bibnamefont
  {Bednik}}, \bibinfo {author} {\bibfnamefont {O.}~\bibnamefont {Pujolas}},\
  and\ \bibinfo {author} {\bibfnamefont {S.}~\bibnamefont {Sibiryakov}},\
  }\bibfield  {title} {\bibinfo {title} {{Emergent Lorentz invariance from
  Strong Dynamics: Holographic examples}},\ }\href
  {https://doi.org/10.1007/JHEP11(2013)064} {\bibfield  {journal} {\bibinfo
  {journal} {JHEP}\ }\textbf {\bibinfo {volume} {11}},\ \bibinfo {pages}
  {064}},\ \Eprint {https://arxiv.org/abs/1305.0011} {arXiv:1305.0011 [hep-th]}
  \BibitemShut {NoStop}%
\bibitem [{\citenamefont {Kharuk}\ and\ \citenamefont
  {Sibiryakov}(2016)}]{Kharuk:2015wga}%
  \BibitemOpen
  \bibfield  {author} {\bibinfo {author} {\bibfnamefont {I.}~\bibnamefont
  {Kharuk}}\ and\ \bibinfo {author} {\bibfnamefont {S.~M.}\ \bibnamefont
  {Sibiryakov}},\ }\bibfield  {title} {\bibinfo {title} {{Emergent Lorentz
  invariance with chiral fermions}},\ }\href
  {https://doi.org/10.1134/S0040577916120084} {\bibfield  {journal} {\bibinfo
  {journal} {Theor. Math. Phys.}\ }\textbf {\bibinfo {volume} {189}},\ \bibinfo
  {pages} {1755} (\bibinfo {year} {2016})},\ \bibinfo {note} {[Teor. Mat. Fiz.
  189, no.3, 405 (2016)]},\ \Eprint {https://arxiv.org/abs/1505.04130}
  {arXiv:1505.04130 [hep-th]} \BibitemShut {NoStop}%
\bibitem [{\citenamefont {Baggioli}\ \emph {et~al.}(2024)\citenamefont
  {Baggioli}, \citenamefont {Pujolas},\ and\ \citenamefont
  {Wu}}]{Baggioli:2024vza}%
  \BibitemOpen
  \bibfield  {author} {\bibinfo {author} {\bibfnamefont {M.}~\bibnamefont
  {Baggioli}}, \bibinfo {author} {\bibfnamefont {O.}~\bibnamefont {Pujolas}},\
  and\ \bibinfo {author} {\bibfnamefont {X.-M.}\ \bibnamefont {Wu}},\
  }\bibfield  {title} {\bibinfo {title} {{Holographic Lifshitz flows}},\ }\href
  {https://doi.org/10.1007/JHEP09(2024)175} {\bibfield  {journal} {\bibinfo
  {journal} {JHEP}\ }\textbf {\bibinfo {volume} {09}},\ \bibinfo {pages}
  {175}},\ \Eprint {https://arxiv.org/abs/2407.11552} {arXiv:2407.11552
  [hep-th]} \BibitemShut {NoStop}%
\bibitem [{\citenamefont {Blas}\ \emph {et~al.}(2010)\citenamefont {Blas},
  \citenamefont {Pujolas},\ and\ \citenamefont {Sibiryakov}}]{Blas:2009qj}%
  \BibitemOpen
  \bibfield  {author} {\bibinfo {author} {\bibfnamefont {D.}~\bibnamefont
  {Blas}}, \bibinfo {author} {\bibfnamefont {O.}~\bibnamefont {Pujolas}},\ and\
  \bibinfo {author} {\bibfnamefont {S.}~\bibnamefont {Sibiryakov}},\ }\bibfield
   {title} {\bibinfo {title} {{Consistent Extension of Ho\v{r}ava Gravity}},\
  }\href {https://doi.org/10.1103/PhysRevLett.104.181302} {\bibfield  {journal}
  {\bibinfo  {journal} {Phys. Rev. Lett.}\ }\textbf {\bibinfo {volume} {104}},\
  \bibinfo {pages} {181302} (\bibinfo {year} {2010})},\ \Eprint
  {https://arxiv.org/abs/0909.3525} {arXiv:0909.3525 [hep-th]} \BibitemShut
  {NoStop}%
\bibitem [{\citenamefont {Blas}\ and\ \citenamefont
  {Lim}(2015)}]{Blas:2014aca}%
  \BibitemOpen
  \bibfield  {author} {\bibinfo {author} {\bibfnamefont {D.}~\bibnamefont
  {Blas}}\ and\ \bibinfo {author} {\bibfnamefont {E.}~\bibnamefont {Lim}},\
  }\bibfield  {title} {\bibinfo {title} {{Phenomenology of theories of gravity
  without Lorentz invariance: the preferred frame case}},\ }\href
  {https://doi.org/10.1142/S0218271814430093} {\bibfield  {journal} {\bibinfo
  {journal} {Int. J. Mod. Phys.}\ }\textbf {\bibinfo {volume} {D23}},\ \bibinfo
  {pages} {1443009} (\bibinfo {year} {2015})},\ \Eprint
  {https://arxiv.org/abs/1412.4828} {arXiv:1412.4828 [gr-qc]} \BibitemShut
  {NoStop}%
\bibitem [{\citenamefont {Emir~G\"umr\"uk\c{c}\"uo\u{g}lu}\ \emph
  {et~al.}(2018)\citenamefont {Emir~G\"umr\"uk\c{c}\"uo\u{g}lu}, \citenamefont
  {Saravani},\ and\ \citenamefont {Sotiriou}}]{EmirGumrukcuoglu:2017cfa}%
  \BibitemOpen
  \bibfield  {author} {\bibinfo {author} {\bibfnamefont {A.}~\bibnamefont
  {Emir~G\"umr\"uk\c{c}\"uo\u{g}lu}}, \bibinfo {author} {\bibfnamefont
  {M.}~\bibnamefont {Saravani}},\ and\ \bibinfo {author} {\bibfnamefont
  {T.~P.}\ \bibnamefont {Sotiriou}},\ }\bibfield  {title} {\bibinfo {title}
  {{Ho\v{r}ava gravity after GW170817}},\ }\href
  {https://doi.org/10.1103/PhysRevD.97.024032} {\bibfield  {journal} {\bibinfo
  {journal} {Phys. Rev. D}\ }\textbf {\bibinfo {volume} {97}},\ \bibinfo
  {pages} {024032} (\bibinfo {year} {2018})},\ \Eprint
  {https://arxiv.org/abs/1711.08845} {arXiv:1711.08845 [gr-qc]} \BibitemShut
  {NoStop}%
\bibitem [{\citenamefont {Bellor\'{\i}n}\ \emph {et~al.}(2022)\citenamefont
  {Bellor\'{\i}n}, \citenamefont {B\'orquez},\ and\ \citenamefont
  {Droguett}}]{Bellorin:2022np}%
  \BibitemOpen
  \bibfield  {author} {\bibinfo {author} {\bibfnamefont {J.}~\bibnamefont
  {Bellor\'{\i}n}}, \bibinfo {author} {\bibfnamefont {C.}~\bibnamefont
  {B\'orquez}},\ and\ \bibinfo {author} {\bibfnamefont {B.}~\bibnamefont
  {Droguett}},\ }\bibfield  {title} {\bibinfo {title} {Cancellation of
  divergences in the nonprojectable ho\ifmmode \check{r}\else \v{r}\fi{}ava
  theory},\ }\href {https://doi.org/10.1103/PhysRevD.106.044055} {\bibfield
  {journal} {\bibinfo  {journal} {Phys. Rev. D}\ }\textbf {\bibinfo {volume}
  {106}},\ \bibinfo {pages} {044055} (\bibinfo {year} {2022})},\ \Eprint
  {https://arxiv.org/abs/2207.08938} {arXiv:2207.08938 [hep-th]} \BibitemShut
  {NoStop}%
\bibitem [{\citenamefont {Barvinsky}\ \emph {et~al.}(2016)\citenamefont
  {Barvinsky}, \citenamefont {Blas}, \citenamefont {Herrero-Valea},
  \citenamefont {Sibiryakov},\ and\ \citenamefont
  {Steinwachs}}]{Barvinsky:2015kil}%
  \BibitemOpen
  \bibfield  {author} {\bibinfo {author} {\bibfnamefont {A.~O.}\ \bibnamefont
  {Barvinsky}}, \bibinfo {author} {\bibfnamefont {D.}~\bibnamefont {Blas}},
  \bibinfo {author} {\bibfnamefont {M.}~\bibnamefont {Herrero-Valea}}, \bibinfo
  {author} {\bibfnamefont {S.~M.}\ \bibnamefont {Sibiryakov}},\ and\ \bibinfo
  {author} {\bibfnamefont {C.~F.}\ \bibnamefont {Steinwachs}},\ }\bibfield
  {title} {\bibinfo {title} {{Renormalization of Ho\v{r}ava gravity}},\ }\href
  {https://doi.org/10.1103/PhysRevD.93.064022} {\bibfield  {journal} {\bibinfo
  {journal} {Phys. Rev.}\ }\textbf {\bibinfo {volume} {D93}},\ \bibinfo {pages}
  {064022} (\bibinfo {year} {2016})},\ \Eprint
  {https://arxiv.org/abs/1512.02250} {arXiv:1512.02250 [hep-th]} \BibitemShut
  {NoStop}%
\bibitem [{\citenamefont {Barvinsky}\ \emph {et~al.}(2018)\citenamefont
  {Barvinsky}, \citenamefont {Blas}, \citenamefont {Herrero-Valea},
  \citenamefont {Sibiryakov},\ and\ \citenamefont
  {Steinwachs}}]{Barvinsky:2017zlx}%
  \BibitemOpen
  \bibfield  {author} {\bibinfo {author} {\bibfnamefont {A.~O.}\ \bibnamefont
  {Barvinsky}}, \bibinfo {author} {\bibfnamefont {D.}~\bibnamefont {Blas}},
  \bibinfo {author} {\bibfnamefont {M.}~\bibnamefont {Herrero-Valea}}, \bibinfo
  {author} {\bibfnamefont {S.~M.}\ \bibnamefont {Sibiryakov}},\ and\ \bibinfo
  {author} {\bibfnamefont {C.~F.}\ \bibnamefont {Steinwachs}},\ }\bibfield
  {title} {\bibinfo {title} {{Renormalization of gauge theories in the
  background-field approach}},\ }\href
  {https://doi.org/10.1007/JHEP07(2018)035} {\bibfield  {journal} {\bibinfo
  {journal} {JHEP}\ }\textbf {\bibinfo {volume} {07}},\ \bibinfo {pages}
  {035}},\ \Eprint {https://arxiv.org/abs/1705.03480} {arXiv:1705.03480
  [hep-th]} \BibitemShut {NoStop}%
\bibitem [{\citenamefont {Barvinsky}\ \emph {et~al.}(2017)\citenamefont
  {Barvinsky}, \citenamefont {Blas}, \citenamefont {Herrero-Valea},
  \citenamefont {Sibiryakov},\ and\ \citenamefont
  {Steinwachs}}]{Barvinsky:2017kob}%
  \BibitemOpen
  \bibfield  {author} {\bibinfo {author} {\bibfnamefont {A.~O.}\ \bibnamefont
  {Barvinsky}}, \bibinfo {author} {\bibfnamefont {D.}~\bibnamefont {Blas}},
  \bibinfo {author} {\bibfnamefont {M.}~\bibnamefont {Herrero-Valea}}, \bibinfo
  {author} {\bibfnamefont {S.~M.}\ \bibnamefont {Sibiryakov}},\ and\ \bibinfo
  {author} {\bibfnamefont {C.~F.}\ \bibnamefont {Steinwachs}},\ }\bibfield
  {title} {\bibinfo {title} {{Ho\v{r}ava Gravity is Asymptotically Free in 2 +
  1 Dimensions}},\ }\href {https://doi.org/10.1103/PhysRevLett.119.211301}
  {\bibfield  {journal} {\bibinfo  {journal} {Phys. Rev. Lett.}\ }\textbf
  {\bibinfo {volume} {119}},\ \bibinfo {pages} {211301} (\bibinfo {year}
  {2017})},\ \Eprint {https://arxiv.org/abs/1706.06809} {arXiv:1706.06809
  [hep-th]} \BibitemShut {NoStop}%
\bibitem [{\citenamefont {Barvinsky}\ \emph {et~al.}(2019)\citenamefont
  {Barvinsky}, \citenamefont {Herrero-Valea},\ and\ \citenamefont
  {Sibiryakov}}]{towards}%
  \BibitemOpen
  \bibfield  {author} {\bibinfo {author} {\bibfnamefont {A.~O.}\ \bibnamefont
  {Barvinsky}}, \bibinfo {author} {\bibfnamefont {M.}~\bibnamefont
  {Herrero-Valea}},\ and\ \bibinfo {author} {\bibfnamefont {S.~M.}\
  \bibnamefont {Sibiryakov}},\ }\bibfield  {title} {\bibinfo {title} {{Towards
  the renormalization group flow of Ho\v rava gravity in $(3+1)$ dimensions}},\
  }\href {https://doi.org/10.1103/PhysRevD.100.026012} {\bibfield  {journal}
  {\bibinfo  {journal} {Phys. Rev. D}\ }\textbf {\bibinfo {volume} {100}},\
  \bibinfo {pages} {026012} (\bibinfo {year} {2019})},\ \Eprint
  {https://arxiv.org/abs/1905.03798} {arXiv:1905.03798 [hep-th]} \BibitemShut
  {NoStop}%
\bibitem [{\citenamefont {Barvinsky}\ \emph {et~al.}(2022)\citenamefont
  {Barvinsky}, \citenamefont {Kurov},\ and\ \citenamefont {Sibiryakov}}]{3+1}%
  \BibitemOpen
  \bibfield  {author} {\bibinfo {author} {\bibfnamefont {A.~O.}\ \bibnamefont
  {Barvinsky}}, \bibinfo {author} {\bibfnamefont {A.~V.}\ \bibnamefont
  {Kurov}},\ and\ \bibinfo {author} {\bibfnamefont {S.~M.}\ \bibnamefont
  {Sibiryakov}},\ }\bibfield  {title} {\bibinfo {title} {{Beta functions of
  (3+1)-dimensional projectable Ho\v{r}ava gravity}},\ }\href
  {https://doi.org/10.1103/PhysRevD.105.044009} {\bibfield  {journal} {\bibinfo
   {journal} {Phys. Rev. D}\ }\textbf {\bibinfo {volume} {105}},\ \bibinfo
  {pages} {044009} (\bibinfo {year} {2022})},\ \Eprint
  {https://arxiv.org/abs/2110.14688} {arXiv:2110.14688 [hep-th]} \BibitemShut
  {NoStop}%
\bibitem [{\citenamefont {Barvinsky}\ \emph {et~al.}(2023)\citenamefont
  {Barvinsky}, \citenamefont {Kurov},\ and\ \citenamefont
  {Sibiryakov}}]{Barvinsky:2023uir}%
  \BibitemOpen
  \bibfield  {author} {\bibinfo {author} {\bibfnamefont {A.~O.}\ \bibnamefont
  {Barvinsky}}, \bibinfo {author} {\bibfnamefont {A.~V.}\ \bibnamefont
  {Kurov}},\ and\ \bibinfo {author} {\bibfnamefont {S.~M.}\ \bibnamefont
  {Sibiryakov}},\ }\bibfield  {title} {\bibinfo {title} {{Asymptotic freedom in
  (3+1)-dimensional projectable Ho\v{r}ava gravity: Connecting the ultraviolet
  and infrared domains}},\ }\href
  {https://doi.org/10.1103/PhysRevD.108.L121503} {\bibfield  {journal}
  {\bibinfo  {journal} {Phys. Rev. D}\ }\textbf {\bibinfo {volume} {108}},\
  \bibinfo {pages} {L121503} (\bibinfo {year} {2023})},\ \Eprint
  {https://arxiv.org/abs/2310.07841} {arXiv:2310.07841 [hep-th]} \BibitemShut
  {NoStop}%
\bibitem [{\citenamefont {Mukohyama}(2009)}]{Mukohyama:2009mz}%
  \BibitemOpen
  \bibfield  {author} {\bibinfo {author} {\bibfnamefont {S.}~\bibnamefont
  {Mukohyama}},\ }\bibfield  {title} {\bibinfo {title} {{Dark matter as
  integration constant in Horava-Lifshitz gravity}},\ }\href
  {https://doi.org/10.1103/PhysRevD.80.064005} {\bibfield  {journal} {\bibinfo
  {journal} {Phys. Rev. D}\ }\textbf {\bibinfo {volume} {80}},\ \bibinfo
  {pages} {064005} (\bibinfo {year} {2009})},\ \Eprint
  {https://arxiv.org/abs/0905.3563} {arXiv:0905.3563 [hep-th]} \BibitemShut
  {NoStop}%
\bibitem [{\citenamefont {Izumi}\ and\ \citenamefont
  {Mukohyama}(2010)}]{Izumi:2009ry}%
  \BibitemOpen
  \bibfield  {author} {\bibinfo {author} {\bibfnamefont {K.}~\bibnamefont
  {Izumi}}\ and\ \bibinfo {author} {\bibfnamefont {S.}~\bibnamefont
  {Mukohyama}},\ }\bibfield  {title} {\bibinfo {title} {{Stellar center is
  dynamical in Horava-Lifshitz gravity}},\ }\href
  {https://doi.org/10.1103/PhysRevD.81.044008} {\bibfield  {journal} {\bibinfo
  {journal} {Phys. Rev. D}\ }\textbf {\bibinfo {volume} {81}},\ \bibinfo
  {pages} {044008} (\bibinfo {year} {2010})},\ \Eprint
  {https://arxiv.org/abs/0911.1814} {arXiv:0911.1814 [hep-th]} \BibitemShut
  {NoStop}%
\bibitem [{\citenamefont {Izumi}\ and\ \citenamefont
  {Mukohyama}(2011)}]{Izumi:2011eh}%
  \BibitemOpen
  \bibfield  {author} {\bibinfo {author} {\bibfnamefont {K.}~\bibnamefont
  {Izumi}}\ and\ \bibinfo {author} {\bibfnamefont {S.}~\bibnamefont
  {Mukohyama}},\ }\bibfield  {title} {\bibinfo {title} {{Nonlinear superhorizon
  perturbations in Ho\v{r}ava-Lifshitz gravity}},\ }\href
  {https://doi.org/10.1103/PhysRevD.84.064025} {\bibfield  {journal} {\bibinfo
  {journal} {Phys. Rev.}\ }\textbf {\bibinfo {volume} {D84}},\ \bibinfo {pages}
  {064025} (\bibinfo {year} {2011})},\ \Eprint
  {https://arxiv.org/abs/1105.0246} {arXiv:1105.0246 [hep-th]} \BibitemShut
  {NoStop}%
\bibitem [{\citenamefont {G{\"u}mr{\"u}k{\c c}{\"u}o{\u g}lu}\ \emph
  {et~al.}(2012)\citenamefont {G{\"u}mr{\"u}k{\c c}{\"u}o{\u g}lu},
  \citenamefont {Mukohyama},\ and\ \citenamefont {Wang}}]{Gumrukcuoglu:2011ef}%
  \BibitemOpen
  \bibfield  {author} {\bibinfo {author} {\bibfnamefont {A.~E.}\ \bibnamefont
  {G{\"u}mr{\"u}k{\c c}{\"u}o{\u g}lu}}, \bibinfo {author} {\bibfnamefont
  {S.}~\bibnamefont {Mukohyama}},\ and\ \bibinfo {author} {\bibfnamefont
  {A.}~\bibnamefont {Wang}},\ }\bibfield  {title} {\bibinfo {title} {{General
  relativity limit of Ho\v{r}ava-Lifshitz gravity with a scalar field in
  gradient expansion}},\ }\href {https://doi.org/10.1103/PhysRevD.85.064042}
  {\bibfield  {journal} {\bibinfo  {journal} {Phys. Rev.}\ }\textbf {\bibinfo
  {volume} {D85}},\ \bibinfo {pages} {064042} (\bibinfo {year} {2012})},\
  \Eprint {https://arxiv.org/abs/1109.2609} {arXiv:1109.2609 [hep-th]}
  \BibitemShut {NoStop}%
\bibitem [{\citenamefont {Bassani}\ \emph {et~al.}(2024)\citenamefont
  {Bassani}, \citenamefont {Magueijo},\ and\ \citenamefont
  {Mukohyama}}]{Bassani:2024lan}%
  \BibitemOpen
  \bibfield  {author} {\bibinfo {author} {\bibfnamefont {P.~M.}\ \bibnamefont
  {Bassani}}, \bibinfo {author} {\bibfnamefont {J.}~\bibnamefont {Magueijo}},\
  and\ \bibinfo {author} {\bibfnamefont {S.}~\bibnamefont {Mukohyama}},\
  }\href@noop {} {\bibinfo {title} {{Violations of energy conservation in
  Horava-Lifshitz gravity: a new ingredient in the dark matter puzzle}}}
  (\bibinfo {year} {2024}),\ \Eprint {https://arxiv.org/abs/2408.03793}
  {arXiv:2408.03793 [gr-qc]} \BibitemShut {NoStop}%
\bibitem [{\citenamefont {Koyama}\ and\ \citenamefont
  {Arroja}(2010)}]{Koyama:2009hc}%
  \BibitemOpen
  \bibfield  {author} {\bibinfo {author} {\bibfnamefont {K.}~\bibnamefont
  {Koyama}}\ and\ \bibinfo {author} {\bibfnamefont {F.}~\bibnamefont
  {Arroja}},\ }\bibfield  {title} {\bibinfo {title} {{Pathological behaviour of
  the scalar graviton in Ho\v{r}ava-Lifshitz gravity}},\ }\href
  {https://doi.org/10.1007/JHEP03(2010)061} {\bibfield  {journal} {\bibinfo
  {journal} {JHEP}\ }\textbf {\bibinfo {volume} {03}},\ \bibinfo {pages}
  {061}},\ \Eprint {https://arxiv.org/abs/0910.1998} {arXiv:0910.1998 [hep-th]}
  \BibitemShut {NoStop}%
\bibitem [{\citenamefont {Blas}\ \emph {et~al.}(2011)\citenamefont {Blas},
  \citenamefont {Pujolas},\ and\ \citenamefont {Sibiryakov}}]{Blas:2010hb}%
  \BibitemOpen
  \bibfield  {author} {\bibinfo {author} {\bibfnamefont {D.}~\bibnamefont
  {Blas}}, \bibinfo {author} {\bibfnamefont {O.}~\bibnamefont {Pujolas}},\ and\
  \bibinfo {author} {\bibfnamefont {S.}~\bibnamefont {Sibiryakov}},\ }\bibfield
   {title} {\bibinfo {title} {{Models of non-relativistic quantum gravity: The
  Good, the bad and the healthy}},\ }\href
  {https://doi.org/10.1007/JHEP04(2011)018} {\bibfield  {journal} {\bibinfo
  {journal} {JHEP}\ }\textbf {\bibinfo {volume} {04}},\ \bibinfo {pages}
  {018}},\ \Eprint {https://arxiv.org/abs/1007.3503} {arXiv:1007.3503 [hep-th]}
  \BibitemShut {NoStop}%
\bibitem [{\citenamefont {Germani}\ \emph {et~al.}(2009)\citenamefont
  {Germani}, \citenamefont {Kehagias},\ and\ \citenamefont
  {Sfetsos}}]{Germani:2009yt}%
  \BibitemOpen
  \bibfield  {author} {\bibinfo {author} {\bibfnamefont {C.}~\bibnamefont
  {Germani}}, \bibinfo {author} {\bibfnamefont {A.}~\bibnamefont {Kehagias}},\
  and\ \bibinfo {author} {\bibfnamefont {K.}~\bibnamefont {Sfetsos}},\
  }\bibfield  {title} {\bibinfo {title} {{Relativistic Quantum Gravity at a
  Lifshitz Point}},\ }\href {https://doi.org/10.1088/1126-6708/2009/09/060}
  {\bibfield  {journal} {\bibinfo  {journal} {JHEP}\ }\textbf {\bibinfo
  {volume} {09}},\ \bibinfo {pages} {060}},\ \Eprint
  {https://arxiv.org/abs/0906.1201} {arXiv:0906.1201 [hep-th]} \BibitemShut
  {NoStop}%
\bibitem [{\citenamefont {Blas}\ \emph {et~al.}(2009)\citenamefont {Blas},
  \citenamefont {Pujolas},\ and\ \citenamefont {Sibiryakov}}]{Blas:2009yd}%
  \BibitemOpen
  \bibfield  {author} {\bibinfo {author} {\bibfnamefont {D.}~\bibnamefont
  {Blas}}, \bibinfo {author} {\bibfnamefont {O.}~\bibnamefont {Pujolas}},\ and\
  \bibinfo {author} {\bibfnamefont {S.}~\bibnamefont {Sibiryakov}},\ }\bibfield
   {title} {\bibinfo {title} {{On the Extra Mode and Inconsistency of
  Ho\v{r}ava Gravity}},\ }\href {https://doi.org/10.1088/1126-6708/2009/10/029}
  {\bibfield  {journal} {\bibinfo  {journal} {JHEP}\ }\textbf {\bibinfo
  {volume} {10}},\ \bibinfo {pages} {029}},\ \Eprint
  {https://arxiv.org/abs/0906.3046} {arXiv:0906.3046 [hep-th]} \BibitemShut
  {NoStop}%
\bibitem [{\citenamefont {Kimpton}\ and\ \citenamefont
  {Padilla}(2010)}]{Kimpton:2010xi}%
  \BibitemOpen
  \bibfield  {author} {\bibinfo {author} {\bibfnamefont {I.}~\bibnamefont
  {Kimpton}}\ and\ \bibinfo {author} {\bibfnamefont {A.}~\bibnamefont
  {Padilla}},\ }\bibfield  {title} {\bibinfo {title} {{Lessons from the
  decoupling limit of Horava gravity}},\ }\href
  {https://doi.org/10.1007/JHEP07(2010)014} {\bibfield  {journal} {\bibinfo
  {journal} {JHEP}\ }\textbf {\bibinfo {volume} {07}},\ \bibinfo {pages}
  {014}},\ \Eprint {https://arxiv.org/abs/1003.5666} {arXiv:1003.5666 [hep-th]}
  \BibitemShut {NoStop}%
\bibitem [{\citenamefont {Abbott}\ \emph {et~al.}(2017)\citenamefont {Abbott}
  \emph {et~al.}}]{Monitor:2017mdv}%
  \BibitemOpen
  \bibfield  {author} {\bibinfo {author} {\bibfnamefont {B.~P.}\ \bibnamefont
  {Abbott}} \emph {et~al.} (\bibinfo {collaboration} {LIGO Scientific, Virgo,
  Fermi-GBM, INTEGRAL}),\ }\bibfield  {title} {\bibinfo {title} {{Gravitational
  Waves and Gamma-rays from a Binary Neutron Star Merger: GW170817 and GRB
  170817A}},\ }\href {https://doi.org/10.3847/2041-8213/aa920c} {\bibfield
  {journal} {\bibinfo  {journal} {Astrophys. J.}\ }\textbf {\bibinfo {volume}
  {848}},\ \bibinfo {pages} {L13} (\bibinfo {year} {2017})},\ \Eprint
  {https://arxiv.org/abs/1710.05834} {arXiv:1710.05834 [astro-ph.HE]}
  \BibitemShut {NoStop}%
\bibitem [{\citenamefont {Kimpton}\ and\ \citenamefont
  {Padilla}(2013)}]{Kimpton:2013zb}%
  \BibitemOpen
  \bibfield  {author} {\bibinfo {author} {\bibfnamefont {I.}~\bibnamefont
  {Kimpton}}\ and\ \bibinfo {author} {\bibfnamefont {A.}~\bibnamefont
  {Padilla}},\ }\bibfield  {title} {\bibinfo {title} {{Matter in
  Ho\v{r}ava-Lifshitz gravity}},\ }\href
  {https://doi.org/10.1007/JHEP04(2013)133} {\bibfield  {journal} {\bibinfo
  {journal} {JHEP}\ }\textbf {\bibinfo {volume} {04}},\ \bibinfo {pages}
  {133}},\ \Eprint {https://arxiv.org/abs/1301.6950} {arXiv:1301.6950 [hep-th]}
  \BibitemShut {NoStop}%
\bibitem [{\citenamefont {Sotiriou}\ \emph {et~al.}(2009)\citenamefont
  {Sotiriou}, \citenamefont {Visser},\ and\ \citenamefont
  {Weinfurtner}}]{Sotiriou:2009gy}%
  \BibitemOpen
  \bibfield  {author} {\bibinfo {author} {\bibfnamefont {T.~P.}\ \bibnamefont
  {Sotiriou}}, \bibinfo {author} {\bibfnamefont {M.}~\bibnamefont {Visser}},\
  and\ \bibinfo {author} {\bibfnamefont {S.}~\bibnamefont {Weinfurtner}},\
  }\bibfield  {title} {\bibinfo {title} {{Phenomenologically viable
  Lorentz-violating quantum gravity}},\ }\href
  {https://doi.org/10.1103/PhysRevLett.102.251601} {\bibfield  {journal}
  {\bibinfo  {journal} {Phys. Rev. Lett.}\ }\textbf {\bibinfo {volume} {102}},\
  \bibinfo {pages} {251601} (\bibinfo {year} {2009})},\ \Eprint
  {https://arxiv.org/abs/0904.4464} {arXiv:0904.4464 [hep-th]} \BibitemShut
  {NoStop}%
\bibitem [{\citenamefont {DeWitt}(1967)}]{PhysRev.162.1195}%
  \BibitemOpen
  \bibfield  {author} {\bibinfo {author} {\bibfnamefont {B.~S.}\ \bibnamefont
  {DeWitt}},\ }\bibfield  {title} {\bibinfo {title} {Quantum theory of gravity.
  ii. the manifestly covariant theory},\ }\href
  {https://doi.org/10.1103/PhysRev.162.1195} {\bibfield  {journal} {\bibinfo
  {journal} {Phys. Rev.}\ }\textbf {\bibinfo {volume} {162}},\ \bibinfo {pages}
  {1195} (\bibinfo {year} {1967})}\BibitemShut {NoStop}%
\bibitem [{\citenamefont {Kallosh}(1974)}]{KALLOSH1974293}%
  \BibitemOpen
  \bibfield  {author} {\bibinfo {author} {\bibfnamefont {R.}~\bibnamefont
  {Kallosh}},\ }\bibfield  {title} {\bibinfo {title} {Renormalization in
  non-abelian gauge theories},\ }\href
  {https://doi.org/https://doi.org/10.1016/0550-3213(74)90284-3} {\bibfield
  {journal} {\bibinfo  {journal} {Nuclear Physics B}\ }\textbf {\bibinfo
  {volume} {78}},\ \bibinfo {pages} {293} (\bibinfo {year} {1974})}\BibitemShut
  {NoStop}%
\bibitem [{\citenamefont {G\"umr\"uk\c{c}\"uo\u{g}lu}\ and\ \citenamefont
  {Mukohyama}(2011)}]{Gumrukcuoglu:2011xg}%
  \BibitemOpen
  \bibfield  {author} {\bibinfo {author} {\bibfnamefont {A.~E.}\ \bibnamefont
  {G\"umr\"uk\c{c}\"uo\u{g}lu}}\ and\ \bibinfo {author} {\bibfnamefont
  {S.}~\bibnamefont {Mukohyama}},\ }\bibfield  {title} {\bibinfo {title}
  {{Ho\v{r}ava-Lifshitz gravity with $\lambda\to\infty$}},\ }\href
  {https://doi.org/10.1103/PhysRevD.83.124033} {\bibfield  {journal} {\bibinfo
  {journal} {Phys. Rev. D}\ }\textbf {\bibinfo {volume} {83}},\ \bibinfo
  {pages} {124033} (\bibinfo {year} {2011})},\ \Eprint
  {https://arxiv.org/abs/1104.2087} {arXiv:1104.2087 [hep-th]} \BibitemShut
  {NoStop}%
\bibitem [{\citenamefont {Radkovski}\ and\ \citenamefont
  {Sibiryakov}(2023)}]{Radkovski:2023cew}%
  \BibitemOpen
  \bibfield  {author} {\bibinfo {author} {\bibfnamefont {J.~I.}\ \bibnamefont
  {Radkovski}}\ and\ \bibinfo {author} {\bibfnamefont {S.~M.}\ \bibnamefont
  {Sibiryakov}},\ }\bibfield  {title} {\bibinfo {title} {{Scattering amplitudes
  in high-energy limit of projectable Ho\v{r}ava gravity}},\ }\href
  {https://doi.org/10.1103/PhysRevD.108.046017} {\bibfield  {journal} {\bibinfo
   {journal} {Phys. Rev. D}\ }\textbf {\bibinfo {volume} {108}},\ \bibinfo
  {pages} {046017} (\bibinfo {year} {2023})},\ \Eprint
  {https://arxiv.org/abs/2306.00102} {arXiv:2306.00102 [hep-th]} \BibitemShut
  {NoStop}%
\bibitem [{\citenamefont {Buchberger}\ and\ \citenamefont
  {Kauers}(2010)}]{Buchberger:2010}%
  \BibitemOpen
  \bibfield  {author} {\bibinfo {author} {\bibfnamefont {B.}~\bibnamefont
  {Buchberger}}\ and\ \bibinfo {author} {\bibfnamefont {M.}~\bibnamefont
  {Kauers}},\ }\bibfield  {title} {\bibinfo {title} {{G}roebner basis},\ }\href
  {https://doi.org/10.4249/scholarpedia.7763} {\bibfield  {journal} {\bibinfo
  {journal} {Scholarpedia}\ }\textbf {\bibinfo {volume} {5}},\ \bibinfo {pages}
  {7763} (\bibinfo {year} {2010})}\BibitemShut {NoStop}%
\bibitem [{\citenamefont {Shafarevich}(1977)}]{Shafarevich}%
  \BibitemOpen
  \bibfield  {author} {\bibinfo {author} {\bibfnamefont {I.~R.}\ \bibnamefont
  {Shafarevich}},\ }\href {https://doi.org/10.1007/978-3-642-96200-4} {\emph
  {\bibinfo {title} {{Basic Algebraic Geometry}}}}\ (\bibinfo  {publisher}
  {Springer-Verlag Berlin},\ \bibinfo {address} {Heidelberg},\ \bibinfo {year}
  {1977})\BibitemShut {NoStop}%
\bibitem [{\citenamefont {{Wolfram Research{,} Inc.}}()}]{Mathematica}%
  \BibitemOpen
  \bibfield  {author} {\bibinfo {author} {\bibnamefont {{Wolfram Research{,}
  Inc.}}},\ }\href {https://www.wolfram.com/mathematica} {\bibinfo {title}
  {{Mathematica, {V}ersion 12.2}}},\ \bibinfo {note} {champaign, IL,
  2020}\BibitemShut {NoStop}%
\bibitem [{\citenamefont {Komargodski}\ and\ \citenamefont
  {Schwimmer}(2011)}]{Komargodski:2011sw}%
  \BibitemOpen
  \bibfield  {author} {\bibinfo {author} {\bibfnamefont {Z.}~\bibnamefont
  {Komargodski}}\ and\ \bibinfo {author} {\bibfnamefont {A.}~\bibnamefont
  {Schwimmer}},\ }\bibfield  {title} {\bibinfo {title} {On renormalization
  group flows in four dimensions},\ }\href
  {https://doi.org/10.1007/JHEP12(2011)099} {\bibfield  {journal} {\bibinfo
  {journal} {JHEP}\ }\textbf {\bibinfo {volume} {2011}}\bibfield  {number}
  {\bibinfo  {number} { (12)},\ \bibinfo {pages} {99}},\ }\Eprint
  {https://arxiv.org/abs/1107.3987} {arXiv:1107.3987 [hep-th]} \BibitemShut
  {NoStop}%
\bibitem [{\citenamefont {Luty}\ \emph {et~al.}(2013)\citenamefont {Luty},
  \citenamefont {Polchinski},\ and\ \citenamefont {Rattazzi}}]{Luty:2013aa}%
  \BibitemOpen
  \bibfield  {author} {\bibinfo {author} {\bibfnamefont {M.~A.}\ \bibnamefont
  {Luty}}, \bibinfo {author} {\bibfnamefont {J.}~\bibnamefont {Polchinski}},\
  and\ \bibinfo {author} {\bibfnamefont {R.}~\bibnamefont {Rattazzi}},\
  }\bibfield  {title} {\bibinfo {title} {The a-theorem and the asymptotics of
  4d quantum field theory},\ }\href {https://doi.org/10.1007/JHEP01(2013)152}
  {\bibfield  {journal} {\bibinfo  {journal} {JHEP}\ }\textbf {\bibinfo
  {volume} {2013}}\bibfield  {number} {\bibinfo  {number} { (1)},\ \bibinfo
  {pages} {152}},\ }\Eprint {https://arxiv.org/abs/1204.5221} {arXiv:1204.5221
  [hep-th]} \BibitemShut {NoStop}%
\bibitem [{\citenamefont {Mack}(1977)}]{Mack:1975je}%
  \BibitemOpen
  \bibfield  {author} {\bibinfo {author} {\bibfnamefont {G.}~\bibnamefont
  {Mack}},\ }\bibfield  {title} {\bibinfo {title} {{All unitary ray
  representations of the conformal group SU(2,2) with positive energy}},\
  }\href {https://doi.org/10.1007/BF01613145} {\bibfield  {journal} {\bibinfo
  {journal} {Commun. Math. Phys.}\ }\textbf {\bibinfo {volume} {55}},\ \bibinfo
  {pages} {1} (\bibinfo {year} {1977})}\BibitemShut {NoStop}%
\bibitem [{\citenamefont {Nishida}\ and\ \citenamefont
  {Son}(2012)}]{Nishida:2010tm}%
  \BibitemOpen
  \bibfield  {author} {\bibinfo {author} {\bibfnamefont {Y.}~\bibnamefont
  {Nishida}}\ and\ \bibinfo {author} {\bibfnamefont {D.~T.}\ \bibnamefont
  {Son}},\ }\bibfield  {title} {\bibinfo {title} {{Unitary Fermi gas, epsilon
  expansion, and nonrelativistic conformal field theories}},\ }\href
  {https://doi.org/10.1007/978-3-642-21978-8_7} {\bibfield  {journal} {\bibinfo
   {journal} {Lect. Notes Phys.}\ }\textbf {\bibinfo {volume} {836}},\ \bibinfo
  {pages} {233} (\bibinfo {year} {2012})},\ \Eprint
  {https://arxiv.org/abs/1004.3597} {arXiv:1004.3597 [cond-mat.quant-gas]}
  \BibitemShut {NoStop}%
\bibitem [{\citenamefont {Goldberger}\ \emph {et~al.}(2015)\citenamefont
  {Goldberger}, \citenamefont {Khandker},\ and\ \citenamefont
  {Prabhu}}]{Goldberger:2014hca}%
  \BibitemOpen
  \bibfield  {author} {\bibinfo {author} {\bibfnamefont {W.~D.}\ \bibnamefont
  {Goldberger}}, \bibinfo {author} {\bibfnamefont {Z.~U.}\ \bibnamefont
  {Khandker}},\ and\ \bibinfo {author} {\bibfnamefont {S.}~\bibnamefont
  {Prabhu}},\ }\bibfield  {title} {\bibinfo {title} {{OPE convergence in
  non-relativistic conformal field theories}},\ }\href
  {https://doi.org/10.1007/JHEP12(2015)048} {\bibfield  {journal} {\bibinfo
  {journal} {JHEP}\ }\textbf {\bibinfo {volume} {12}},\ \bibinfo {pages}
  {048}},\ \Eprint {https://arxiv.org/abs/1412.8507} {arXiv:1412.8507 [hep-th]}
  \BibitemShut {NoStop}%
\bibitem [{\citenamefont {Hogervorst}\ \emph {et~al.}(2016)\citenamefont
  {Hogervorst}, \citenamefont {Rychkov},\ and\ \citenamefont {van
  Rees}}]{Hogervorst:2015akt}%
  \BibitemOpen
  \bibfield  {author} {\bibinfo {author} {\bibfnamefont {M.}~\bibnamefont
  {Hogervorst}}, \bibinfo {author} {\bibfnamefont {S.}~\bibnamefont
  {Rychkov}},\ and\ \bibinfo {author} {\bibfnamefont {B.~C.}\ \bibnamefont {van
  Rees}},\ }\bibfield  {title} {\bibinfo {title} {{Unitarity violation at the
  Wilson-Fisher fixed point in 4-$\epsilon$ dimensions}},\ }\href
  {https://doi.org/10.1103/PhysRevD.93.125025} {\bibfield  {journal} {\bibinfo
  {journal} {Phys. Rev. D}\ }\textbf {\bibinfo {volume} {93}},\ \bibinfo
  {pages} {125025} (\bibinfo {year} {2016})},\ \Eprint
  {https://arxiv.org/abs/1512.00013} {arXiv:1512.00013 [hep-th]} \BibitemShut
  {NoStop}%
\bibitem [{\citenamefont {Gorbenko}\ \emph {et~al.}(2018)\citenamefont
  {Gorbenko}, \citenamefont {Rychkov},\ and\ \citenamefont
  {Zan}}]{Gorbenko:2018ncu}%
  \BibitemOpen
  \bibfield  {author} {\bibinfo {author} {\bibfnamefont {V.}~\bibnamefont
  {Gorbenko}}, \bibinfo {author} {\bibfnamefont {S.}~\bibnamefont {Rychkov}},\
  and\ \bibinfo {author} {\bibfnamefont {B.}~\bibnamefont {Zan}},\ }\bibfield
  {title} {\bibinfo {title} {{Walking, Weak first-order transitions, and
  Complex CFTs}},\ }\href {https://doi.org/10.1007/JHEP10(2018)108} {\bibfield
  {journal} {\bibinfo  {journal} {JHEP}\ }\textbf {\bibinfo {volume} {10}},\
  \bibinfo {pages} {108}},\ \Eprint {https://arxiv.org/abs/1807.11512}
  {arXiv:1807.11512 [hep-th]} \BibitemShut {NoStop}%
\bibitem [{\citenamefont {Gromov}\ \emph {et~al.}(2018)\citenamefont {Gromov},
  \citenamefont {Kazakov}, \citenamefont {Korchemsky}, \citenamefont {Negro},\
  and\ \citenamefont {Sizov}}]{Gromov:2017cja}%
  \BibitemOpen
  \bibfield  {author} {\bibinfo {author} {\bibfnamefont {N.}~\bibnamefont
  {Gromov}}, \bibinfo {author} {\bibfnamefont {V.}~\bibnamefont {Kazakov}},
  \bibinfo {author} {\bibfnamefont {G.}~\bibnamefont {Korchemsky}}, \bibinfo
  {author} {\bibfnamefont {S.}~\bibnamefont {Negro}},\ and\ \bibinfo {author}
  {\bibfnamefont {G.}~\bibnamefont {Sizov}},\ }\bibfield  {title} {\bibinfo
  {title} {{Integrability of Conformal Fishnet Theory}},\ }\href
  {https://doi.org/10.1007/JHEP01(2018)095} {\bibfield  {journal} {\bibinfo
  {journal} {JHEP}\ }\textbf {\bibinfo {volume} {01}},\ \bibinfo {pages}
  {095}},\ \Eprint {https://arxiv.org/abs/1706.04167} {arXiv:1706.04167
  [hep-th]} \BibitemShut {NoStop}%
\bibitem [{\citenamefont {Jepsen}\ \emph {et~al.}(2021)\citenamefont {Jepsen},
  \citenamefont {Klebanov},\ and\ \citenamefont {Popov}}]{Jepsen:2020czw}%
  \BibitemOpen
  \bibfield  {author} {\bibinfo {author} {\bibfnamefont {C.~B.}\ \bibnamefont
  {Jepsen}}, \bibinfo {author} {\bibfnamefont {I.~R.}\ \bibnamefont
  {Klebanov}},\ and\ \bibinfo {author} {\bibfnamefont {F.~K.}\ \bibnamefont
  {Popov}},\ }\bibfield  {title} {\bibinfo {title} {{RG limit cycles and
  unconventional fixed points in perturbative QFT}},\ }\href
  {https://doi.org/10.1103/PhysRevD.103.046015} {\bibfield  {journal} {\bibinfo
   {journal} {Phys. Rev. D}\ }\textbf {\bibinfo {volume} {103}},\ \bibinfo
  {pages} {046015} (\bibinfo {year} {2021})},\ \Eprint
  {https://arxiv.org/abs/2010.15133} {arXiv:2010.15133 [hep-th]} \BibitemShut
  {NoStop}%
\end{thebibliography}%

\end{document}